\documentclass[12pt,letterpaper]{article}
\usepackage{jheppub}
\pdfoutput=1

\usepackage{amsmath, amsthm, amsfonts, amssymb}
\usepackage{graphicx}

\newcommand\Backlund {B\"{a}cklund }
\newcommand\Schrodinger {Schr\"{o}dinger }

\newcommand\Poincare {Poincar\'e\ }
\newcommand\Lame {Lam\'e\ }

\newcommand\csch {\mathrm{csch}}
\newcommand\diag {\mathrm{diag}}

\title{Static Elliptic Minimal Surfaces in AdS$_4$}
\author[a,b]{Georgios Pastras}
\affiliation[a]{Department of Physics, School of Applied Mathematics and Physical Sciences, National Technical University, Athens 15780, Greece}
\affiliation[b]{NCSR ``Demokritos'', Institute of Nuclear and Particle Physics
15310 Aghia Paraskevi, Attiki, Greece}
\emailAdd{pastras@mail.ntua.gr}
\emailAdd{pastras@inp.demokritos.gr}

\abstract{The Ryu-Takayanagi conjecture connects the entanglement entropy in the boundary CFT to the area of open co-dimension two minimal surfaces in the bulk. Especially in AdS$_4$, the latter are two-dimensional surfaces, and, thus, solutions of a Euclidean non-linear sigma model on a symmetric target space that can be reduced to an integrable system via Pohlmeyer reduction. In this work, we invert Pohlmeyer reduction to construct static minimal surfaces in AdS$_4$ that correspond to elliptic solutions of the reduced system, namely the cosh-Gordon equation. The constructed minimal surfaces comprise a two-parameter family of surfaces that include helicoids and catenoids in H$^3$ as special limits. Minimal surfaces that correspond to identical boundary conditions are discovered within the constructed family of surfaces and the relevant geometric phase transitions are studied.}

\arxivnumber{}

\dedicated{In memoriam and gratitude of Prof. Ioannis Bakas}

\begin{document}

\thispagestyle{empty}

\maketitle

\setcounter{footnote}{0}

\def\thefootnote{\arabic{footnote}}

\section{Introduction}

The AdS/CFT correspondence \cite{Maldacena:1997re,Gubser:1998bc,Witten:1998qj} is a framework that connects theories which include gravitational dynamics in spacetimes with AdS asymptotics to conformal field theories defined on the AdS boundary. The similarity of the laws of black hole physics to those of thermodynamics suggests that the emergence of gravity in the bulk theory incorporates thermodynamics of some underlying degrees of freedom \cite{Verlinde:2010hp}, which in the context of AdS/CFT correspondence can be naturally selected to be the degrees of freedom of the boundary theory. Then, gravity can be understood as an emergent entropic force originating from the strongly coupled CFT degrees of freedom.

The more recent point of view of gravity as an emergent entropic force suggests that gravity is not related with thermal statistics but rather with quantum entanglement statistics. The original proposal was made by Ryu and Takayanagi \cite{Ryu:2006bv,Ryu:2006ef} and its basic element is the RT conjecture. This states that the entanglement entropy of a subsystem of the degrees of freedom of the boundary CFT is mapped through the holographic correspondence to the area of an open minimal co-dimension two surface $\left(A^{{\rm{extr}}} \right)$ in the bulk geometry, anchored to the entangling surface in the boundary, i.e. the surface separating the subsystem from its environment,
\begin{equation}
{S_{EE}} = {\frac{1}{4{G_N}}} \, {\rm Area} \left(A^{{\rm{extr}}} \right) \, .
\label{eq:RT_conjecture}
\end{equation}

This program has advanced a lot ever since \cite{Hubeny:2007xt,Nishioka:2009un,VanRaamsdonk:2009ar,VanRaamsdonk:2010pw,Takayanagi:2012kg,Casini:2011kv}, including an understanding of Einstein equations at linear order as directly emerging from the first law of entanglement thermodynamics \cite{Lashkari:2013koa,Faulkner:2013ica,Bakas:2015opa}. A major difficulty in these calculations is the specification of the minimal surface for an arbitrary entangling surface, which arises from the non-linearity of the relevant equations. More specifically, closed forms for the minimal surface in more than three spacetime dimensions are known only for the case that the bulk geometry is pure AdS and furthermore the entangling surface encloses a region with the shape of a disk or an infinite strip.

Both the disk and strip minimal surfaces are anchored to entangling surfaces characterized by trivial curvature. Furthermore, disk minimal surfaces are special in the sense that they have vanishing Gaussian curvature. The discovery of more general minimal surfaces, apart being interesting from a purely mathematical point of view, can provide a useful tool for the study of holographic entanglement entropy and its dependence on the geometry of the entangling surface. It can also provide a non-trivial check of the connection between Einstein equations and entanglement thermodynamics through the RT conjecture.

In the special case of AdS$_4$, the co-dimension two minimal surfaces that are related to the entanglement entropy through the RT conjecture are two-dimensional surfaces. Consequently, their area can be considered as the Euclidean analogue of the Nambu-Goto action describing strings propagating in AdS$_4$. As such, minimal surfaces correspond to solutions of a Euclidean non-linear sigma model (NLSM) in a symmetric space that can be reduced to an integrable Hamiltonian system through Pohlmeyer reduction.

The oldest known reduction of this kind is the correspondence of the O$(3)$ NLSM to the sine-Gordon equation \cite{Pohlmeyer:1975nb,Zakharov:1973pp}, which is now known to be generalizable to sigma models defined on any symmetric space, such as O$(N)$ sigma models \cite{Eichenherr:1979yw,Pohlmeyer:1979ch} and CP$(N)$ models \cite{Eichenherr:1979uk}. Although Pohlmeyer reduction incorporates a non-local connection between the degrees of freedom of the initial sigma model and the reduced integrable system, it can be shown that the dynamics of the reduced system emerge from a local, systematically derivable Lagrangian density \cite{Bakas:1993xh,Bakas:1995bm,FernandezPousa:1996hi,Miramontes:2008wt}. Pohlmeyer reduction has been extended to sigma models describing strings propagating in symmetric spacetime geometries \cite{Barbashov:1980kz,DeVega:1992xc,Larsen:1996gn}, including spacetimes particularly interesting in the context of holography, such as AdS$_5 \times$S$^5$ \cite{Grigoriev:2007bu,Mikhailov:2007xr,Grigoriev:2008jq} or AdS$_4 \times$CP$^3$ \cite{Rashkov:2008rm}.

Although the integrability of the reduced system can be used to derive several of its solutions, the non-locality of the relation between the original and the reduced degrees of freedom incommodes the inversion of the reduction and thus, the use of Pohlmeyer reduction for the discovery of solutions of the original sigma model. In recent literature, a method has been developed for the inversion of Pohlmeyer reduction in the specific case of elliptic solutions of the reduced system, leading to the construction of a class of classical string solutions in AdS$_3$ and dS$_3$ \cite{Bakas:2016jxp}. This class of string solutions includes the known family of spiky string solutions in AdS$_3$ \cite{Kruczenski:2004wg}, as well as several new ones. In this paper, we exploit these techniques to construct new static minimal surfaces in AdS$_4$, corresponding to elliptic solutions of the reduced system.

In section \ref{sec:Pohlmeyer}, we review the Pohlmeyer reduction of minimal surfaces in AdS$_4$, as well as the limits of the reduced integrable system for planar and static minimal surfaces. In section \ref{sec:Elliptic}, we study the elliptic solutions of the reduced system, the emergence of effective \Schrodinger problems in the process of inverting Pohlmeyer reduction and the construction of the minimal surfaces adopting the techniques of \cite{Bakas:2016jxp}. In section \ref{sec:Surfaces}, we study basic properties of the derived minimal surfaces, interesting limits of them, as well as their area and consequently the corresponding entanglement entropy. In section \ref{sec:Phase_Transitions}, we study the global stability of the elliptic minimal surfaces and possible geometric phase transitions between them. Finally, there is an appendix including useful properties of Weierstrass functions that are used throughout the text.

\section{Polhmayer Reduction of Minimal Surfaces in AdS$_4$}
\label{sec:Pohlmeyer}

Pohlmeyer reduction relates in a non-local way the action of a NLSM defined in a symmetric target space to integrable systems of the family of the sine-Gorgon equation. In this section, following the literature \cite{Larsen:1996gn,Rashkov:2008rm,Bakas:2016jxp}, we review the Pohlmeyer reduction of space-like minimal surfaces in AdS$_4$, resulting in a two-component integrable system, which in the specific case of static minimal surfaces it is reduced to the cosh-Gordon equation.

Pohlmeyer reduction of NLSMs defined on a symmetric target space is based on the study of the embedding of the two-dimensional NLSM solution into the symmetric target space, in turn embedded into an enhanced higher-dimensional flat space. The AdS$_4$ can be implemented as a submanifold in an enhanced five-dimensional flat space with an extra time-like dimension, i.e. $\mathbb{R}^{(2,3)}$. We denote the coordinates in this enhanced space as $Y^{-1}$, $Y^0$, $Y^1$, $Y^2$ and $Y^3$. Then, AdS$_4$ is the submanifold
\begin{equation}
Y \cdot Y = - \Lambda^2.
\label{eq:Pohlmeyer_submanifold}
\end{equation}

Furthermore, in the following we will use the notation
\begin{equation}
{A^\mu }{B_\mu } \equiv A \cdot B ,
\end{equation}
where $g_{\mu\nu} = \diag \{-1, -1, +1, +1, +1 \}$.

\subsection{Action, Equations of Motion and Virasoro Conditions}

A two-dimensional surface in AdS$_4$ can be parametrized with two space-like parameters $\sigma_1$ and $\sigma_2$. Introducing an auxiliary metric $\gamma$, the area that can be written in the form of a Polyakov action as
\begin{equation}
A = \frac{1}{2}\int {d{\sigma _1}d{\sigma _2}\sqrt \gamma \left( {\gamma ^{ab}}{\partial _a}Y \cdot {\partial _b}Y + \lambda \left( {Y \cdot Y + {\Lambda ^2}} \right) \right) } .
\end{equation}
Selecting the conformal gauge ${\gamma _{ab}} = {e^\omega }{\delta _{ab}}$ and introducing the complex coordinate $z = \left( {{\sigma _1} + i{\sigma _2}} \right) / 2$, the action is written as
\begin{equation}
A = \int {dzd\bar z \left( \partial Y \cdot \bar \partial Y + \lambda \left( {Y \cdot Y + {\Lambda ^2}} \right) \right)} . 
\label{eq:reduction_area}
\end{equation}

The equations of motion for the fields $Y$ are
\begin{equation}
\partial \bar \partial Y = \lambda Y,
\label{eq:reduction_eom}
\end{equation}
while the equation of motion for the Lagrange multiplier $\lambda$ is the geometric constraint \eqref{eq:Pohlmeyer_submanifold}. We can eliminate the Lagrange multiplier $\lambda$ from the equations of motion of the fields $Y$ \eqref{eq:reduction_eom} using the geometric constraint. We find
\begin{equation}
\partial \bar \partial Y = \frac{1}{{{\Lambda ^2}}}\left( {\partial Y \cdot \bar \partial Y} \right)Y .
\end{equation}

The stress-energy tensor takes the form
\begin{align}
{T_{zz}} &= \partial Y \cdot \partial Y ,\\
{T_{\bar z\bar z}} &= \bar \partial Y \cdot \bar \partial Y ,\\
{T_{z\bar z}} &= 0 .
\end{align}
Thus, the Virasoro constraints take the form
\begin{equation}
\partial Y \cdot \partial Y = 0 .
\label{eq:Pohlmeyer_Virasoro}
\end{equation}

\subsection{The Reduced Integrable System}
We would like to introduce a basis in the enhanced five-dimensional flat space, which includes the vectors $Y$, $\partial Y$ and $\bar \partial Y$. For the purposes of this section, we will name these vectors as $v_1$, $v_2$ and $v_3$ and introduce two more vectors $v_4$ and $v_5$ to form the basis in $R^{(2,3)}$,
\begin{equation}
{v_i} = \left\{ {Y,\partial Y,\bar \partial Y,{v_4},{v_5}} \right\} .
\end{equation}
The vectors $\partial Y$ and $\bar \partial Y$ span the tangent space of the two-dimensional surface we study. Although Virasoro conditions suggest that they are both null, this is due to the fact that they are actually complex. Since we study a space-like surface, the tangent space contains two space-like directions and consequently real linear combinations of $\partial Y$ and $\bar \partial Y$ will be space-like. It follows that one of the vectors $v_4$ and $v_5$ has to be space-like and the other time-like, as the basis should contain two time-like and three space-like vectors and furthermore $v_1$ is time-like as implied by the geometric constraint \eqref{eq:Pohlmeyer_submanifold}. We choose $v_4$ to be space-like and $v_5$ to be time-like and we demand that $v_4$ has constant norm equal to one, $v_5$ has constant norm equal to minus one and they are both orthogonal to $v_1$, $v_2$ and $v_3$ and to each other,
\begin{align}
{v_4} \cdot {v_5} &= {v_{4/5}} \cdot Y = {v_{4/5}} \cdot \partial Y = {v_{4/5}} \cdot \bar \partial Y = 0 , \\
{v_4} \cdot {v_4} &= 1,\quad {v_5} \cdot {v_5} =  - 1 .
\end{align}

We define the reduced field $a$ as
\begin{equation}
{e^a} : = \partial Y \cdot \bar \partial Y ,
\label{eq:Pohlmeyer_field_definition}
\end{equation}
since $\partial Y \cdot \bar \partial Y$ is a positive quantity.

The form of the inner products of the basis vectors implies that a general vector $X$ in the enhanced space can be decomposed in the basis $v_i$ as
\begin{equation}
X = - \frac{1}{{{\Lambda ^2}}}\left( {X \cdot {v_1}} \right){v_1} + {e^{ - a}}\left( {X \cdot {v_3}} \right){v_2} + {e^{ - a}}\left( {X \cdot {v_2}} \right){v_3} + \left( {X \cdot {v_4}} \right){v_4} - \left( {X \cdot {v_5}} \right){v_5} .
\end{equation}

We would like to decompose the rate of change of the basis vectors with the complex coordinate $z$ in the basis $v_i$ itself. We do so defining the complex $5\times 5$ matrix $A$ as
\begin{equation}
\partial {v_i} = {A_{ij}}{v_j} .
\end{equation}

By definition we have $\partial {v_1} = {v_2}$, while the equations of motions for the field $Y$ imply that
\begin{equation}
\partial {v_3} = \partial \bar \partial Y = \frac{1}{{{\Lambda ^2}}}{e^a} Y = \frac{1}{{{\Lambda ^2}}}{e^a} {v_1}.
\end{equation}
For the derivative of $v_2$, in general we have
\begin{equation}
\partial {v_2} = {\partial ^2}Y = {a_0}{v_1} + {a_ + }{v_1} + {a_ - }{v_2} + {a_4}{v_4} + {a_5}{v_5} ,
\end{equation}
The geometric constraint \eqref{eq:Pohlmeyer_submanifold} upon differentiation yields $\partial Y \cdot Y = 0$, while upon another differentiation and the use of the Virasoro constraints \eqref{eq:Pohlmeyer_Virasoro}, we get ${\partial ^2}Y \cdot Y = 0$. The latter implies that ${a_0} = 0$. Differentiation of the Virasoro constraint \eqref{eq:Pohlmeyer_Virasoro} yields ${\partial ^2}Y \cdot \partial Y = 0$, implying that $ {a_ - } = 0$. Finally, differentiating the definition of the reduced field $a$ \eqref{eq:Pohlmeyer_field_definition} and using the equations of motion and the fact that $Y$ is orthogonal to its derivative, we get ${\partial ^2}Y \cdot \bar \partial Y = \partial a{e^a} $,
implying that ${a_ + } = {\partial}a $. Summing up, we have shown that
\begin{equation}
\partial {v_2} = \partial a{v_2} + {a_4}{v_4} + {a_5}{v_5} .
\end{equation}

Finally, the orthogonality conditions for the vectors $v_4$ and $v_5$ yield their derivatives as follows
\begin{align*}
{v_{4/5}} \cdot {v_{4/5}} &=  \pm 1 \Rightarrow \partial {v_4} \cdot {v_4} = \partial {v_5} \cdot {v_5} = 0 ,\\
{v_4} \cdot {v_5} &= 0 \Rightarrow \partial {v_4} \cdot {v_5} =  - \partial {v_5} \cdot {v_4} \equiv f ,\\
{v_{4/5}} \cdot Y &= 0 \Rightarrow \partial {v_{4/5}} \cdot Y =  - {v_{4/5}} \cdot \partial Y = 0 ,\\
{v_{4/5}} \cdot \partial Y &= 0 \Rightarrow \partial {v_{4/5}} \cdot \partial Y =  - {v_{4/5}} \cdot {\partial ^2}Y =  \mp {a_{4/5}} ,\\
{v_{4/5}} \cdot \bar \partial Y &= 0 \Rightarrow \partial {v_{4/5}} \cdot \bar \partial Y =  - {v_{4/5}} \cdot \partial \bar \partial Y = 0 .
\end{align*}
Putting everything together, the derivatives of $v_4$ and $v_5$ equal
\begin{align}
\partial {v_4} &=  - {a_4}{e^{ - a}}{v_3} - f{v_5} ,\\
\partial {v_5} &= {a_5}{e^{ - a}}{v_3} - f{v_4} .
\end{align}

It is important to notice that $\bar \partial {v_i} \ne {{\bar A}_{ij}}{v_j} $, since the basis vectors $v_2$ and $v_3$ are not real but rather they are complex conjugates of each other. As a result, it is true that
\begin{equation}
\bar \partial {v_i} = {{\tilde A}_{ij}}{v_j} ,
\end{equation}
where ${\tilde A}$ is the complex conjugate of the matrix that is produced after the interchange of the second and third lines and rows of $A$. Thus, the matrices $A$ and ${\tilde A}$ are equal to
\begin{equation}
A = \left( {\begin{array}{*{20}{c}}
0&1&0&0&0\\
0&{\partial a}&0&{{a_4}}&{{a_5}}\\
{ \frac{1}{{{\Lambda ^2}}}{e^a}}&0&0&0&0\\
0&0&{ - {a_4}{e^{ - a}}}&0&{ - f}\\
0&0&{{a_5}{e^{ - a}}}&{ - f}&0
\end{array}} \right) , \quad {\tilde A} = \left( {\begin{array}{*{20}{c}}
0&0&1&0&0\\
{ \frac{1}{{{\Lambda ^2}}}{e^a}}&0&0&0&0\\
0&0&{\bar \partial a}&{{{\bar a}_4}}&{{{\bar a}_5}}\\
0&{ - {{\bar a}_4}{e^{ - a}}}&0&0&{ - \bar f}\\
0&{{{\bar a}_5}{e^{ - a}}}&0&{ - \bar f}&0
\end{array}} \right) .
\end{equation}
The above matrices have to obey the compatibility condition
\begin{equation}
\bar \partial \left( {{A_{ij}}{e_j}} \right) = \partial \left( {{\tilde A}{_{ij}}{e_j}} \right) \Rightarrow \left( {\bar \partial {A_{ij}}} \right){e_j} + {A_{ik}}{{\tilde A}_{kj}}{e_j} = \left( {\partial {{\tilde A}_{ij}}} \right){e_j} + {{\tilde A}_{ik}}{A_{kj}}{e_j} ,
\end{equation}
which in matrix form can be written as the zero-curvature condition
\begin{equation}
\bar \partial A - \partial {\tilde A} + \left[ {A,{\tilde A}} \right] = 0 .
\end{equation}

It is a matter of algebra to show that the zero-curvature condition implies the equations
\begin{align}
\partial \bar \partial a &= \left( {{{\left| {{a_4}} \right|}^2} - {{\left| {{a_5}} \right|}^2}} \right){e^{ - a}} + \frac{1}{{{\Lambda ^2}}}{e^a} , \label{eq:spacelike_zcc_1}\\
\partial \bar f - \bar \partial f &= {e^{ - a}}\left( {{a_4}{{\bar a}_5} - {{\bar a}_4}{a_5}} \right) , \label{eq:spacelike_zcc_2}\\
\bar \partial {a_4} &= {a_5}\bar f , \label{eq:spacelike_zcc_3}\\
\bar \partial {a_5} &= {a_4}\bar f . \label{eq:spacelike_zcc_4}
\end{align}
Equations \eqref{eq:spacelike_zcc_3} and \eqref{eq:spacelike_zcc_4} yield
\begin{equation}
\bar \partial \left( {a_4^2 - a_5^2} \right) = 0 ,
\end{equation}
which allows the following two inequivalent parametrizations of $a_4$ and $a_5$,
\begin{equation}
\begin{aligned}
{a_4} &= g\left( z \right)\cosh \theta \left( {z,\bar z} \right) ,\\
{a_5} &= g\left( z \right)\sinh \theta \left( {z,\bar z} \right) ,
\end{aligned}  \quad \mathrm{or} \quad  
\begin{aligned}
{a_4} &= g\left( z \right)\sinh \theta \left( {z,\bar z} \right) ,\\
{a_5} &= g\left( z \right)\cosh \theta \left( {z,\bar z} \right) .
\end{aligned}
\end{equation}
The relative magnitude of $a_4$ and $a_5$ determines which parametrization is appropriate. In both cases,
\begin{equation}
\bar f = \frac{{\bar \partial {a_4}}}{{{a_5}}} = \bar \partial \theta .
\end{equation}
Then equations \eqref{eq:spacelike_zcc_1} and \eqref{eq:spacelike_zcc_2} take the form
\begin{align}
\partial a &= \pm {\left| {g\left( z \right)} \right|^2}\cosh \left( {\theta  - \bar \theta } \right){e^{ - a}} + \frac{1}{{{\Lambda ^2}}}{e^a} ,
\label{eq:spacelike_equation_1a} \\
\partial \bar \partial \left( {\theta  - \bar \theta } \right) &= \mp g\left( z \right)\bar g\left( {\bar z} \right)\sinh \left( {\theta  - \bar \theta } \right){e^{ - a}} , 
\label{eq:spacelike_equation_2a}
\end{align}
where the sign depends on the relative magnitude of $a_4$ and $a_5$. We define the fields $\alpha$ and $\beta$ as
\begin{align}
\alpha  &:= a - \ln \left( {\Lambda \left| {g\left( z \right)} \right|} \right) , \label{eq:reduction_a_redef}\\
\beta  &:= \frac{1}{2}{\mathop{\rm Im}\nolimits} \theta  \label{eq:reduction_b_redef}
\end{align}
and furthermore we define the complex coordinate $z' = z'\left( z \right)$ so that
\begin{equation}
\frac{{dz'}}{{dz}} = \sqrt {\Lambda g\left( z \right)} .
\label{eq:reduction_z_redef}
\end{equation}
Then, the reduced equations take the form
\begin{align}
\partial \bar \partial \alpha  &= \frac{1}{{{\Lambda ^2}}}\left( {\pm \cos \beta {e^{ - \alpha }} + {e^\alpha }} \right) , \\
\partial \bar \partial \beta  &= \mp \frac{1}{{{\Lambda ^2}}}\sin \beta {e^{ - \alpha }} .
\end{align}
The above equations are derivable from the Lagrangian density
\begin{equation}
\mathcal{L} = \frac{1}{2}\partial \alpha \bar \partial \alpha  - \frac{1}{2}\partial \beta \bar \partial \beta  + \frac{1}{{{\Lambda ^2}}}\left( { \mp \cos \beta {e^{ - \alpha }} + {e^\alpha }} \right) .
\end{equation}

Finally, none of the above parametrizations can describe the special case, where $a_4$ and $a_5$ have the same magnitude. In this case, we should parametrize $a_4$ and $a_5$ as
\begin{align}
{a_4} &= g\left( z \right){{\mathop{\rm e}\nolimits} ^{i\beta \left( {z,\bar z} \right)}} , \\
{a_5} &= g\left( z \right){{\mathop{\rm e}\nolimits} ^{ - i\beta \left( {z,\bar z} \right)}} 
\end{align}
and it is straightforward to show that equations \eqref{eq:spacelike_zcc_1} and \eqref{eq:spacelike_zcc_2} are directly written as
\begin{align}
\partial \bar \partial \alpha &= \frac{1}{{{\Lambda ^2}}}{e^a} ,\\
\partial \bar \partial \beta  &= 0 ,
\end{align}
which are derivable from the Lagrangian density
\begin{equation}
\mathcal{L} = \frac{1}{2}\partial \alpha \bar \partial \alpha  - \frac{1}{2}\partial \beta \bar \partial \beta + \frac{1}{{{\Lambda ^2}}}{e^\alpha } .
\end{equation}

\subsection{The Area of the Minimal Surface in the Reduced Problem}
The area of the minimal surface is given by the action \eqref{eq:reduction_area}, as the integral of the conformal factor $e^a$.
\begin{equation}
A = \int {d{z}d{\bar z}{e^a}}.
\end{equation}
However, the reduced integrable system equations are expressed in terms of the fields $\alpha$ and $\beta$ defined in equations \eqref{eq:reduction_a_redef} and \eqref{eq:reduction_b_redef} and the coordinate $z'$ defined in \eqref{eq:reduction_z_redef}. We are interested in acquiring a simple expression for the area of the minimal surface, in terms of the reduced degrees of freedom to facilitate the calculation of the area of the constructed minimal surfaces later. It turns out that the redefinitions \eqref{eq:reduction_a_redef}, \eqref{eq:reduction_b_redef} and \eqref{eq:reduction_z_redef} are nothing more than a conformal transformation, thus, they leave the expression for the area invariant,
\begin{equation}
A = \int {d{z}d{\bar z}{e^a}}  = \int {\frac{{d{z}'}}{{\sqrt {\Lambda f\left( {{z}} \right)} }}\frac{{d{\bar z}'}}{{\sqrt {\Lambda g\left( {{\bar z}} \right)} }}\Lambda \sqrt {f\left( {{z}} \right)g\left( {{\bar z}} \right)} {e^\alpha }}  = \int {d{z}'d{\bar z}'{e^\alpha }}  .
\end{equation}
Defining the real and imaginary parts of the complex number $z'$ as $z' = \left( {u + iv} \right)/2$, the area formula takes the form
\begin{equation}
A = \frac{1}{2}\int {dudv{e^\alpha }} .
\label{eq:reduction_area_formula}
\end{equation}

\subsection{Restricting to AdS$_3$ or H$^3$}
We may restrict our attention to ``flat'' space-like surfaces constrained in the $Y_3 = 0$ plane, i.e. minimal surfaces in AdS$_3$. Such surfaces correspond to $a_4 = 0$ implying that ${a_5} = g\left( z \right)$. After the appropriate redefinition of the fields and the complex coordinate, we result in the equation
\begin{equation}
\partial \bar \partial \alpha  = \frac{2}{{{\Lambda ^2}}} \sinh \alpha ,
\end{equation}
which is the Euclidean sinh-Gordon equation.

Similarly, we may restrict our attention to static surfaces in AdS$_4$, i.e. minimal surfaces in the hyperboloid H$^3$. In this case, we have $a_5 = 0$ implying that ${a_4} = g\left( z \right)$ yielding the equation
\begin{equation}
\partial \bar \partial \alpha  = \frac{2}{{{\Lambda ^2}}} \cosh \alpha ,
\label{eq:Pohlmeyer_cosh_eq}
\end{equation}
which is the Euclidean cosh-Gordon equation.

For both ``flat'' and static minimal surfaces, the special case $g\left( z \right) = 0$ results in the Euclidean Liouville equation
\begin{equation}
\partial \bar \partial \alpha  = \frac{1}{{{\Lambda ^2}}}{e^\alpha }.
\end{equation}

\section{Static Elliptic Minimal Surfaces}
\label{sec:Elliptic}

In this section, we restrict our attention to static minimal surfaces in AdS$_4$, which are mapped through Pohlmeyer reduction to solutions of equation \eqref{eq:Pohlmeyer_cosh_eq}, namely the Euclidean cosh-Gordon equation, as shown in section \ref{sec:Pohlmeyer}. Furthermore, we focus on a specific class of solutions of the cosh-Gordon equation, having the property that they depend on either the real or the imaginary part of the complex coordinate. It turns out that such solutions can be expressed in terms of elliptic functions. For these elliptic solutions, it is possible to invert Pohlmeyer reduction and find analytic expressions for the corresponding minimal surfaces. The derivation closely follows \cite{Bakas:2016jxp}, which applies similar techniques for the construction of classical string solutions, so the reader is encouraged to recur there for more details. 

It has to be noticed that it is not possible to find non-trivial solutions of the reduced system using \Backlund transformations. Although the cosh-Gordon equation possesses \Backlund transformations similar to those of the sinh-Gordon, it lacks a vacuum solution to serve as the seed solution.

\subsection{Elliptic Solutions of the Cosh-Gordon Equation}
\label{subsec:Elliptic}

The Pohlmeyer reduced system equation of interest \eqref{eq:Pohlmeyer_cosh_eq}, i.e. the Euclidean cosh-Gordon equation can be expressed in terms of the real and imaginary parts of the complex coordinate $z' = \left( {u + iv} \right)/2$ as
\begin{equation}
\frac{{{\partial ^2}\alpha }}{{\partial {u^2}}} + \frac{{{\partial ^2}\alpha }}{{\partial {v^2}}} = \frac{2}{{{\Lambda ^2}}}\cosh \alpha .
\label{eq:elliptic_cosh_equation}
\end{equation}
We restrict our attention to solutions of \eqref{eq:elliptic_cosh_equation} that depend solely on either $u$ or $v$. Without loss of generality, we assume that they depend on $u$,
\begin{equation}
\alpha \left( {u,v} \right) = \alpha \left( u \right) .
\end{equation}
Then, the Euclidean cosh-Gordon equation reduces to the following ordinary differential equation,
\begin{equation}
\frac{{{d^2}\alpha }}{{d{u^2}}} = \frac{2}{{{\Lambda ^2}}}\cosh \alpha ,
\end{equation}
which can be easily integrated once to yield
\begin{equation}
\frac{1}{2}{\left( {\frac{{d\alpha }}{{du}}} \right)^2} - \frac{2}{{{\Lambda ^2}}}\sinh \alpha = E .
\label{eq:Elliptic_energy_conservation}
\end{equation}
Defining
\begin{equation}
{e^\alpha} = 2{\Lambda ^2}\left( {y - \frac{E}{6}} \right) ,
\label{eq:Elliptic_ydef}
\end{equation}
equation \eqref{eq:Elliptic_energy_conservation} takes the form
\begin{equation}
{\left( {\frac{{dy}}{{du}}} \right)^2} = 4{y^3} - \left( {\frac{{{E^2}}}{3} + \frac{1}{{{\Lambda ^4}}}} \right)y + \frac{E}{3}\left( {\frac{{{E^2}}}{9} + \frac{1}{{2{\Lambda ^4}}}} \right) ,
\end{equation}
which is the standard form of the Weierstrass equation
\begin{equation}
{\left( {\frac{{dy}}{{dz}}} \right)^2} = 4{y^3} - {g_2}y - {g_3} ,
\label{eq:Elliptic_Weierstrass_equation}
\end{equation}
where the moduli $g_2$ and $g_3$ take the values
\begin{align}
{g_2} &= \frac{{{E^2}}}{3} + \frac{1}{{{\Lambda ^4}}} , \label{eq:Elliptic_g2}\\
{g_3} &=  - \frac{E}{3}\left( {\frac{{{E^2}}}{9} + \frac{1}{{2{\Lambda ^4}}}} \right) . \label{eq:Elliptic_g3}
\end{align}

In the complex domain, equation \eqref{eq:Elliptic_Weierstrass_equation} is solved by the Weierstrass elliptic function $\wp \left( z ; g_2 , g_3 \right)$. Several of its properties depend on the reality of the roots $e_{1,2,3}$ of the polynomial
\begin{equation}
Q \left( z \right) = 4 z^3 - g_2 z -g_3.
\label{eq:Elliptic_polynomial}
\end{equation}
In our case, $g_2$ and $g_3$ are not arbitrary, but they are given by the expressions \eqref{eq:Elliptic_g2} and \eqref{eq:Elliptic_g3}, which imply that all three roots are real independently of the value of the integration constant $E$. It turns out that the roots are given by simple expressions, which read
\begin{align}
{e_1} &=  - \frac{E}{{12}} + \frac{1}{4}\sqrt {{E^2} + \frac{4}{{{\Lambda ^4}}}} , \label{eq:Elliptic_e1}\\
{e_2} &= \frac{E}{6} , \label{eq:Elliptic_e2}\\
{e_3} &=  - \frac{E}{{12}} - \frac{1}{4}\sqrt {{E^2} + \frac{4}{{{\Lambda ^4}}}} .\label{eq:Elliptic_e3}
\end{align}
The three roots are defined so that $e_1 > e_2 > e_3$.

Since the polynomial \eqref{eq:Elliptic_polynomial} has three real roots, the fundamental periods of Weierstrass elliptic function $\wp \left( z ; g_2 , g_3 \right)$ are a real one $2 \omega_1$ and a purely imaginary one $2 \omega_2$, given by equations \eqref{eq:Weierstrass_period_wp}. Furthermore, in this case, equation \eqref{eq:Elliptic_Weierstrass_equation} has two distinct real solutions in the real domain, namely,
\begin{align}
y &= \wp \left( {x;{g_2},{g_3}} \right) , \label{eq:elliptic_unbound_sol}\\
y &= \wp \left( {x + {\omega _2};{g_2},{g_3}} \right) . \label{eq:elliptic_bound_sol}
\end{align}
Both solutions are periodic with period equal to $2 \omega_1$. Solution \eqref{eq:elliptic_unbound_sol} ranges from $e_1$ to infinity, while the second one \eqref{eq:elliptic_bound_sol} ranges between $e_3$ and $e_2$. Equations \eqref{eq:Elliptic_ydef} and \eqref{eq:Elliptic_e2} imply that the bounded solution \eqref{eq:elliptic_bound_sol} does not correspond to a real solution for the reduced field $\alpha$ and consequently, the only acceptable solution for the Pohlmeyer reduced field $\alpha$ that depends solely on variable $u$ is
\begin{equation}
\alpha = \ln \left[ {2{\Lambda ^2}\left( {\wp \left( {u;{g_2},{g_3}} \right) - {e_2}} \right)} \right] ,
\label{eq:Elliptic_asols}
\end{equation}
where $g_2$ and $g_3$ are given by equations \eqref{eq:Elliptic_g2} and \eqref{eq:Elliptic_g3} and $e_2$ is given by equation \eqref{eq:Elliptic_e2}.

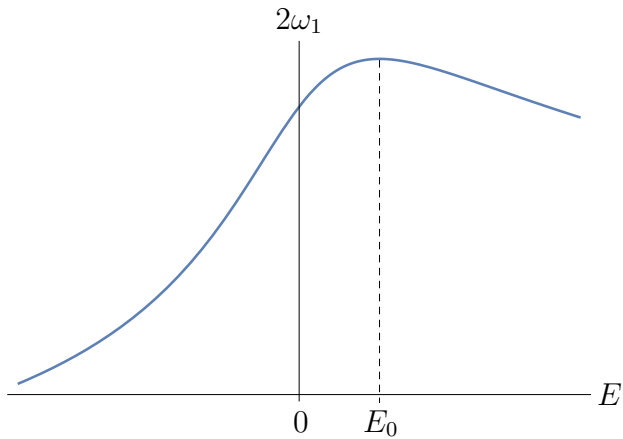
\begin{figure}[ht]
\centering
\begin{picture}(100,37)
\put(25,2.5){\includegraphics[width = 0.5\textwidth]{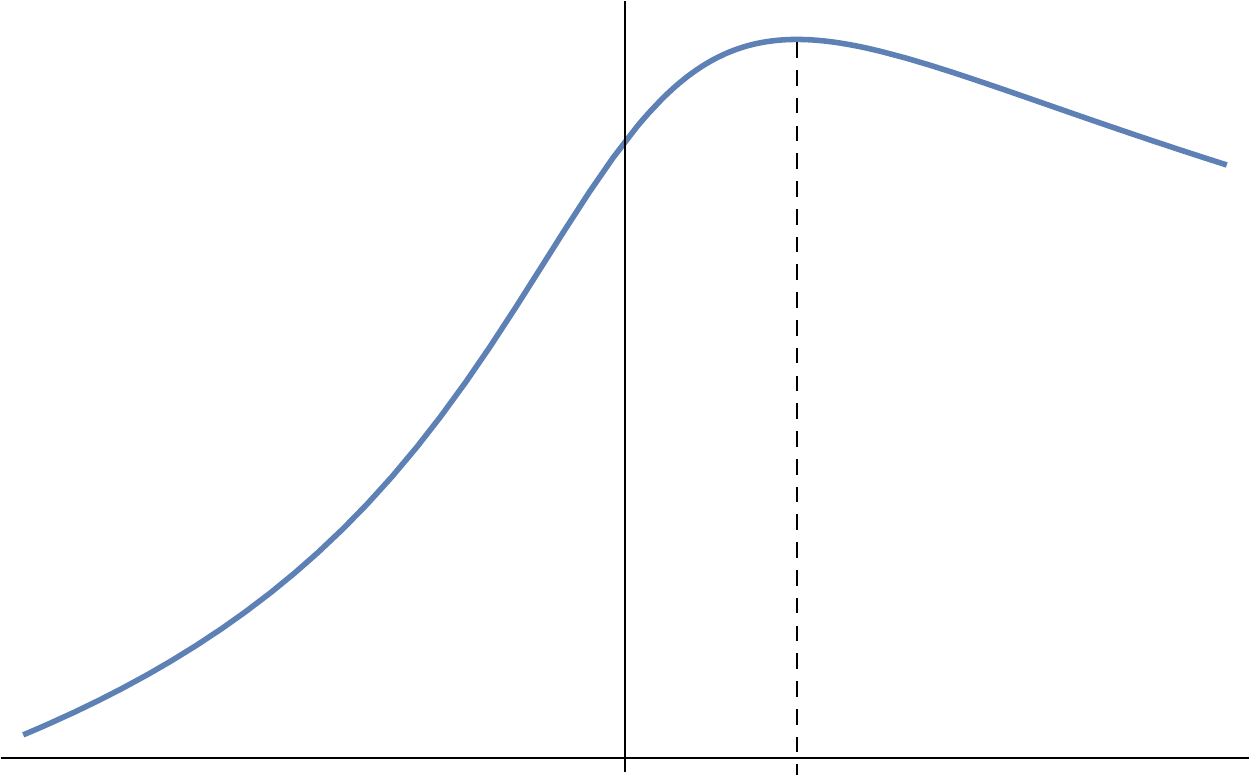}}
\put(75.5,2.25){$E$}
\put(49.5,0){$0$}
\put(55.5,0){$E_0$}
\put(48,34.5){$2 \omega_1$}
\end{picture}
\vspace{-20pt}
\caption{The period of the elliptic solution as function of the integration constant $E$}
\label{fig:period}
\end{figure}

The dependence of the period $2\omega_1$ with the integration constant $E$ is plotted in figure \ref{fig:period}.
There is a global maximum at a positive value of the constant $E = E_0 $. In later sections, we will show that the existence of this maximum is related with the stability properties of the elliptic minimal surfaces. Using formula \eqref{eq:elliptic_periods_D_pos}, one can show that $E_0$ obeys,
\begin{equation}
K\left( {k_0 } \right) = 2E\left( {k_0 } \right) , \quad k_0 = \sqrt {\frac{{{e_2 \left(E_0\right)} - {e_3 \left(E_0\right)}}}{{{e_1 \left(E_0\right)} - {e_3 \left(E_0\right)}}}}
\label{eq:elliptic_E0_analytical}
\end{equation}
resulting in
\begin{equation}
E_0 \simeq 1.72087 \Lambda^{-2} .
\label{eq:elliptic_E0}
\end{equation}

One can acquire a qualitative picture for the existence of this maximum. Equation \eqref{eq:Elliptic_energy_conservation} can be understood as the energy conservation for an one-dimensional effective mechanical problem of one point particle, where $\alpha$ plays the role of the position coordinate, $u$ plays the role of time, $E$ plays the role of energy and the potential is given by $V = - \left( 2 / \Lambda^2 \right) \sinh \alpha$. All solutions for this problem are scattering solutions coming from and going to plus infinity and for all of them the ``time of flight'' $2 \omega_1$ is finite due to the exponential fall of the potential at $+ \infty$. The flattest region of the potential is the region around $u = 0$ and this is the reason the maximum ``time of flight'' corresponds to a given positive value of the energy constant. For this energy, the point particle spends a relatively large amount of time with small velocity at the flat region, where it is not violently repelled. For energies smaller than this critical value, it does not reach the flat region, while for larger energies, the point particle does reach the flat region, but it passes through with a larger velocity and then it gets violently reflected in a region where the potential has a steeper slope, thus spending less time with relatively small velocities.

\subsection{The Effective \Schrodinger Problems}
\label{subsec:Schrodinger}

Given a solution $\alpha$ of the cosh-Gordon equation, the construction of the minimal surface is a non-trivial procedure, due to the non-local nature of the transformation relating the embedding functions $Y^\mu$ with the reduced field $\alpha$ \eqref{eq:Pohlmeyer_field_definition}. Such construction requires the solution of the equations of motion
\begin{equation}
\frac{{{\partial ^2}Y^\mu}}{{\partial {u^2}}} + \frac{{{\partial ^2}Y^\mu}}{{\partial {v^2}}}  = \frac{1}{{{\Lambda ^2}}}{e^\alpha}{Y^\mu } ,
\label{eq:Schrodinger_eom}
\end{equation}
simultaneously taking care that the embedding functions ${Y^\mu }$ obey the geometric and Virasoro constraints
\begin{align}
Y \cdot Y &= - {\Lambda ^2} ,\label{eq:Schrodinger_geometric}\\
{\partial }Y \cdot {\partial }Y &= 0 .\label{eq:Schrodinger_Virasoro}
\end{align}

In section \ref{subsec:Elliptic}, we focused on solutions of the reduced system that depend on only one of the two variables. This choice was not arbitrary; for solutions of this kind, equations \eqref{eq:Schrodinger_eom} can be solved using separation of variables. Defining
\begin{equation}
Y^\mu \left( u , v \right) = U^\mu \left( u \right) V^\mu \left( v \right) 
\end{equation}
and using the explicit form of the elliptic solutions for the reduced field \eqref{eq:Elliptic_asols}, we arrive at four pairs of effective \Schrodinger problems with opposite eigenvalues, each pair being of the form
\begin{align}
 - U'' + 2\left( {\wp \left( {u;{g_2}, {g_3}} \right) - {e_2}} \right)U = \kappa U , \label{eq:Schrodinger_uproblem}\\
 - \ddot V =  - \kappa V , \label{eq:Schrodinger_vproblem}
\end{align}
where the prime stands for differentiation with respect to $u$, while the dot stands for differentiation with respect to $v$. We have dropped the indices $\mu$ for simplicity, but in general the eigenvalue $\kappa$ may have a different value for each component. Actually, each component $Y^\mu$ may in general be equal to a linear combination of solutions corresponding to various eigenvalues $\kappa$, however, in this work we focus on constructions made of solutions corresponding to a single eigenvalue for each component.

Taking advantage of the geometric constraint, the real and imaginary parts of the Virasoro constraint \eqref{eq:Schrodinger_Virasoro} can be written in the form
\begin{align}
\left( {\frac{{{\partial ^2}Y}}{{\partial {u^2}}} - \frac{{{\partial ^2}Y}}{{\partial {v^2}}}} \right) \cdot Y &= 0 , \label{eq:Schrodinger_Virasoro_1}\\
\frac{{{\partial ^2}Y}}{{\partial u\partial v}} \cdot Y &= 0 ,\label{eq:Schrodinger_Virasoro_2}
\end{align}
which are easier to deal in the language of the effective \Schrodinger problems.

Trivially, the flat potential problem \eqref{eq:Schrodinger_vproblem}, for positive eigenvalues $\kappa  = {\ell ^2}$, has hyperbolic solutions of the form
\begin{equation}
V\left( v \right) = {c_1}\cosh \ell v + {c_2}\sinh \ell v ,
\end{equation}
whereas for negative eigenvalues $\kappa  = - {\ell ^2}$, it has trigonometric solutions of the form
\begin{equation}
V\left( v \right) = {c_1}\cos \ell v + {c_2}\sin \ell v .
\end{equation}

\subsection{The $n = 1$ \Lame Effective \Schrodinger Problem}
\label{subsec:Lame}

The periodic potentials of the class
\begin{equation}
V \left( x \right) = n\left( {n + 1} \right)\wp \left( x \right) ,
\label{eq:lame_n}
\end{equation}
are called \Lame potentials. For any integer $n$, it is possible to analytically find the band structure of the problem and it turns out that it contains up to $n$ finite bands plus an infinite band extending to infinite positive energies.

For the elliptic solutions of the Pohlmeyer reduced problem found in section \ref{subsec:Elliptic}, the equations for the embedding functions of the minimal surface take the form of effective \Schrodinger problems of the form \eqref{eq:lame_n} with $n=1$,
\begin{equation}
- \frac{{{d^2}y}}{{d{x^2}}} + 2\wp \left( x \right)y = \lambda y .
\label{eq:lame_n1_problem}
\end{equation}
It is known that the eigenfunctions of the $n=1$ \Lame problem are given by
\begin{equation}
{y_ \pm } \left( {x ; a} \right) = \frac{{\sigma \left( {x \pm a} \right)}}{{\sigma \left( x \right) \sigma \left( \pm a \right)}}{e^{ - \zeta \left( \pm \alpha  \right)x}} 
\label{eq:lame_eigenstates}
\end{equation}
with eigenvalues
\begin{equation}
\lambda = - \wp \left( a \right) .
\label{eq:lame_eigenvalues}
\end{equation}
These eigenfunctions are linearly independent, as long as the modulus $a$ does not coincide with any of the half-periods of the Weierstrass function appearing in the potential, and, thus, they provide the general solution of the problem. In the degenerate case, both solutions tend to
\begin{equation}
{y_ \pm }\left( {x;{\omega _{1,2,3}}} \right) = \sqrt {\wp \left( x \right) - {e_{1,3,2}}} \, ,
\label{eq:lame_special_eigenstates}
\end{equation}
where $\omega_3 = \omega_1 + \omega_2$ and there is another linearly independent solution,
\begin{equation}
\tilde y\left( {x;{\omega _{1,2,3}}} \right) = \sqrt {\wp \left( x \right) - {e_{1,3,2}}} \left( {\zeta \left( {x + {\omega _{1,2,3}}} \right) + {e_{1,3,2}}x} \right) . 
\end{equation} 

The special functions $\zeta \left( z \right)$ and $\sigma \left( z \right)$ appearing in the formulas above are the Weierstrass zeta and sigma functions respectively, which are defined as
\begin{equation}
\frac{{d\zeta }}{{dz}} = - \wp , \quad \frac{1}{\sigma }\frac{{d\sigma }}{{dz}}= \zeta .
\label{eq:lame_zeta_sigma_defs}
\end{equation}
The functions $\zeta$ and $\sigma$, unlike the elliptic function $\wp$, are not periodic. More information is provided in the appendix.

It can be shown that the band structure of the $n=1$ \Lame potential is directly connected with the roots of the cubic polynomial associated with the Weierstrass function appearing in the potential. In the case there are three real roots, which is the case of interest in this study, there is a finite ``valence'' band for $-e_1 < \lambda < -e_2$ and an infinite ``conduction'' band for $\lambda > -e_3$. The eigenfunctions ${y_ \pm }$ for eigenvalues within the bands are complex conjugate to each other and upon a shift of their argument by the period $2 \omega_1$ they acquire a complex phase as expected by Bloch's theorem. On the contrary, for eigenvalues within the gaps of the spectrum, they are both real and upon a shift of their argument by the period $2 \omega_1$ they get multiplied by a real number, in general different than one, and consequently they diverge exponentially at either plus or minus infinity. Exceptionally, the eigenfunctions \eqref{eq:lame_special_eigenstates} corresponding to the boundaries of the bands are both real and periodic. These eigenfunctions do not have the physical interpretation of a wavefunction and consequently they do not have to obey any specific normalization conditions. As a result none of them is excluded.

The eigenfunctions \eqref{eq:lame_eigenstates} obey a set of properties that will become useful later,
\begin{align}
{y_ + }{y_ - } &= \wp \left( x \right) - \wp \left( a \right) , \label{eq:lame_eigen_property_1}\\
{y_ + }'{y_ - } - {y_ + }{y_ - }' &= - \wp '\left( a \right) . \label{eq:lame_eigen_property_2}
\end{align}

\subsection{Construction of Elliptic Minimal Surfaces}
\label{subsec:Construction}

At this point, there is only one step left to complete the process of the inversion of Pohlmeyer reduction for elliptic solutions of the cosh-Gordon equation. The equations of motions are satisfied by the solutions of the effective \Schrodinger problems, but one needs to make an appropriate arrangement of such solutions in the components of the embedding functions, so that the geometric and Virasoro constraints are also satisfied.

Since we are constrained to static minimal surfaces, we may select to set $Y^0 = 0$. In the following, we neglect the $Y^0$ component when we express $Y$ in a matrix form. 

Similarly to the construction of classical string solutions in \cite{Bakas:2016jxp}, one can show that it is not possible to construct a minimal surface using solutions of the effective \Schrodinger problems corresponding to a single eigenvalue. It turns out that the simplest possible construction involves at least two distinct eigenvalues. The form of the metric of the enhanced space suggests that these eigenvalues should be selected to have opposite signs. Let these eigenvalues be equal to
\begin{align}
{\kappa _1} &= \ell _1^2 =  - \wp \left( {{a_1}} \right) - 2{e_2} , \label{eq:construction_cosh_a1}\\
{\kappa _2} &=  - \ell _2^2 =  - \wp \left( {{a_2}} \right) - 2{e_2} , \label{eq:construction_cosh_a2}
\end{align}
where $a_1$ and $a_2$ are the moduli appearing in the corresponding solutions of the $n=1$ \Lame effective \Schrodinger problem. The form of the eigenvalues restricts $a_1$ and $a_2$ so that
\begin{equation}
\wp \left( {{a_2}} \right) > \wp \left( {{a_1}} \right) .
\end{equation}

The form of the enhanced metric and the geometric and Virasoro constraints favour an ansatz of the form
\begin{equation}
Y = \left( {\begin{array}{*{20}{c}}
{c_1^ + U_1^ + \left( u \right)\cosh {\ell _1}v + c_1^ - U_1^ - \left( u \right)\sinh {\ell _1}v}\\
{c_1^ + U_1^ + \left( u \right)\sinh {\ell _1}v + c_1^ - U_1^ - \left( u \right)\cosh {\ell _1}v}\\
{c_2^ + U_2^ + \left( u \right)\cos {\ell _2}v + c_2^ - U_2^ - \left( u \right)\sin {\ell _2}v}\\
{c_2^ + U_2^ + \left( u \right)\sin {\ell _2}v - c_2^ - U_2^ - \left( u \right)\cos {\ell _2}v}
\end{array}} \right) ,
\label{eq:construction_cosh_ansatz}
\end{equation}
where $U_1^ \pm \left( u \right)$ and $U_2^ \pm \left( u \right)$ are in general linear combinations of the eigenfunctions of the $n=1$ \Lame problem, ${{y_ \pm }\left( {u;{a_1}} \right)}$ and ${{y_ \pm }\left( {u;{a_2}} \right)}$ respectively. The geometric and Virasoro constraints \eqref{eq:Schrodinger_geometric}, \eqref{eq:Schrodinger_Virasoro_1} and \eqref{eq:Schrodinger_Virasoro_2} take the form
\begin{align}
{ - {{\left( {c_1^ + U_1^ +} \right)}^2} + {{\left( {c_1^ - U_1^ - } \right)}^2}} + {{{\left( {c_2^ + U_2^ + } \right)}^2} + {{\left( {c_2^ - U_2^ - } \right)}^2}} &=  - {\Lambda ^2} , \label{eq:construction_cosh_geometric} \\
 - \left[ { - {{\left( {c_1^ + U_1^ + } \right)}^2} + {{\left( {c_1^ - U_1^ - } \right)}^2}} \right]\ell _1^2 + \left[ {{{\left( {c_2^ + U_2^ + } \right)}^2} + {{\left( {c_2^ - U_2^ - } \right)}^2}} \right]\ell _2^2 &= \left( {\wp \left( u \right) - {e_2}} \right){\Lambda ^2} , \label{eq:construction_cosh_Virasoro_1}\\
c_1^ + c_1^ - \left( {{U_1^ +} ' U_1^ -  - {U_1^ -} ' U_1^ + } \right){\ell _1} - c_2^ + c_2^ - \left( {{U_2^ +} ' U_2^ -  - {U_2^ -} ' U_2^ + } \right){\ell _2} &= 0 . \label{eq:construction_cosh_Virasoro_2}
\end{align}

The geometric constraint \eqref{eq:construction_cosh_geometric}, combined with property \eqref{eq:lame_eigen_property_1} of the $n=1$ \Lame eigenfunctions suggests that we have to select
\begin{align}
c_1^ +  &= c_1^ - ,\quad U_1^ + \left( u \right) = \frac{1}{2}\left( {{y_ + }\left( {u;{a_1}} \right) + {y_ - }\left( {u;{a_1}} \right)} \right),\quad U_1^ - \left( u \right) = \frac{1}{2}\left( {{y_ + }\left( {u;{a_1}} \right) - {y_ - }\left( {u;{a_1}} \right)} \right) ,\\
c_2^ +  &= c_2^ - ,\quad U_2^ + \left( u \right) = \frac{1}{2}\left( {{y_ + }\left( {u;{a_2}} \right) + {y_ - }\left( {u;{a_2}} \right)} \right),\quad U_2^ - \left( u \right) = \frac{1}{{2i}}\left( {{y_ + }\left( {u;{a_2}} \right) - {y_ - }\left( {u;{a_2}} \right)} \right) .
\end{align}
Reality of the solution implies that ${{y_ \pm }\left( {u;{a_1}} \right)}$ are non-normalizable eigenstates corresponding to the gaps of the \Lame spectrum, whereas ${{y_ \pm }\left( {u;{a_2}} \right)}$ are Bloch waves, lying within the allowed bands of the \Lame spectrum. Since $\wp \left( {{a_1}} \right) < \wp \left( {{a_2}} \right)$, they necessarily lie within the finite gap and in the finite ``valence'' band of the \Lame spectrum respectively. Consequently,
\begin{equation}
{e_3} < \wp \left( {{a_1}} \right) < {e_2}\quad \mathrm{and} \quad {e_2} < \wp \left( {{a_2}} \right) < {e_1} .
\label{eq:construction_cosh_a2_bands}
\end{equation}

Inserting the above selections into the geometric constraint \eqref{eq:construction_cosh_geometric} yields the equation
\begin{equation}
\left( { - c_1^2 + c_2^2} \right)\wp \left( u \right) + \left( {c_1^2\wp \left( {{a_1}} \right) - c_2^2\wp \left( {{a_2}} \right)} \right) =  - {\Lambda ^2} ,
\end{equation}
which in turn implies that
\begin{equation}
c_1^2 = c_2^2 \equiv {c^2} = \frac{{{\Lambda ^2}}}{{\wp \left( {{a_2}} \right) - \wp \left( {{a_1}} \right)}} .
\label{eq:construction_cosh_c}
\end{equation}
Then, after some algebra, the Virasoro constraint \eqref{eq:construction_cosh_Virasoro_1} yields
\begin{equation}
\wp \left( {{a_1}} \right) + \wp \left( {{a_2}} \right) =  - {e_2} ,
\label{eq:construction_cosh_Virasoro_final}
\end{equation}
while the Virasoro constraint \eqref{eq:construction_cosh_Virasoro_2}, using property \eqref{eq:lame_eigen_property_2} takes the form
\begin{equation}
\wp '\left( {{a_1}} \right){\ell _1} = i\wp '\left( {{a_2}} \right){\ell _2} .
\label{eq:construction_Virasoro_2_final}
\end{equation}
The last equation is always satisfied, as long as \eqref{eq:construction_cosh_Virasoro_final} is satisfied. To verify this, one can make use of Weierstrass equation to rewrite \eqref{eq:construction_Virasoro_2_final} as
\begin{multline}
4\left( {\wp \left( {{a_1}} \right) + 2{e_2}} \right)\left( {\wp \left( {{a_1}} \right) - {e_1}} \right)\left( {\wp \left( {{a_1}} \right) - {e_2}} \right)\left( {\wp \left( {{a_1}} \right) - {e_3}} \right)\\
 = 4\left( {\wp \left( {{a_2}} \right) + 2{e_2}} \right)\left( {\wp \left( {{a_2}} \right) - {e_1}} \right)\left( {\wp \left( {{a_2}} \right) - {e_2}} \right)\left( {\wp \left( {{a_2}} \right) - {e_3}} \right) .
\end{multline}
Then, equation \eqref{eq:construction_cosh_Virasoro_final} connects the factors of the left and right hand sides one by one.

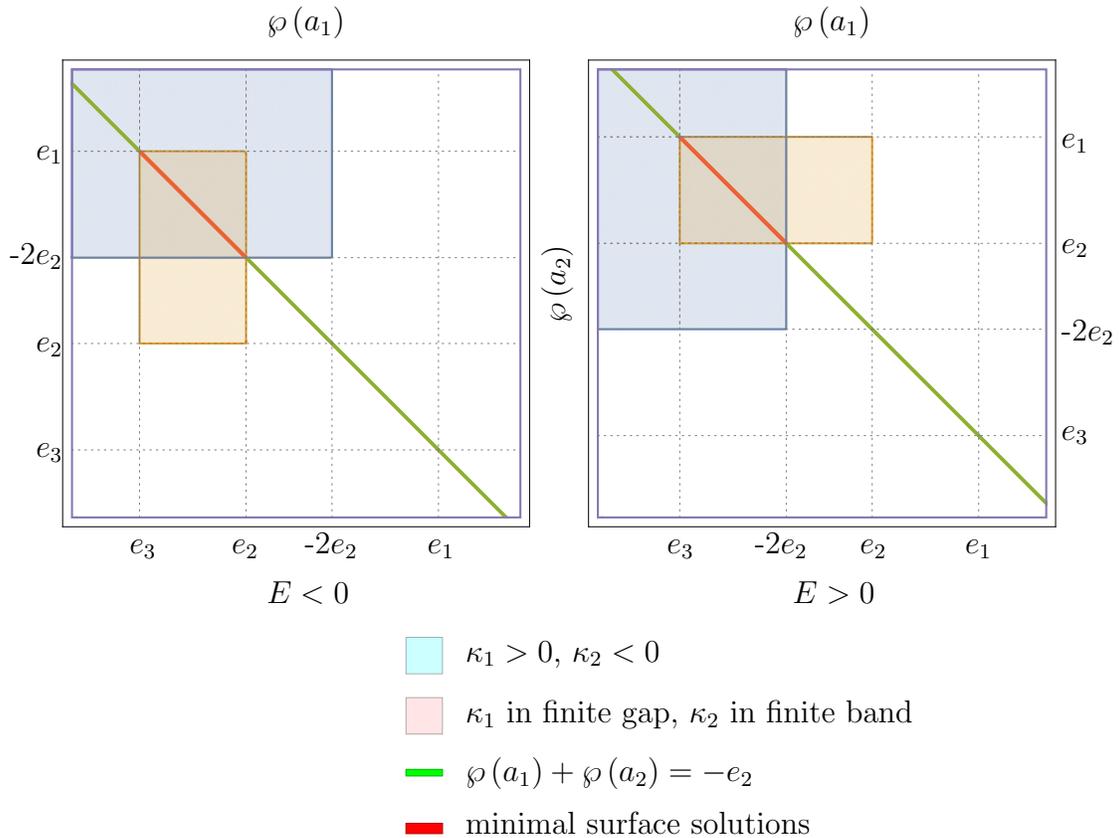
\begin{figure}[ht]
\centering
\begin{picture}(100,75)
\put(7.5,30){\includegraphics[width = 0.4\textwidth]{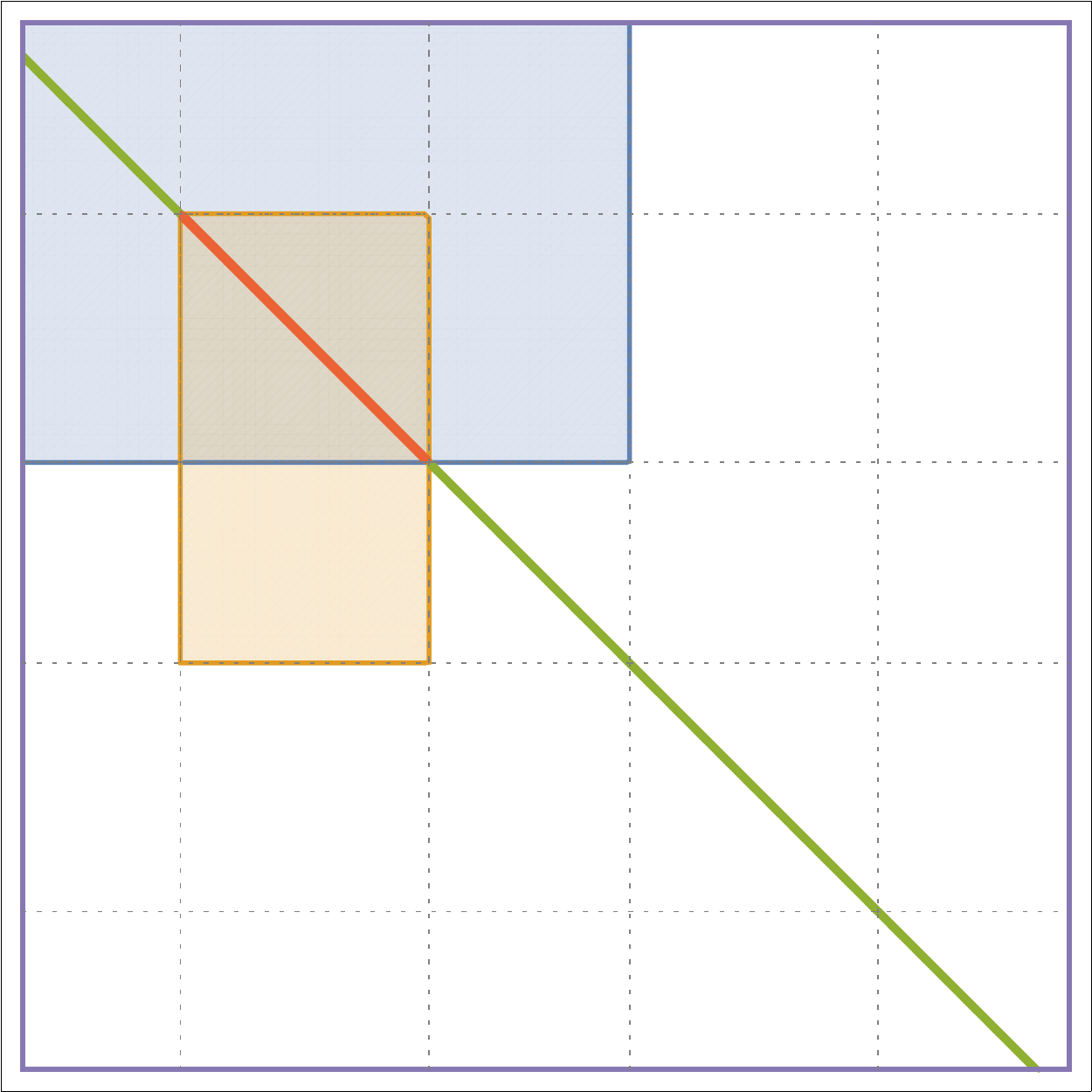}}
\put(52.5,30){\includegraphics[width = 0.4\textwidth]{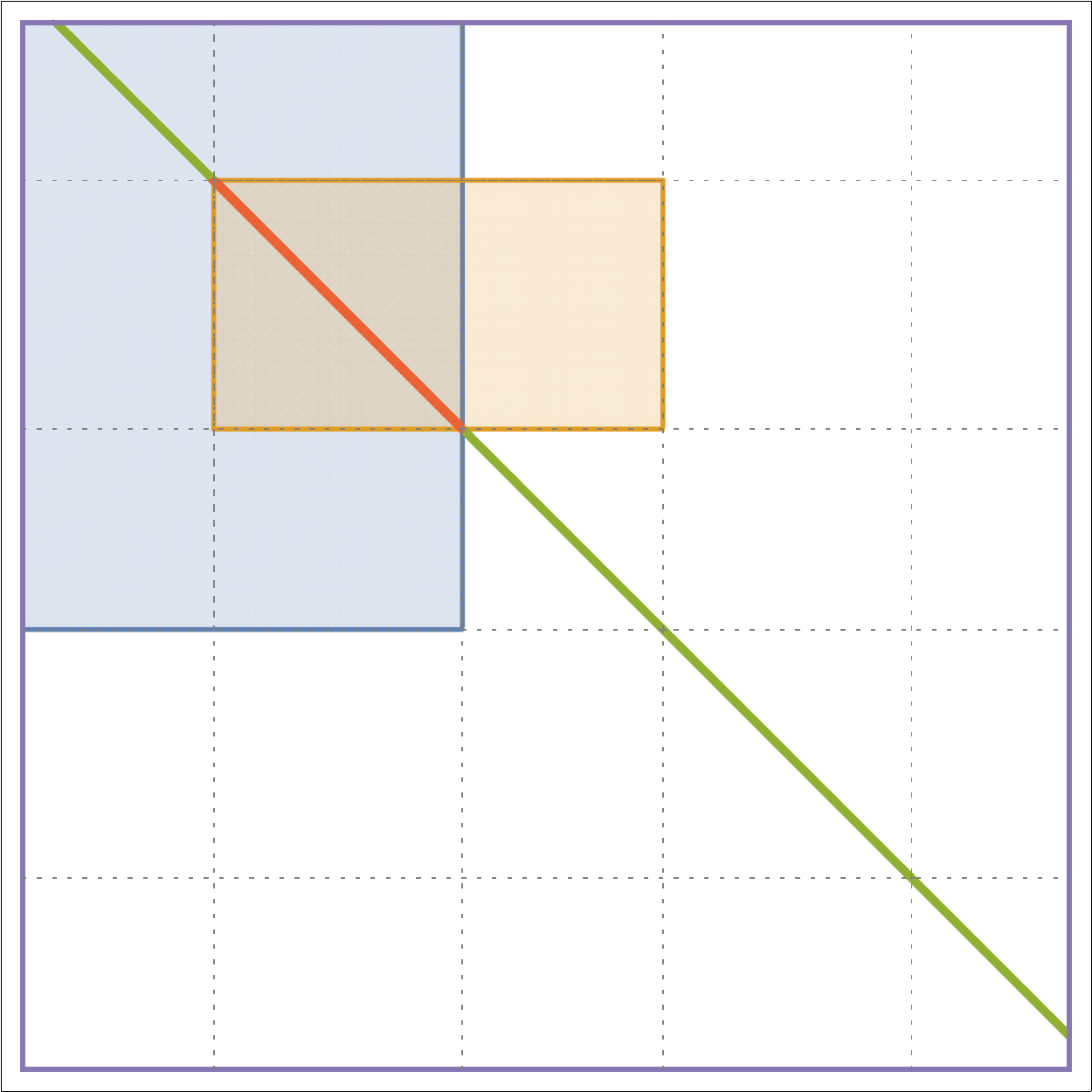}}
\put(36,1){\includegraphics[width = 0.08\textwidth]{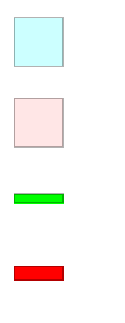}}
\put(5.25,61.5){$e_1$}
\put(3,52.5){-$2e_2$}
\put(5.25,45){$e_2$}
\put(5.25,36){$e_3$}
\put(93,62.5){$e_1$}
\put(93,53.5){$e_2$}
\put(93,46){-$2e_2$}
\put(93,37.5){$e_3$}
\put(13.25,27.75){$e_3$}
\put(22,27.75){$e_2$}
\put(28.25,27.75){-$2e_2$}
\put(38.75,27.75){$e_1$}
\put(59.25,27.75){$e_3$}
\put(66.75,27.75){-$2e_2$}
\put(75.75,27.75){$e_2$}
\put(84.75,27.75){$e_1$}
\put(25,72.5){$\wp \left( a_1 \right)$}
\put(70,72.5){$\wp \left( a_1 \right)$}
\put(48.5,47.5){\rotatebox{90}{$\wp \left( a_2 \right)$}}
\put(25,23.5){$E < 0$}
\put(70,23.5){$E > 0$}
\put(42,18.5){$\kappa_1 > 0$, $\kappa_2 < 0$}
\put(42,13.25){$\kappa_1$ in finite gap, $\kappa_2$ in finite band}
\put(42,8.25){$\wp \left( a_1 \right)+\wp \left( a_2 \right) = -e_2$}
\put(42,3.75){minimal surface solutions}
\end{picture}
\vspace{-20pt}
\caption{The pairs of $\wp\left( a_1 \right)$ and $\wp\left( a_2 \right)$ that generate minimal surface solutions built from eigenstates of the effective \Schrodinger problems corresponding to two distinct eigenvalues}
\label{fig:region}
\end{figure}

Putting everything together, the construction of a minimal surface solution, built from eigenstates of the effective \Schrodinger problems corresponding to two distinct eigenvalues, is equivalent to the specification of $\wp \left( {{a_1}} \right)$ and $\wp \left( {{a_2}} \right)$ so that obey \eqref{eq:construction_cosh_a2_bands}, \eqref{eq:construction_cosh_Virasoro_final} and simultaneously are such that the eigenvalues $\kappa_{1,2}$ as specified by \eqref{eq:construction_cosh_a1} and \eqref{eq:construction_cosh_a2} have the appropriate sign. As shown in figure \ref{fig:region}, for any value of the integration constant $E$, there is a set of appropriate selections of $\wp \left( {{a_1}} \right)$ and $\wp \left( {{a_2}} \right)$, constituting a linear segment in the $\left( \wp \left( {{a_1}} \right),\wp \left( {{a_2}} \right) \right)$ plane. One of the two endpoints of this linear segment is always $\left( e_3 , e_1 \right)$, while the other one is $\left( e_2 , - 2 e_2 \right)$ when $E < 0$ and $\left( - 2 e_2 , e_2 \right)$ when $E > 0$.

\section{Properties of the Elliptic Minimal Surfaces}
\label{sec:Surfaces}

In this section, we study basic geometric properties of the minimal surfaces constructed in section \ref{subsec:Construction}. We are particularly interesting in the form of their trace at the AdS$_4$ boundary and their area, since in the language of the RT conjecture they play the role of the entangling curve and the corresponding entanglement entropy. Furthermore, we will investigate the specific forms of the minimal surfaces corresponding to the endpoints of the linear segments depicted in figure \ref{fig:region} and identify them with well known minimal surfaces in H$^3$, such as helicoids and catenoids.

\subsection{Parameter Space of Elliptic Minimal Surfaces}
\label{subsec:moduli}
\begin{figure}[ht]
\centering
\begin{picture}(100,40)
\put(4.5,2.5){\includegraphics[width = 0.7\textwidth]{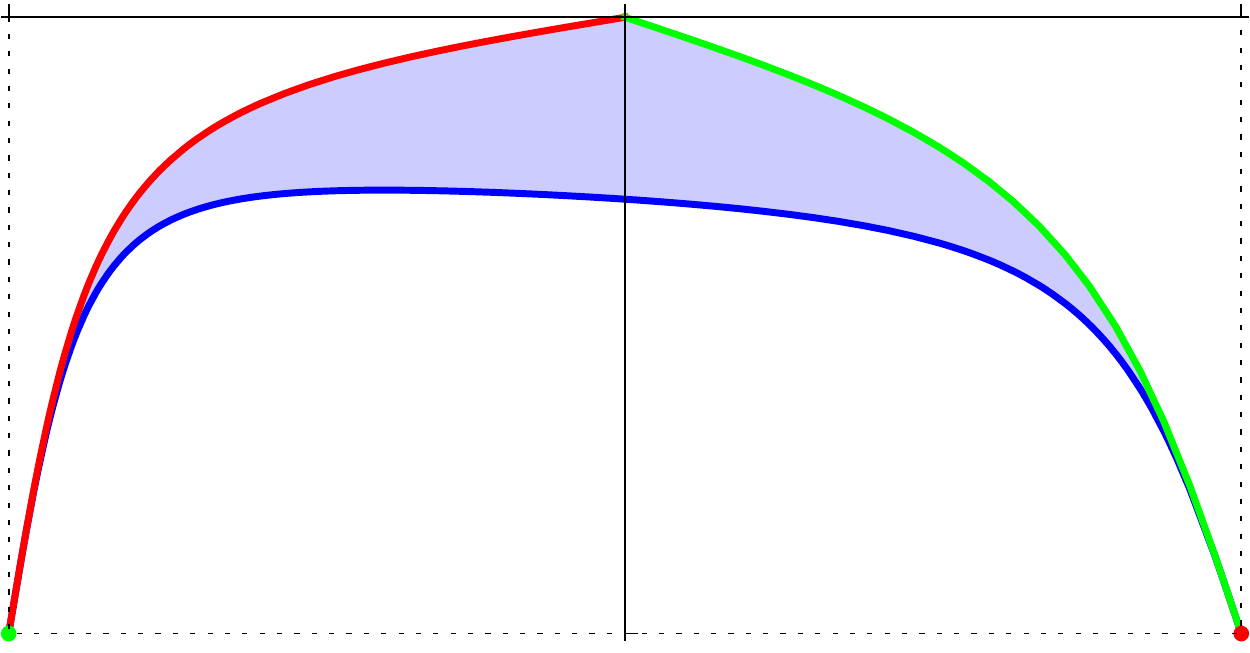}}
\put(75,10){\includegraphics[width = 0.08\textwidth]{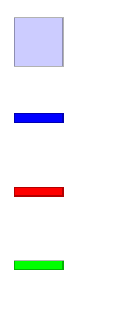}}
\put(2,39.5){$- \pi / 2$}
\put(72,39.5){$\pi / 2$}
\put(38.75,39.5){$0$}
\put(39.5,4){$- \pi / 2$}
\put(35,0){$\arctan E$}
\put(0.5,14){\rotatebox{90}{$\arctan \wp \left( a_1 \right)$}}
\put(81,27){allowed $E$, $\wp \left( a_1 \right)$}
\put(81,22.25){$\wp \left( a_1 \right) = e_3$}
\put(81,17.5){$\wp \left( a_1 \right) = e_2$}
\put(81,12.75){$\wp \left( a_1 \right) = - 2 e_2$}
\end{picture}
\vspace{-20pt}
\caption{The parameter space of elliptic minimal surface solutions}
\label{fig:parameters}
\end{figure}
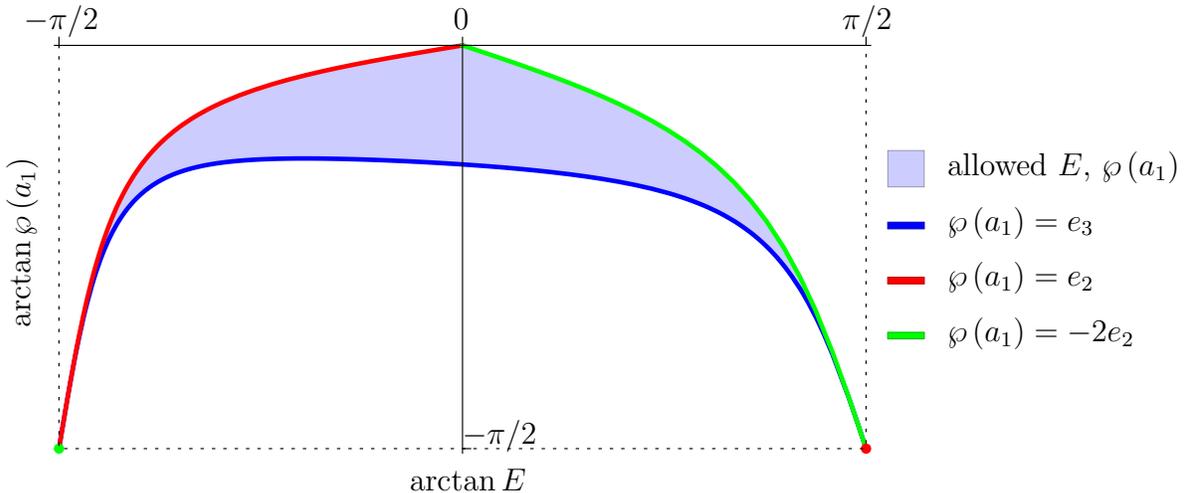

The family of elliptic minimal surfaces that we have constructed contains two free parameters. One of those is the constant of integration $E$, which alters the moduli of the Weierstrass functions and consequently the roots of the associated cubic polynomial and may take any real value. The other one is the parameter $\wp \left( a_1 \right)$, which takes values between $e_3$ and $\min \left( e_2 , -2 e_2 \right)$, as shown in figure \ref{fig:region}. Notice that all minimal surfaces corresponding to the same value of $E$ are mapped to the same solution of the cosh-Gordon equation through Pohlmeyer reduction, independently of the value of $\wp \left( a_1 \right)$. The space of parameters for the elliptic minimal surfaces is plotted in figure \ref{fig:parameters}.
The allowed parameters comprise a connected region in the parameter space, bounded by the non-smooth union of three smooth curves. Later on, we will see that these boundary curves correspond to three qualitatively distinct and interesting limits of the solutions.

We may simplify the expressions for the elliptic minimal surfaces that we constructed in section \ref{subsec:Construction} taking advantage of the fact that the functions ${y_ \pm }\left( {u;{a_1}} \right)$ are real, whereas the functions ${y_ \pm }\left( {u;{a_2}} \right)$ are complex conjugate to each other. We define
\begin{align}
{y_ \pm }\left( {u;{a_1}} \right) &:= {r_1}\left( u \right){e^{ \pm {\varphi _1}\left( u \right)}} ,\\
{y_ \pm }\left( {u;{a_2}} \right) &:= {r_2}\left( u \right){e^{ \pm i{\varphi _2}\left( u \right)}} .
\end{align}
A direct application of property \eqref{eq:lame_eigen_property_1} yields
\begin{align}
{r_1^2}\left( u \right) &= {\wp \left( u \right) - \wp \left( {{a_1}} \right)} ,\\
{r_2^2}\left( u \right) &= {\wp \left( u \right) - \wp \left( {{a_2}} \right)} ,
\end{align}
while the explicit form of the $n=1$ \Lame eigenfunctions \eqref{eq:lame_eigenstates} implies that
\begin{align}
{\varphi _1}\left( u \right) &= \frac{1}{2}\ln \left( { - \frac{{\sigma \left( {u + {a_1}} \right)}}{{\sigma \left( {u - {a_1}} \right)}}} \right) - \zeta \left( {{a_1}} \right)u ,\\
{\varphi _2}\left( u \right) &=  - \frac{i}{2}\ln \left( { - \frac{{\sigma \left( {u + {a_2}} \right)}}{{\sigma \left( {u - {a_2}} \right)}}} \right) + i\zeta \left( {{a_2}} \right)u .
\end{align}

Using the above definitions, the minimal surface solution takes the form
\begin{equation}
Y = \frac{\Lambda }{{\sqrt {\wp \left( {{a_2}} \right) - \wp \left( {{a_1}} \right)} }}\left( {\begin{array}{*{20}{c}}
{\sqrt {\wp \left( u \right) - \wp \left( {{a_1}} \right)} \cosh \left( {{\ell _1}v + {\varphi _1}\left( u \right)} \right)}\\
{\sqrt {\wp \left( u \right) - \wp \left( {{a_1}} \right)} \sinh \left( {{\ell _1}v + {\varphi _1}\left( u \right)} \right)}\\
{\sqrt {\wp \left( u \right) - \wp \left( {{a_2}} \right)} \cos \left( {{\ell _2}v - {\varphi _2}\left( u \right)} \right)}\\
{\sqrt {\wp \left( u \right) - \wp \left( {{a_2}} \right)} \sin \left( {{\ell _2}v - {\varphi _2}\left( u \right)} \right)}
\end{array}} \right) .
\label{eq:properties_elliptic_solution}
\end{equation}

In order to better visualize the form of the derived minimal surfaces, we will use two common set of coordinates in a constant time slice of AdS$_4$ (i.e. the H$^3$), the global spherical coordinates $\left( r, \theta, \varphi \right)$, defined as
\begin{equation}
Y = \left( {\begin{array}{*{20}{c}}
{\Lambda \sqrt {1 + \frac{{{r^2}}}{{{\Lambda ^2}}}} }\\
{r\cos \theta }\\
{r\sin \theta \cos \varphi }\\
{r\sin \theta \sin \varphi }
\end{array}} \right) 
\end{equation}
and the \Poincare coordinates $\left( z, r, \varphi \right)$, defined as
\begin{equation}
Y = \left( {\begin{array}{*{20}{c}}
{\frac{1}{{2z}}\left( {{z^2} + {r^2} + {\Lambda ^2}} \right)}\\
{\frac{1}{{2z}}\left( {{z^2} + {r^2} - {\Lambda ^2}} \right)}\\
{\frac{\Lambda }{z}r\cos \varphi }\\
{\frac{\Lambda }{z}r\sin \varphi }
\end{array}} \right).
\end{equation}
In global coordinates, the metric takes the form
\begin{equation}
d{s^2} = {\left( {1 + \frac{{{r^2}}}{{{\Lambda ^2}}}} \right)^{ - 1}}d{r^2} + {r^2}\left( {d{\theta ^2} + {{\sin }^2}\theta d{\varphi ^2}} \right) ,
\end{equation}
whereas in \Poincare coordinates it takes the form
\begin{equation}
d{s^2} = \frac{{{\Lambda ^2}}}{{{z^2}}}\left( {d{z^2} + d{r^2} + {r^2}d{\varphi ^2}} \right) .
\end{equation}

In global coordinates, the elliptic minimal surfaces take the parametric form
\begin{align}
r &= \Lambda \sqrt {\frac{{\wp \left( u \right) - \wp \left( {{a_1}} \right)}}{{\wp \left( {{a_2}} \right) - \wp \left( {{a_1}} \right)}}{{\cosh }^2}\left( {{\ell _1}v + {\varphi _1}\left( u \right)} \right) - 1} , \label{eq:Properties_radial_global} \\
\theta  &= {\tan ^{ - 1}}\left( {\sqrt {\frac{{\wp \left( u \right) - \wp \left( {{a_1}} \right)}}{{\wp \left( u \right) - \wp \left( {{a_2}} \right)}}} \csch\left( {{\ell _1}v + {\varphi _1}\left( u \right)} \right)} \right) ,\\
\varphi  &= {\ell _2}v - {\varphi _2}\left( u \right) .
\end{align}
It is simple to eliminate $v$ to show that the minimal surface acquires an expression of the form
\begin{equation}
f\left( {\varphi  - \frac{{{\ell _2}}}{{{\ell _1}}}{{\tanh }^{ - 1}}\frac{{r\cos \theta }}{{\sqrt {{r^2} + {\Lambda ^2}} }},r\sin \theta } \right) = 0 .
\label{eq:Properties_rigid_global}
\end{equation}
Similarly, in \Poincare coordinates we get the parametric form
\begin{align}
z = \Lambda \sqrt {\frac{{\wp \left( {{a_2}} \right) - \wp \left( {{a_1}} \right)}}{{\wp \left( u \right) - \wp \left( {{a_1}} \right)}}} {e^{{\ell _1}v + {\varphi _1}\left( u \right)}} , \label{eq:Properties_radial_Poincare} \\
r = \Lambda \sqrt {\frac{{\wp \left( u \right) - \wp \left( {{a_2}} \right)}}{{\wp \left( u \right) - \wp \left( {{a_1}} \right)}}} {e^{{\ell _1}v + {\varphi _1}\left( u \right)}} ,\\
\varphi  = {\ell _2}v - {\varphi _2}\left( u \right) .
\end{align}
Once again, it is trivial to eliminate $v$ to show that the minimal surface is described in closed form as
\begin{equation}
f\left( {\varphi  - \frac{{{\ell _2}}}{{{\ell _1}}}{{\tanh }^{ - 1}}\frac{{{z^2} + {r^2} - {\Lambda ^2}}}{{{z^2} + {r^2} + {\Lambda ^2}}},\frac{r}{z}} \right) = 0 .
\label{eq:Properties_rigid_Poincare}
\end{equation}

Equations \eqref{eq:Properties_radial_global} and \eqref{eq:Properties_radial_Poincare} provide a geometric explanation to the exclusion of the bounded real solution of the Pohlmeyer reduced system. Solutions built on the bounded solution of the reduced problem would not be anchored at the boundary, and, thus, would be compact surfaces shrinkable to a point. As such, they could not be minimal surfaces.

\begin{figure}[ht]
\centering
\begin{picture}(100,52)
\put(0,0){\includegraphics[width = 0.5\textwidth]{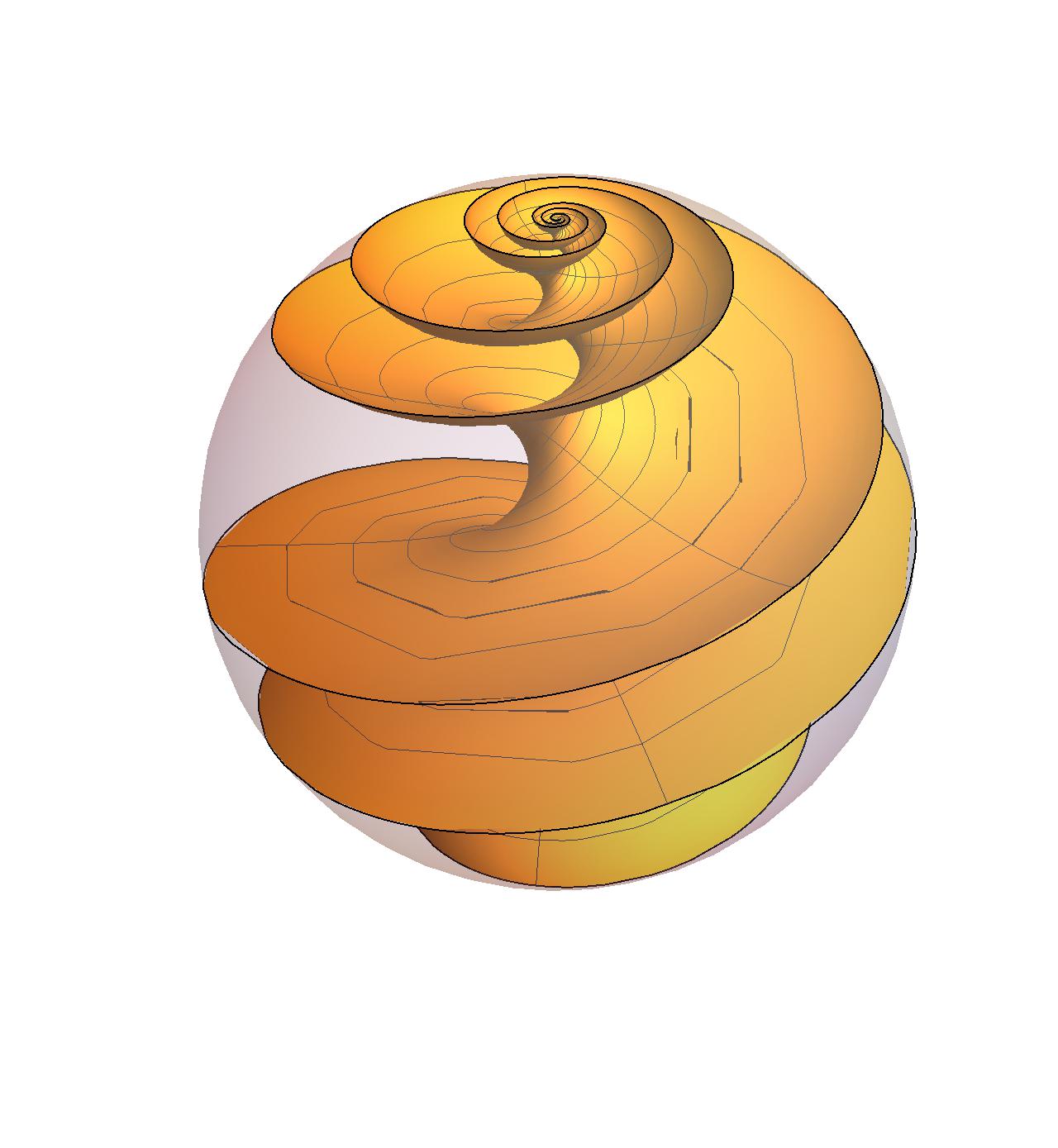}}
\put(50,0){\includegraphics[width = 0.5\textwidth]{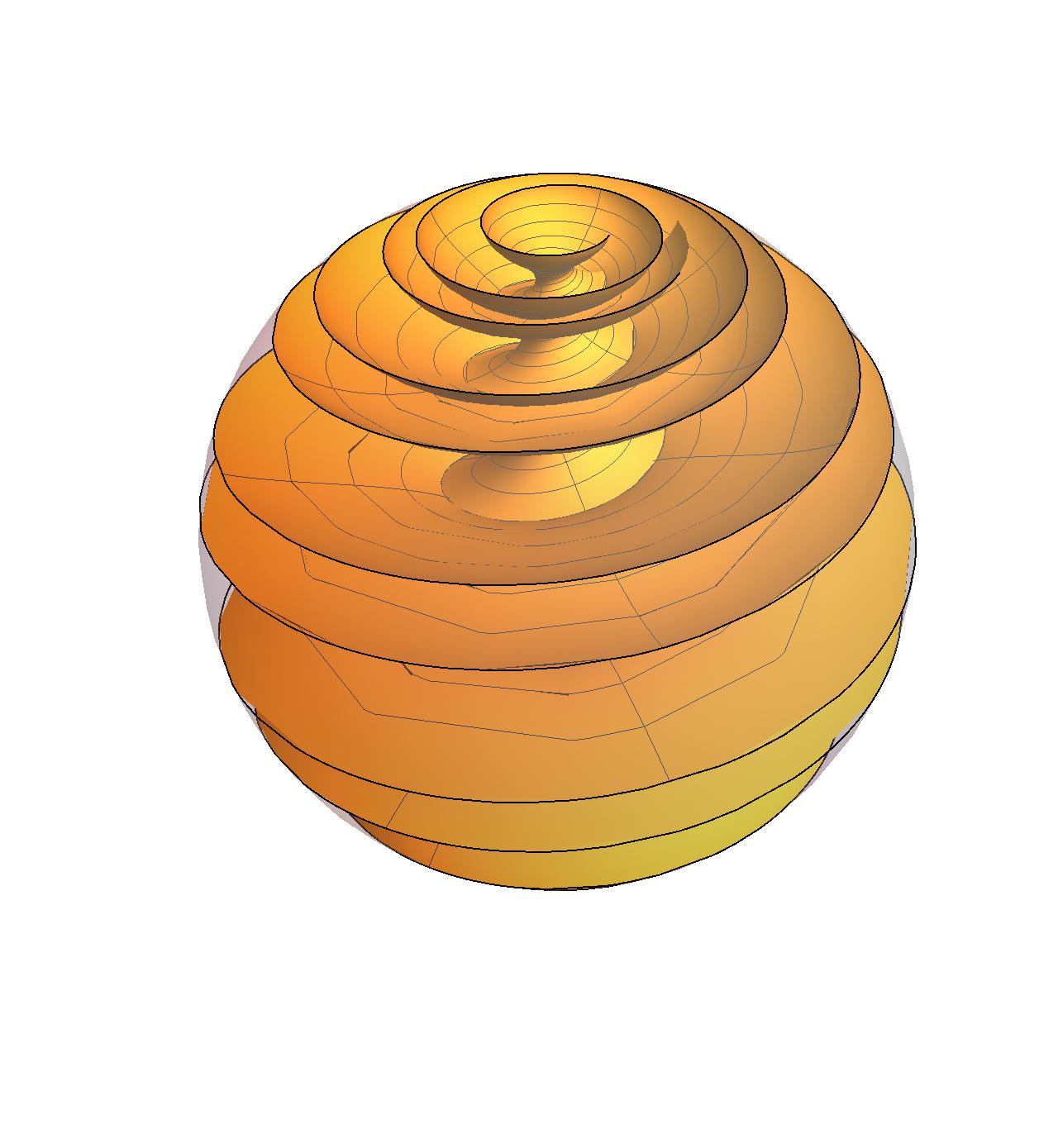}}
\end{picture}
\vspace{-45pt}
\caption{The elliptic minimal surface in global coordinates for two distinct selections of the parameters $E$ and $\wp \left( a_1 \right)$}
\label{fig:minimal_surface}
\end{figure}

The form of the expressions \eqref{eq:Properties_rigid_global} and \eqref{eq:Properties_rigid_Poincare} is not unexpected. The analogue of the elliptic minimal surfaces in NLSMs describing string propagation in AdS$_3$ are several classes of solutions, some of them describing rigidly rotating strings, as shown in \cite{Bakas:2016jxp}. Equations \eqref{eq:Properties_rigid_global} and \eqref{eq:Properties_rigid_Poincare} are the analogue of the rigid rotation condition for the elliptic minimal surfaces. It follows that the constructed elliptic minimal surfaces have in general a ``helicoid'' shape, as shown in figure \ref{fig:minimal_surface}.
In this figure, as well as in all following figures depicting elliptic minimal surfaces in global coordinates, the plotted radial coordinate is proportional to the tortoise coordinate $r^* = \arctan r$, so that the opaque sphere in the graphs depicts the H$^3$ boundary.

Furthermore, figure \ref{fig:minimal_surface} indicates that the elliptic minimal surfaces may or may not have self-intersections. Later on, we will specify the condition that the free parameters $E$ and $\wp \left( a_1 \right)$ must obey in order to generate an embedding minimal surface.

Finally, surfaces characterized by the same integration constant $E$, apart from having the property of having the same counterpart in the Pohlmeyer reduced theory, they comprise an associate (Bonnet) family of minimal surfaces. It is a matter of simple algebra to show that parametrizing the minimal surface with $u$ and $v$, the first and second fundamental forms  acquire the simple expressions
\begin{align}
I &= {\Lambda ^2}\left( {\wp \left( u \right) - {e_2}} \right)\diag\left\{ {1,1} \right\} ,\\
II &= \frac{1}{{2\Lambda }}\diag\left\{ {1, - 1} \right\} .
\end{align}
The latter imply that the principal curvatures are equal to
\begin{equation}
{\kappa _{1,2}} =  \pm \frac{1}{{2{\Lambda ^3}\left( {\wp \left( u \right) - {e_2}} \right)}} ,
\end{equation}
Therefore, the principal curvatures do not depend on the value of $\wp \left( a_1 \right)$. As a result, changing the value of  
$\wp \left( a_1 \right)$ keeping $E$ constant has the effect of a local rotation of the principal curvature directions of the minimal surface.

\subsection{The Trace of the Minimal Surfaces at the Boundary}
\label{subsec:Boundary_Regions}
The Ryu-Takayanagi conjecture relates the area of a co-dimension two minimal surface anchored at the boundary of AdS on an entangling surface to the entanglement entropy of the boundary CFT with respect to the regions separated by the entangling surface (entangling curve in AdS$_4$). Therefore, it is very important to specify what is the entangling curve for the constructed elliptical minimal surfaces.

The minimal surface \eqref{eq:properties_elliptic_solution} intersects the boundary at the points where the Weierstrass elliptic function diverges, namely $u = 2 n \omega_1$. Thus, an appropriately anchored at the boundary minimal surface is spanned by
\begin{equation}
u \in \left(2 n \omega_1 , 2 \left( n + 1 \right) \omega_1 \right) ,\quad
v \in \mathbb{R} ,
\label{eq:boundary_spanning}
\end{equation}
where $n \in \mathbb{Z}$.

The trace of the minimal surfaces on the boundary, i.e. the entangling curve, can be found by taking the limit $u \to 2 n \omega_1$. Properties \eqref{eq:Weierstrass_period_zeta} and \eqref{eq:Weierstrass_period_sigma} of Weierstrass functions imply that
\begin{equation}
\mathop {\lim }\limits_{u \to 2n{\omega _1}^ \pm } {y_ \pm }\left( {u;a} \right) = {e^{ \pm 2n\left( {\zeta \left( {{\omega _1}} \right)a - \zeta \left( a \right){\omega _1}} \right)}}\mathop {\lim }\limits_{u \to {0^ \pm }} \frac{1}{{\sigma \left( u \right)}} .
\label{eq:properties_Lame_limit}
\end{equation}
Notice that $\sigma \left( u \right) = u + \mathcal{O} \left( u^5 \right)$, consequently, the limit $\mathop {\lim }\limits_{u \to 2n{\omega _1}^ \pm } {y_ \pm }\left( {u;a} \right)$ depends on whether is is taken from smaller or larger values than $2n{\omega _1}$. Applying \eqref{eq:properties_Lame_limit} to \eqref{eq:properties_elliptic_solution} yields
\begin{equation}
\mathop {\lim }\limits_{u \to 2n{\omega _1}^ \pm } Y = \frac{\Lambda }{{\sqrt {\wp \left( {{a_2}} \right) - \wp \left( {{a_1}} \right)} }}\left( {\begin{array}{*{20}{c}}
{\cosh \left( {{\ell _1}v + 2n\left( {\zeta \left( {{\omega _1}} \right){a_1} - \zeta \left( {{a_1}} \right){\omega _1}} \right)} \right)}\\
{\sinh \left( {{\ell _1}v + 2n\left( {\zeta \left( {{\omega _1}} \right){a_1} - \zeta \left( {{a_1}} \right){\omega _1}} \right)} \right)}\\
{\cos \left( {{\ell _2}v + i2n\left( {\zeta \left( {{\omega _1}} \right){a_2} - \zeta \left( {{a_2}} \right){\omega _1}} \right)} \right)}\\
{\sin \left( {{\ell _2}v + i2n\left( {\zeta \left( {{\omega _1}} \right){a_2} - \zeta \left( {{a_2}} \right){\omega _1}} \right)} \right)}
\end{array}} \right)\mathop {\lim }\limits_{u \to {0^ \pm }} \frac{1}{{\sigma \left( u \right)}} .
\end{equation}

It is convenient to define
\begin{align}
{\delta _1} &\equiv \zeta \left( {{\omega _1}} \right){a_1} - \zeta \left( {{a_1}} \right){\omega _1} ,\\
{\delta _2} &\equiv \zeta \left( {{\omega _1}} \right){a_2} - \zeta \left( {{a_2}} \right){\omega _1} .
\end{align}
The quantities $\delta_1$ and $\delta_2$ obey the following properties
\begin{align}
{\mathop{\rm Im}\nolimits} {\delta _1} = \frac{\pi }{2},&\quad \mathop {\lim }\limits_{\wp \left( {{a_1}} \right) \to {e_3}} {\mathop{\rm Re}\nolimits} {\delta _1} = 0, \label{eq:boundary_df_properties_1}\\
{\mathop{\rm Re}\nolimits} {\delta _2} = 0,&\quad \mathop {\lim }\limits_{\wp \left( {{a_2}} \right) \to {e_1}} {\mathop{\rm Im}\nolimits} {\delta _2} = 0,\quad \mathop {\lim }\limits_{\wp \left( {{a_2}} \right) \to {e_2}} {\mathop{\rm Im}\nolimits} {\delta _2} = \frac{\pi }{2} . \label{eq:boundary_df_properties_2}
\end{align}

Denoting as $\theta_\pm \left( v \right)$ and $\varphi_\pm \left( v \right)$ the angular coordinates of the trace of the extremal surface at the boundary sphere as $u \to {2 n \omega_1} ^+$ and as $u \to {2 \left( n + 1 \right) \omega_1} ^-$ respectively, we find
\begin{align}
\cot {\theta _ + } &=  \pm \sinh \left( {\omega\left( {{\varphi _ + } + {\varphi _0}} \right)} \right) ,\\
\cot {\theta _ - } &=  \pm \sinh \left( {\omega\left( {{\varphi _ - } + {\varphi _0} - \delta \varphi } \right)} \right) ,
\end{align}
where 
\begin{align}
\omega &= \frac{{{\ell _1}}}{{{\ell _2}}}, \label{eq:boundary_omega}\\
\delta \varphi &= \pi  - 2\left( {{\mathop{\rm Im}\nolimits} {\delta _2} + \frac{{{\ell _2}}}{{{\ell _1}}}{\mathop{\rm Re}\nolimits} {\delta _1}} \right) . \label{eq:boundary_df}
\end{align}
Each of these curves have a spiral form and endpoints at the north and south poles of the boundary sphere. As such, one of them cannot split the boundary to two regions, but the union of the two spirals does so.

Similarly, converting to \Poincare coordinates and denoting as $r_\pm \left( v \right)$ and $\varphi_\pm \left( v \right)$ the polar coordinates of the trace of the extremal surface at the boundary plane as $u \to {2 n \omega_1} ^+$ and as $u \to {2 \left( n + 1 \right) \omega_1} ^-$ respectively, we find
\begin{align}
{r_ + } &= \Lambda {e^{\omega \left( {{\varphi _ + } + {\varphi _0}} \right)}} ,\\
{r_ - } &= \Lambda {e^{\omega \left( {\varphi  + {\varphi _0} - \delta \varphi } \right)}} .
\end{align}
So in \Poincare coordinates, the trace of the minimal surface in the boundary is the union of two logarithmic spirals with the same exponential coefficient. The two curves comprising the trace of the minimal surface at the boundary are connected through a rotation of the angle $\varphi$ by $\delta \varphi$.

\begin{figure}[ht]
\centering
\begin{picture}(100,52)
\put(0,0){\includegraphics[width = 0.5\textwidth]{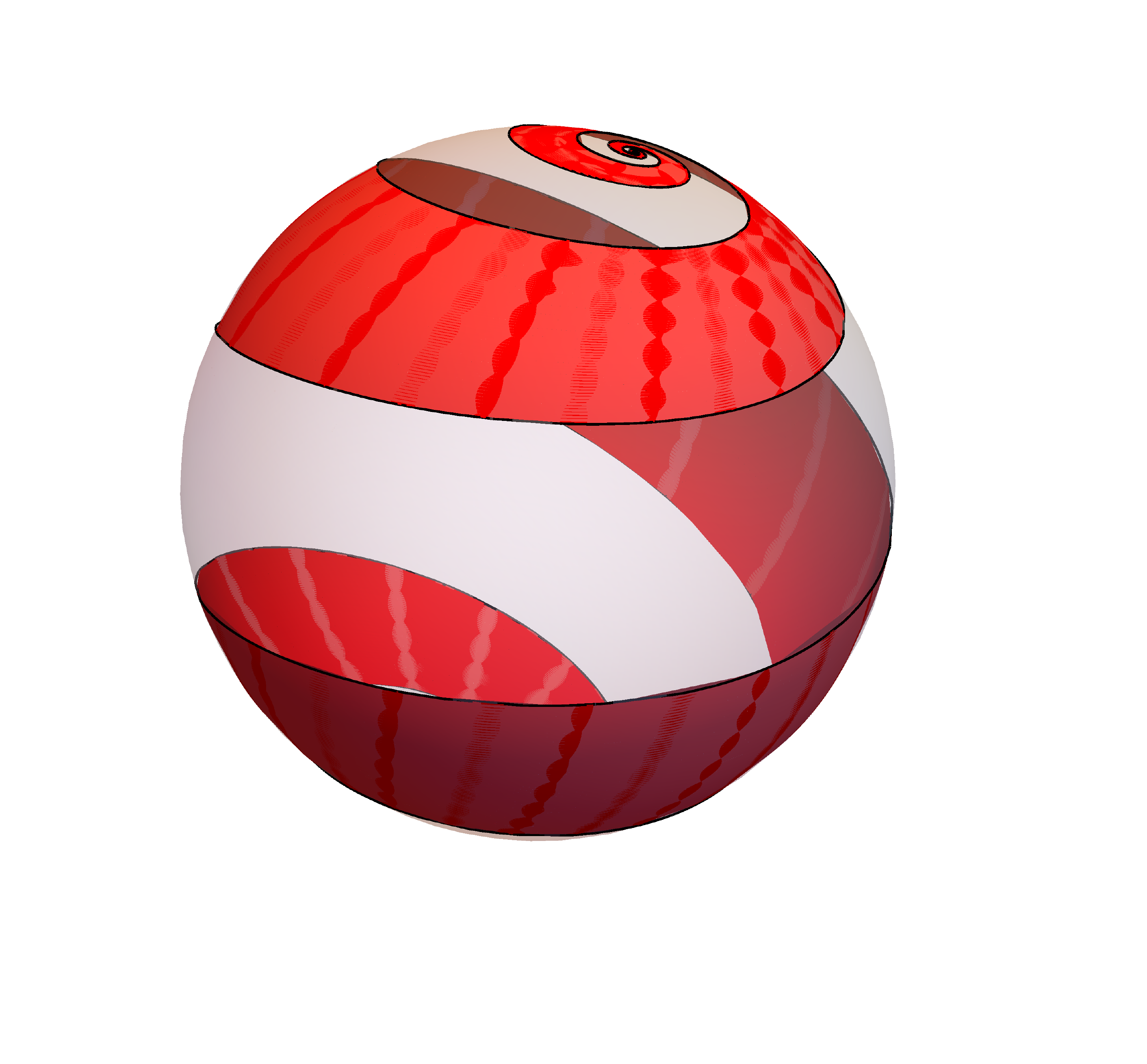}}
\put(50,10){\includegraphics[width = 0.5\textwidth]{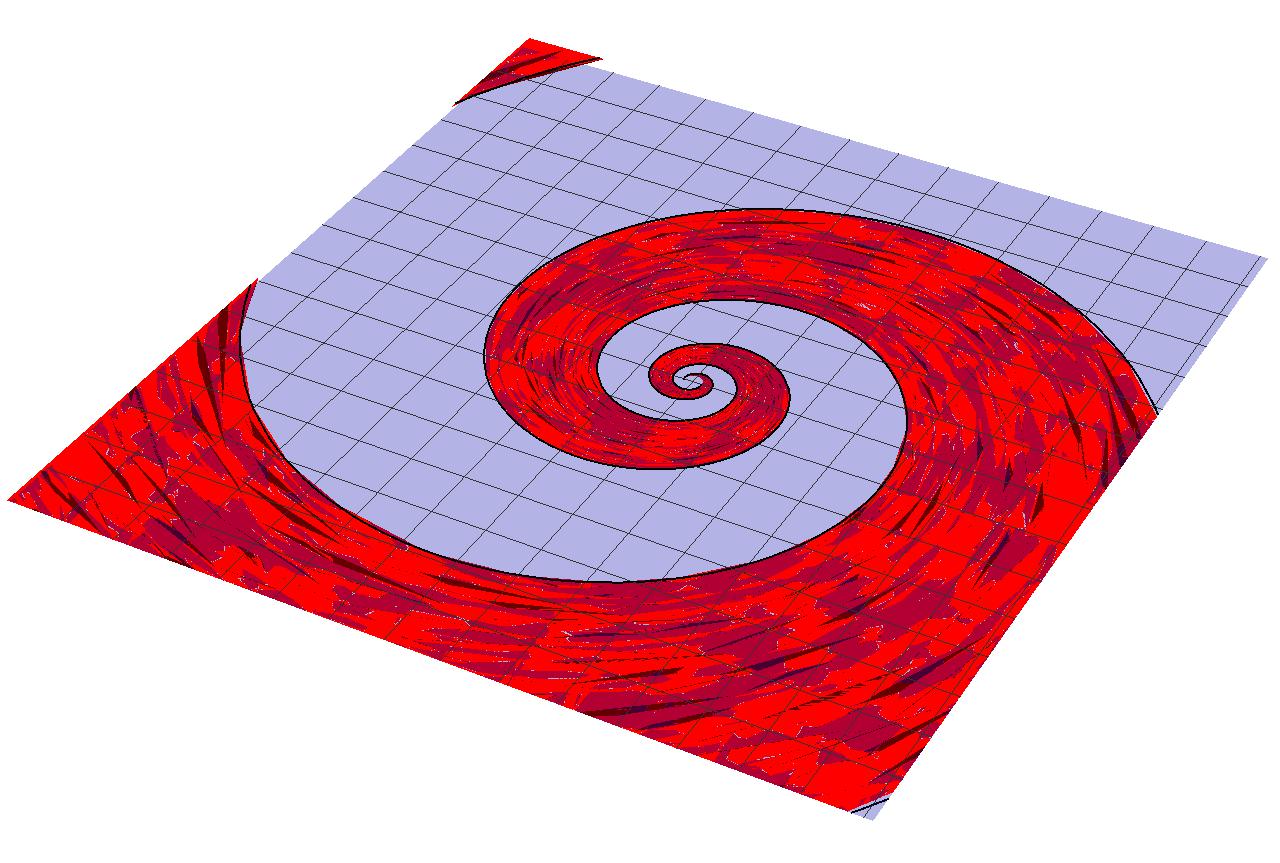}}
\end{picture}
\vspace{-45pt}
\caption{The entangling curve and the corresponding boundary regions in global and \Poincare coordinates}
\label{fig:boundary_region}
\end{figure}

In general, the entangling curve separates the boundary to two regions of unequal size. The ratio of the area of two regions is simply $ \left( \delta \varphi \mod 2 \pi \right) / \left(2 \pi - \delta \varphi \mod 2 \pi \right)$. The form of the entangling curve and the corresponding boundary regions in both global and \Poincare coordinates are displayed in figure \ref{fig:boundary_region}.

The entangling curve, which sets the Plateau problem for the minimal surface, is solely determined by the parameters $\omega$ and $\delta \varphi$. It is an interesting question, whether in the family of elliptic minimal surfaces \eqref{eq:properties_elliptic_solution}, there are different solutions corresponding to the same pair of $\omega$ and $\delta \varphi$ and thus to the same entangling curve. (Surfaces corresponding to the same $\omega$ and two $\delta \varphi$ that sum to $2 \pi$ also correspond to the same entangling curve.)

The parameter $\delta \varphi$ determines whether the minimal surface has self-intersections. Embedding minimal surfaces have the property
\begin{equation}
\delta \varphi < 2\pi.
\end{equation}
For $E>0$ the parameter $\omega$ can become arbitrarily close to zero as $\wp \left( a_1 \right)$ approaches $-2e_2$. Equation \eqref{eq:boundary_df} implies that at the same limit the angle $\delta \varphi$ becomes arbitrarily large and consequently larger than $2 \pi$. The special case $\omega = 0$ is an exception to this rule as we will show later.

\subsection{Area and Entanglement Entropy}
\label{subsec:Entanglement}
The minimal surface is spanned for $u$ and $v$ taking values in the ranges defined in equation \eqref{eq:boundary_spanning}. Consequently, the area of the minimal surface can be directly calculated with the use of formula \eqref{eq:reduction_area_formula},
\begin{equation}
A = {\Lambda ^2}\int_{ - \infty }^{ + \infty } {dv\int_{{2 n \omega _1}}^{{2 \left( n + 1 \right) \omega _1}} {du\left( {\wp \left( u \right) - {e_2}} \right)} } .
\end{equation}
The length of the entangling curve, as $u$ and $v$ are isothermal coordinates, can be expressed as
\begin{equation}
L = \mathop {\lim }\limits_{u \to {2 n \omega _1}^+} \Lambda \int_{ - \infty }^{ + \infty } {dv\sqrt {\wp \left( u \right) - {e_2}} } + \mathop {\lim }\limits_{u \to {2 \left( n + 1 \right) \omega _1}^-} \Lambda \int_{ - \infty }^{ + \infty } {dv\sqrt {\wp \left( u \right) - {e_2}} }.
\end{equation}
Straightforward application of relations \eqref{eq:Weierstrass_zeta} and \eqref{eq:Weierstrass_period_zeta} yields
\begin{equation}
\begin{split}
A &= {\Lambda ^2}\int_{ - \infty }^{ + \infty } {dv\left( {\mathop {\lim }\limits_{u \to {0^ + }} \zeta \left( u \right) - \mathop {\lim }\limits_{u \to {0^ - }} \zeta \left( u \right) - 2 \zeta \left( \omega_1 \right) - 2{e_2}{\omega _1}} \right)} \\
 &= 2{\Lambda ^2}\left( {\mathop {\lim }\limits_{u \to {0^ + }} \frac{1}{u} - \zeta \left( \omega_1 \right) - {e_2}{\omega _1}} \right) \int_{ - \infty }^{ + \infty } {dv},
\end{split}
\end{equation}
while
\begin{equation}
L = 2\Lambda \int_{ - \infty }^{ + \infty } {dv\mathop {\lim }\limits_{u \to 0^+} \frac{1}{u}} .
\end{equation}
Thus, we recover the usual ``area law''
\begin{equation}
A = \Lambda L - 2{\Lambda ^2}\left( \zeta \left( \omega_1 \right) + {e_2}{\omega _1}\right) \int_{ - \infty }^{ + \infty } {dv} .
\end{equation}

The universal constant term diverges. In global coordinates the divergence can be attributed to the non-smoothness of the entangling curve at the poles \cite{Bueno:2015xda}. In \Poincare coordinates, additionally, the entangling curve is infinite, similarly to the case of the minimal surface corresponding to an infinite strip. This divergence introduces a subtlety in the comparison of the areas of two distinct surfaces corresponding to the same entangling curve, as one may rescale $v$ for each of those at will. An appropriate regularization of the universal constant term must enforce that $v$ is connected to the physical position of a given point on the entangling curve. In both sets of coordinates, the azimuthal angle $\varphi$, which specifies uniquely a point on the spiral entangling curve, is given by $\varphi = \ell_2 v +\varphi_0$. Consequently, an appropriate redefinition of the parameter $v$ is
\begin{equation}
v = \frac{\varphi}{\ell_2} ,
\end{equation}
leading to
\begin{equation}
A = \Lambda L - {\sqrt 2 }{\Lambda ^2} \sqrt {\frac{ {1 - {\omega ^2}} }{E} } \left( {\frac{E}{3} \omega _1} + 2 \zeta \left( \omega_1 \right) \right) \int_{ - \infty }^{ + \infty } {d\varphi}.
\label{eq:properties_area}
\end{equation}

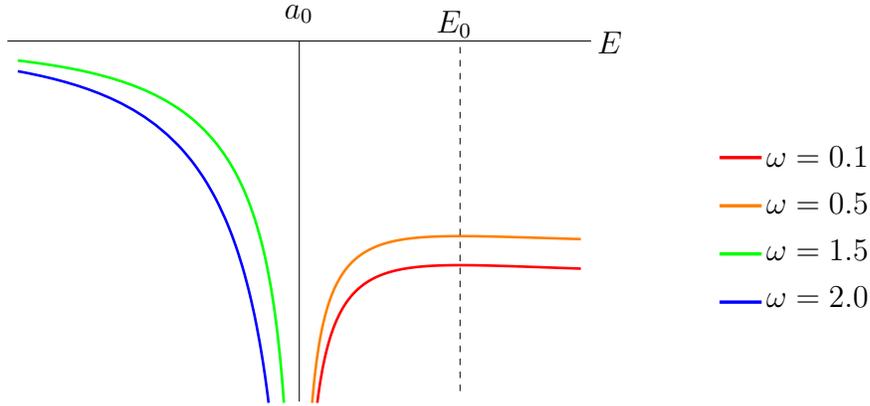
\begin{figure}[ht]
\centering
\begin{picture}(100,37)
\put(15,0.5){\includegraphics[width = 0.5\textwidth]{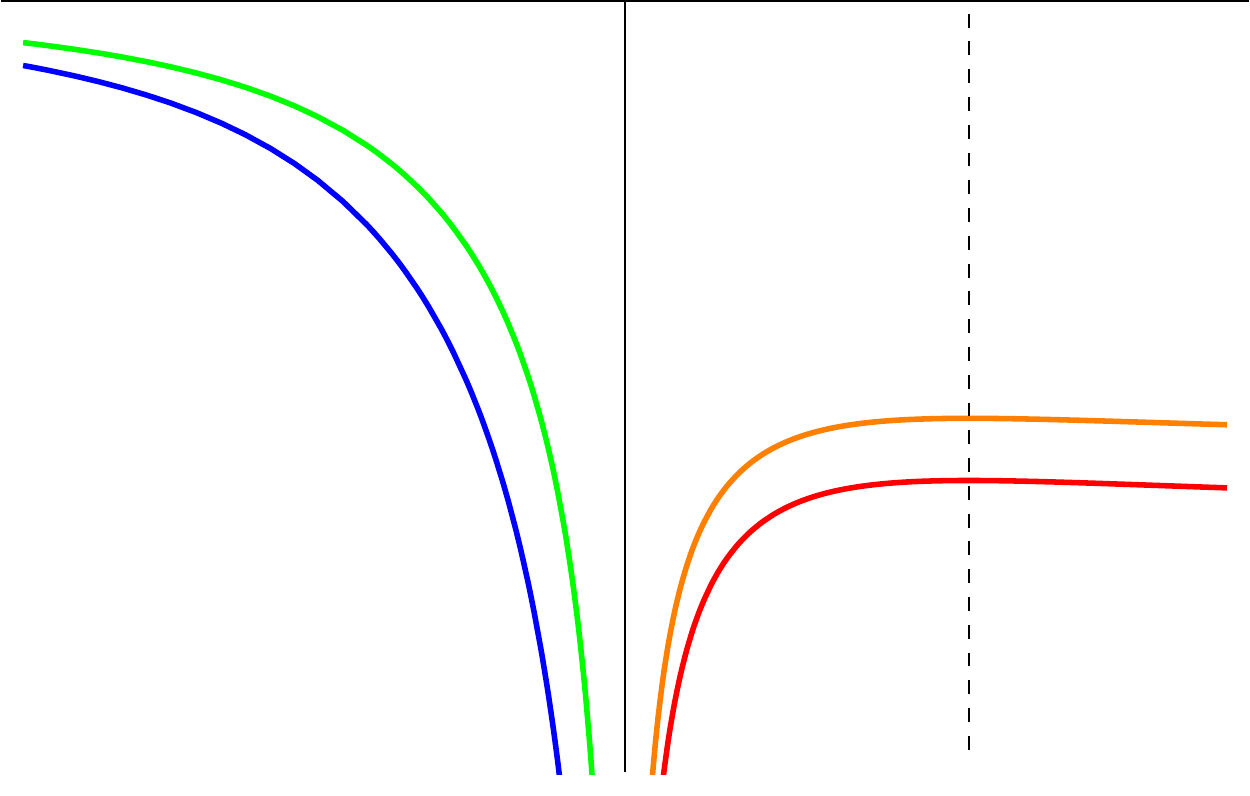}}
\put(75,7){\includegraphics[width = 0.07\textwidth]{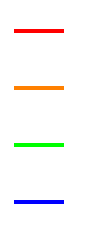}}
\put(65.5,31.25){$E$}
\put(38.75,34.25){$a_0$}
\put(51.75,33){$E_0$}
\put(80,21.5){$\omega = 0.1$}
\put(80,17.5){$\omega = 0.5$}
\put(80,13.5){$\omega = 1.5$}
\put(80,9.5){$\omega = 2.0$}
\end{picture}
\vspace{-20pt}
\caption{The coefficient $a_0$ as function of the integration constant $E$ for various $\omega$}
\label{fig:helicoid_a0}
\end{figure}

We define
\begin{equation}
a_0 \left( E , \omega \right) := - {\sqrt 2 }{\Lambda ^2} \sqrt {\frac{ {1 - {\omega ^2}} }{E} } \left( {\frac{E}{3} \omega _1 \left(E\right)} + 2 \zeta \left( \omega_1 \left(E\right) \right) \right),
\label{eq:properties_a0}
\end{equation}
which can be used as a measure of comparison for the areas corresponding to the same entangling curve. It can be shown that $a_0$ is always negative, it diverges to minus infinity at $E \to 0$ and
\begin{equation}
\begin{split}
\frac{\partial }{{\partial E}}a_0 \left( {E,\omega } \right) &< 0, \quad \mathrm{for} \quad E < 0 ,\\
\frac{\partial }{{\partial E}}a_0 \left( {E,\omega } \right) &> 0,\quad \mathrm{for} \quad 0 < E < {E_0} ,\\
\frac{\partial }{{\partial E}}a_0 \left( {E,\omega } \right) &< 0,\quad \mathrm{for} \quad E > {E_0} ,
\end{split}
\end{equation}
where $E_0$ is the energy constant maximizing the real period of the Weierstrass function, given by equation \eqref{eq:elliptic_E0}. Figure \ref{fig:helicoid_a0} depicts the dependence of $a_0$ on the energy constant $E$.

\subsection{Interesting Limits}
\label{subsec:limits}
In section \ref{subsec:moduli}, we showed that the space of allowed parameters $E$ and $\wp \left( a_1 \right)$ for elliptic minimal surfaces is bounded by three curves, $\wp \left( a_1 \right) = e_3 \left( E \right)$, $\wp \left( a_1 \right) = e_2 \left( E \right)$ and $\wp \left( a_1 \right) = - 2 e_2 \left( E \right)$. These curves correspond to interesting limits of the elliptic minimal surfaces.
\subsubsection{The Helicoid Limit}
\label{subsec:Helicoid}
One of the boundaries of the moduli space is $\wp \left( a_1 \right) = e_3$ and $\wp \left( a_2 \right) = e_1$, for all values of $E$. Comparing with the elliptic classical string solutions presented in \cite{Bakas:2016jxp}, this special limit is the analogue to the Gubser-Klebanov-Polyakov limit \cite{Gubser:2002tv} of the spiky string solutions.

At this special point, the solution acquires a simpler form. All wavefunctions of the $n=1$ \Lame problem that appear in the minimal surface solution become simultaneously real and periodic as they correspond to eigenvalues at the edges of the bands of the \Lame spectrum. Consequently, both functions $\varphi_1$ and $\varphi_2$ vanish identically and the solution acquires the simple form
\begin{equation}
Y = \frac{\Lambda }{{\sqrt {{e_1} - {e_3}} }}\left( {\begin{array}{*{20}{c}}
{\sqrt {\wp \left( u \right) - {e_3}} \cosh \left( {\sqrt {{e_1} - {e_2}} v} \right)}\\
{\sqrt {\wp \left( u \right) - {e_3}} \sinh \left( {\sqrt {{e_1} - {e_2}} v} \right)}\\
{\sqrt {\wp \left( u \right) - {e_1}} \cos \left( {\sqrt {{e_2} - {e_3}} v} \right)}\\
{\sqrt {\wp \left( u \right) - {e_1}} \sin \left( {\sqrt {{e_2} - {e_3}} v} \right)}
\end{array}} \right) ,
\end{equation}
which has the form of a helicoid in H$^3$. It is not surprising that the minimal surface being the analogue of the GKP solution, i.e. a rigidly rotating rod, is a ruled surface.

Converting to global coordinates on the hyperboloid H$^3$, the helicoid minimal surface take the following parametric form
\begin{equation}
\begin{split}
r &= \frac{\Lambda }{{\sqrt {{e_1} - {e_3}} }}\sqrt {\left( {\wp \left( u \right) - {e_3}} \right){{\cosh }^2}\sqrt {{e_1} - {e_2}} v - \left( {{e_1} - {e_3}} \right)} , \\
\theta  &= \tan^{-1} \left( \sqrt {\frac{{\wp \left( u \right) - {e_1}}}{{\wp \left( u \right) - {e_3}}}} \csch \sqrt {{e_1} - {e_2}} v \right) ,\\
\varphi  &= \sqrt {{e_2} - {e_3}} v ,
\end{split}
\end{equation}
which can be written in closed form as
\begin{equation}
r = \Lambda \sqrt {\frac{1}{{{{\cos }^2}\theta \csch^2 \omega \varphi  - {{\sin }^2}\theta }}} .
\label{eq:KGP_closed_form}
\end{equation}
where
\begin{equation}
{\omega ^2} = \frac{{{e_1} - {e_2}}}{{{e_2} - {e_3}}} = \frac{{ - E + \sqrt {{E^2} + 4{\Lambda ^{ - 4}}} }}{{ E + \sqrt {{E^2} + 4{\Lambda ^{ - 4}}} }}.
\label{eq:helicoid_omega}
\end{equation}

\begin{figure}[ht]
\centering
\begin{picture}(100,52)
\put(0,0){\includegraphics[width = 0.5\textwidth]{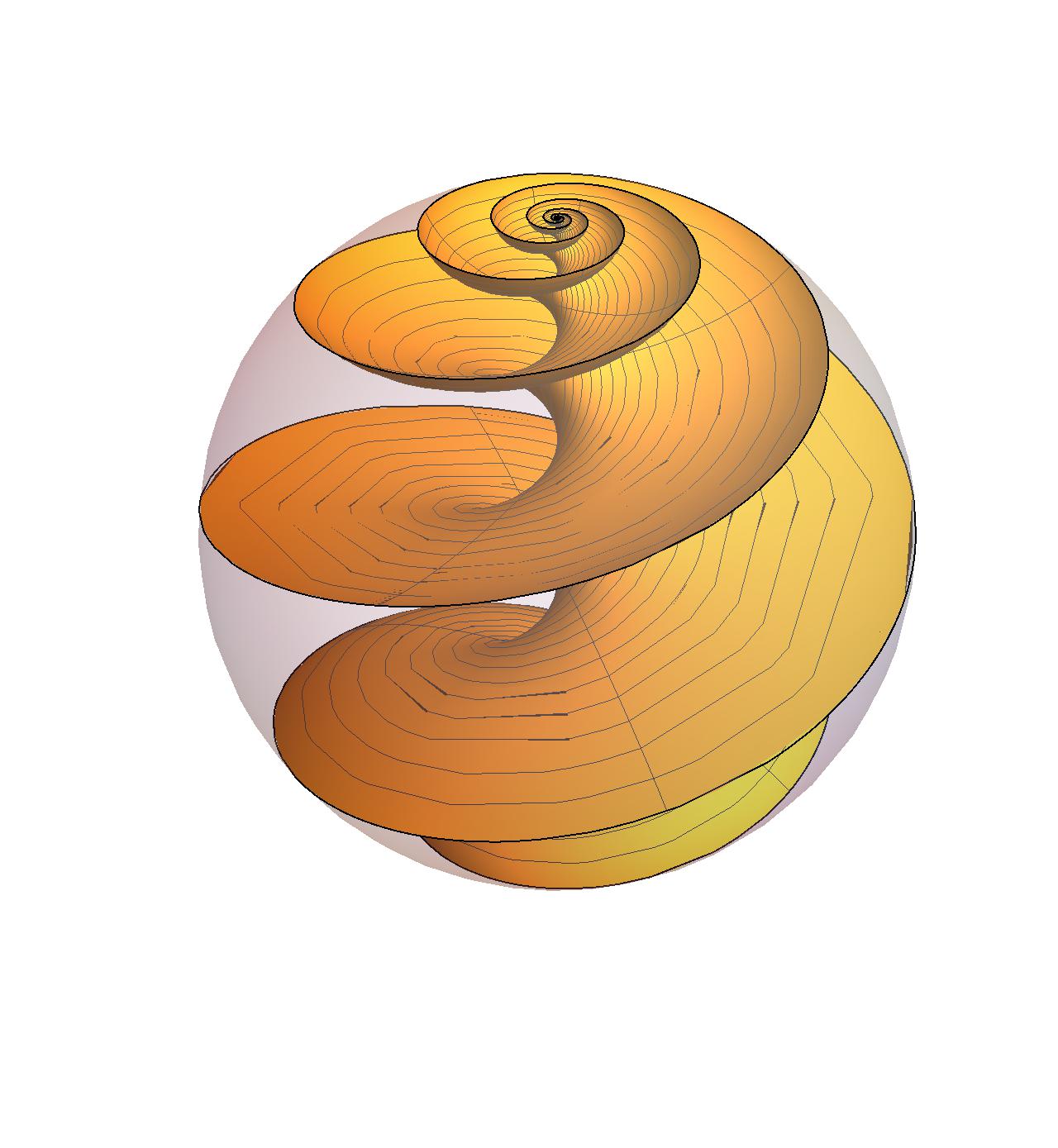}}
\put(50,10){\includegraphics[width = 0.5\textwidth]{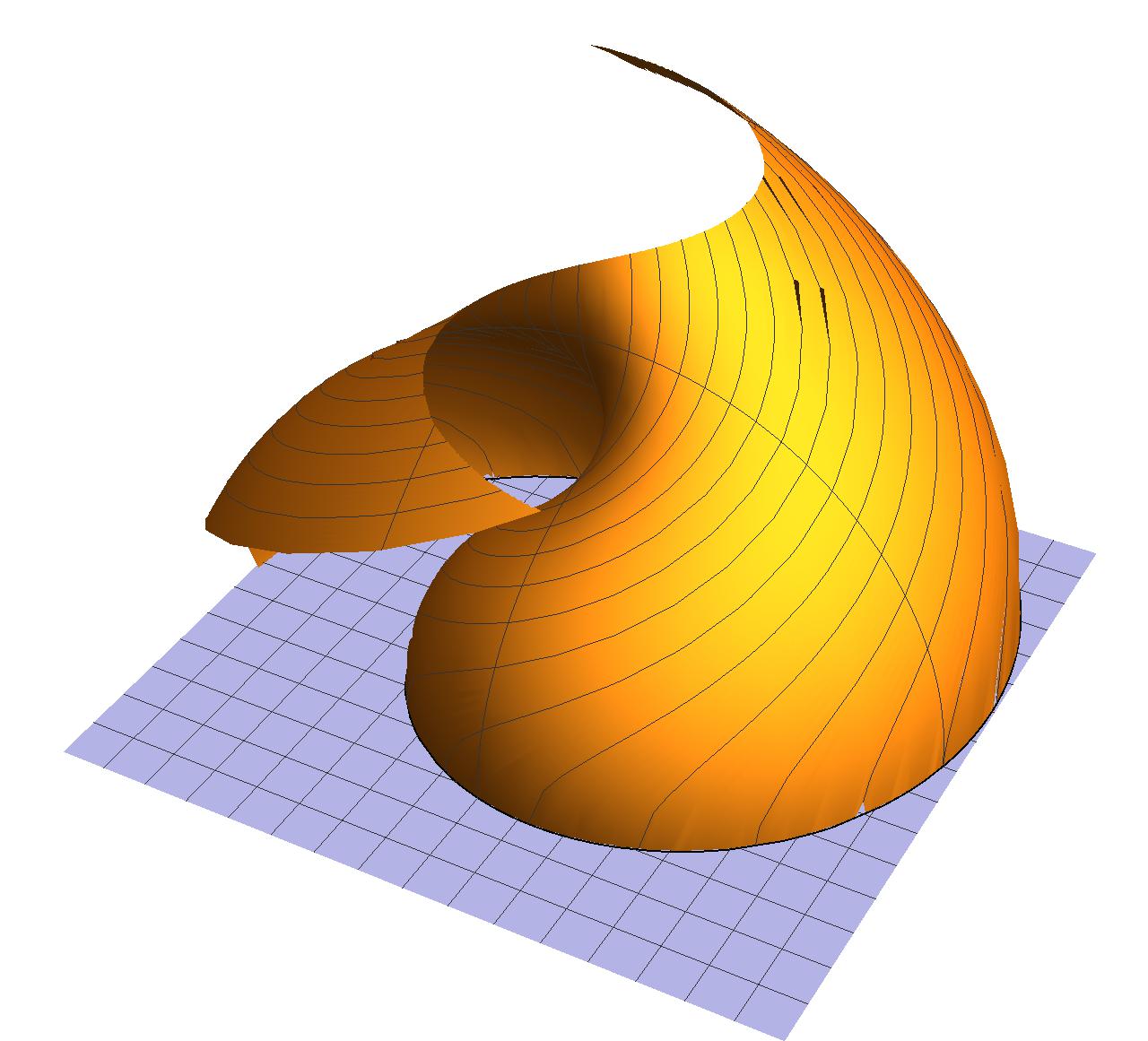}}
\end{picture}
\vspace{-45pt}
\caption{The helicoid minimal surface in global and \Poincare coordinates}
\label{fig:helicoid}
\end{figure}

Similarly, in \Poincare coordinates, the helicoid takes the parametric form
\begin{align}
z &= \Lambda \sqrt {\frac{{{e_1} - {e_3}}}{{\wp \left( u \right) - {e_3}}}} {e^{ - \sqrt {{e_1} - {e_2}} v}} ,\\
r &= \Lambda \sqrt {\frac{{\wp \left( u \right) - {e_1}}}{{\wp \left( u \right) - {e_3}}}} {e^{ - \sqrt {{e_1} - {e_2}} v}} ,\\
\varphi &= \sqrt {{e_2} - {e_3}} v ,
\end{align}
which can be written in closed form as
\begin{equation}
z = \sqrt {{\Lambda ^2}{e^{ - 2\omega \varphi }} - {r^2}} .
\end{equation}
The helicoid minimal surface in both sets of coordinates is depicted in figure \ref{fig:helicoid}.

As the integration constant $E$ tends to plus infinity, $\omega$ tends to zero and the minimal surface becomes the equatorial plane $\theta=\pi / 2$ in global coordinates. In \Poincare coordinates, in this limit the logarithmic spiral degenerates to a circle of radius $\Lambda$ and the minimal surface to the usual ``semi-sphere'' $z = \sqrt{\Lambda^2 - r^2} $. As the energy constant $E$ tends to minus infinity, $\omega$ tends to infinity and the minimal surface tends to the meridian plane $\tan \varphi = \tan \varphi_0$ in global coordinates, while in the \Poincare coordinates the entangling curve degenerates to a straight line passing through the origin and the minimal surface to the infinite semi-plane $\tan \varphi = \tan \varphi_0$.

Finally, properties \eqref{eq:boundary_df_properties_1} and \eqref{eq:boundary_df_properties_2} imply that the parameter $\delta \varphi$ for all helicoid minimal surfaces equals
\begin{equation}
\delta \varphi_{\mathrm{helicoid}} = \pi .
\end{equation}
Therefore, the entangling curve corresponding to a helicoid minimal surface separates the boundary to two regions of equal area.

\subsubsection{The Catenoid Limit}
\label{subsec:Catenoid}

The second boundary of the moduli space that we are going to consider is $\wp \left( a_1 \right) = - 2 e_2$ and $\wp \left( a_2 \right) = e_2$ for $E > 0$. In this case, only one of the two $n=1$ \Lame eigenfunctions that appear in the solution corresponds to the edge of a band of the spectrum and thus, it becomes simultaneously periodic and real, allowing the solution to reduce to
\begin{equation}
Y = \frac{\Lambda }{{\sqrt {3{e_2}} }}\left( {\begin{array}{*{20}{c}}
{\sqrt {\wp \left( u \right) + 2{e_2}} \cosh \left( {{\varphi _1}\left( u ; a_1 \right)} \right)}\\
{\sqrt {\wp \left( u \right) + 2{e_2}} \sinh \left( {{\varphi _1}\left( u ; a_1 \right)} \right)}\\
{\sqrt {\wp \left( u \right) - {e_2}} \cos \left( {\sqrt {3{e_2}} v} \right)}\\
{\sqrt {\wp \left( u \right) - {e_2}} \sin \left( {\sqrt {3{e_2}} v} \right)}
\end{array}} \right) ,
\end{equation}
where $\wp \left( a_1 \right) = - 2 e_2$.

\begin{figure}[ht]
\centering
\begin{picture}(100,52)
\put(0,0){\includegraphics[width = 0.5\textwidth]{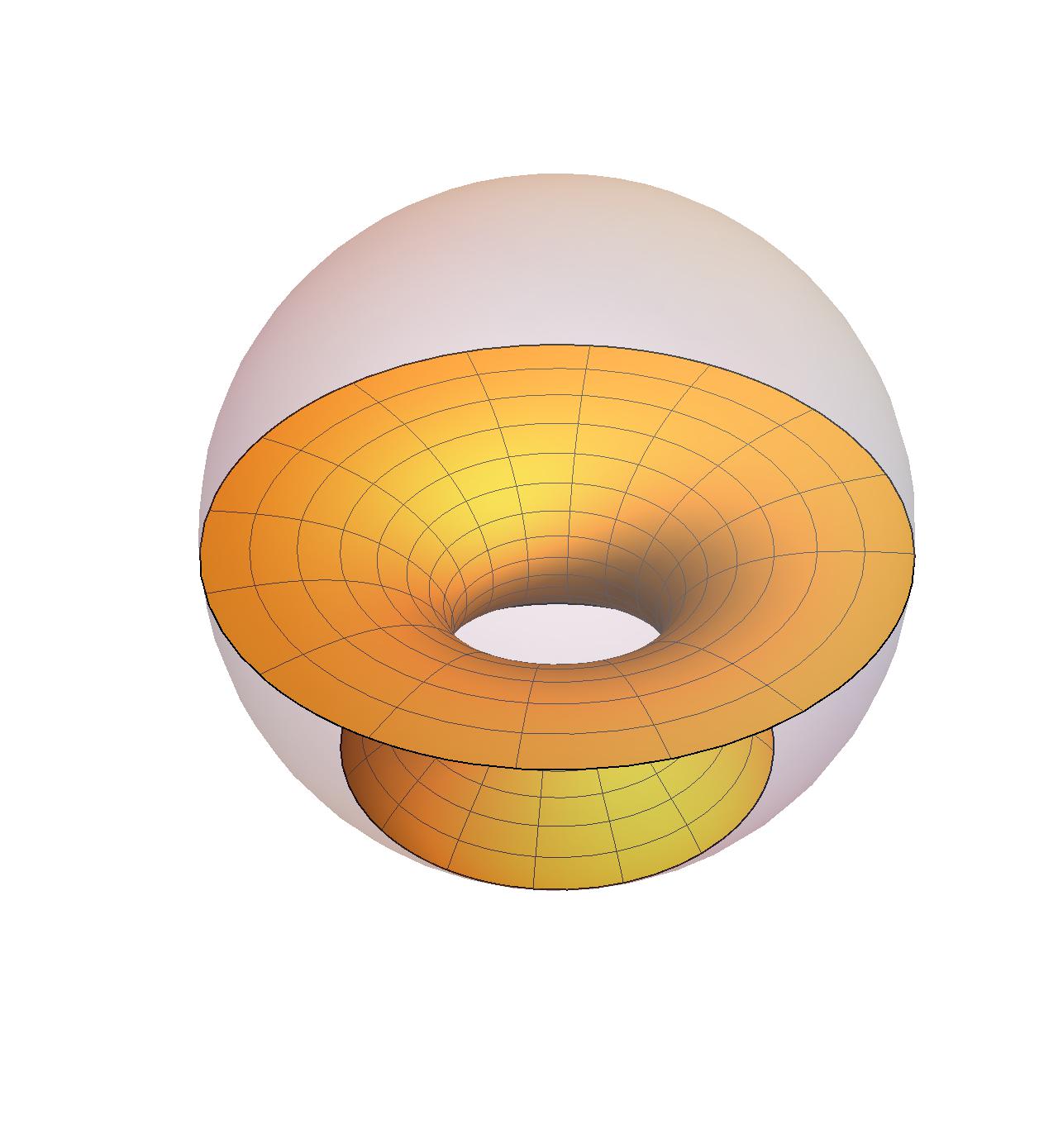}}
\put(50,10){\includegraphics[width = 0.5\textwidth]{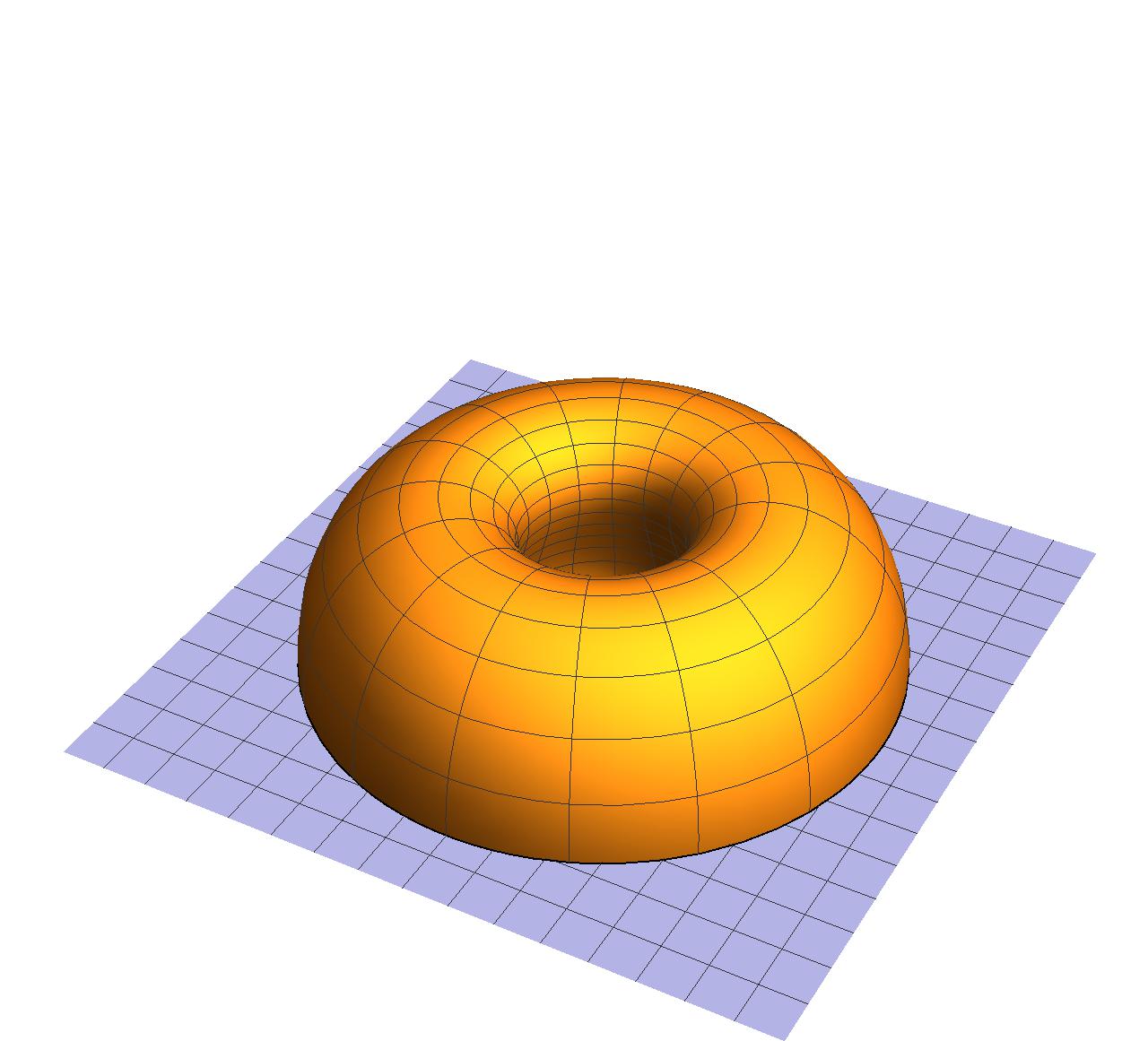}}
\end{picture}
\vspace{-45pt}
\caption{The catenoid minimal surface in global and \Poincare coordinates}
\label{fig:catenoid}
\end{figure}

Although, in this case we cannot acquire a closed form for the solution, the shape of the minimal surface can be understood by simple observations. Converting to global coordinates we find
\begin{align}
r &= \frac{\Lambda }{{\sqrt {3{e_2}} }}\sqrt {\wp \left( u \right){{\cosh }^2}\left( {{\varphi _1}\left( {u;{a_1}} \right)} \right) + \left( {\cosh \left( {2{\varphi _1}\left( {u;{a_1}} \right)} \right) - 2} \right){e_2}} = r\left( u \right) ,\\
\theta  &= \tan^{-1} \left( {\sqrt {\frac{{\wp \left( u \right) - {e_2}}}{{\wp \left( u \right) + 2{e_2}}}} \csch \left( {{\varphi _1}\left( {u;{a_1}} \right)} \right)} \right) = \theta \left( u \right) ,\\
\varphi &= \sqrt {3{e_2}} v = \varphi \left( v \right) .
\end{align}
Therefore, the surface can be expressed in the form
\begin{equation}
f \left( r , \theta \right) = 0
\end{equation}
and consequently it is a surface of revolution. Since it is both a minimal surface and a surface of revolution, it is by definition a catenoid in H$^3$. Similarly in \Poincare coordinates, we acquire the expression
\begin{align}
z &= \Lambda \sqrt {\frac{{3{e_2}}}{{\wp \left( u \right) + 2{e_2}}}} {e^{ - {\varphi _1}\left( {u;{a_1}} \right)}} = z \left( u \right),\\
r &= \Lambda \sqrt {\frac{{\wp \left( u \right) - {e_2}}}{{\wp \left( u \right) + 2{e_2}}}} {e^{ - {\varphi _1}\left( {u;{a_1}} \right)}} = r \left( u \right) ,\\
\varphi &= \sqrt {3{e_2}} v = \varphi \left( v \right),
\end{align}
having the same interpretation of a surface by revolution. Figure \ref{fig:catenoid} depicts such a catenoid in global and \Poincare coordinates.

\begin{figure}[ht]
\centering
\begin{picture}(100,37)
\put(25,0.5){\includegraphics[width = 0.5\textwidth]{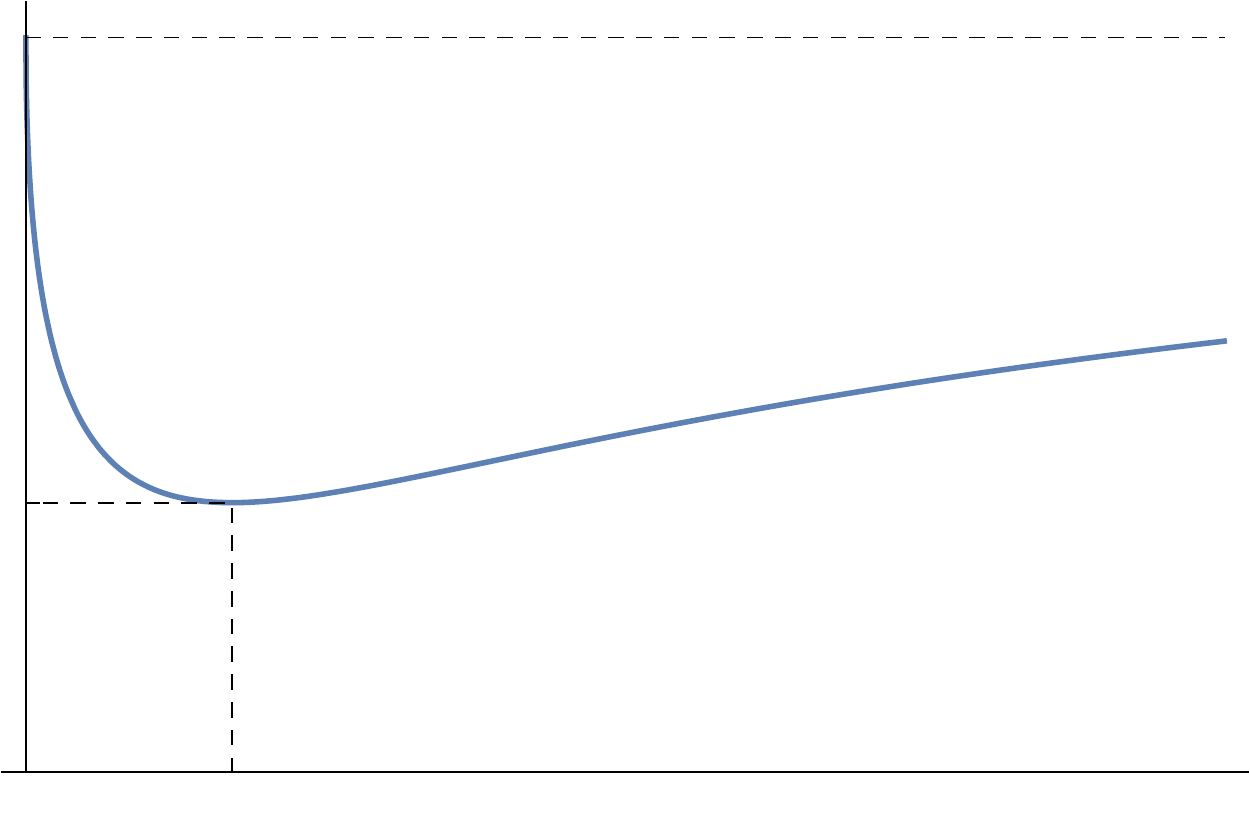}}
\put(75.5,1.5){$E$}
\put(25.5,0){$0$}
\put(32.75,0){$E_0$}
\put(24.75,35.25){$\frac{r_-}{r_+}$}
\put(18.5,12.5){$\left(\frac{r_-}{r_+}\right)_0$}
\put(24.25,31){$1$}
\end{picture}
\vspace{-20pt}
\caption{The ratio of the radii of the circles comprising the entangling curve of a catenoid in \Poincare coordinates, as function of the integration constant $E$}
\label{fig:catenoid_radii}
\end{figure}
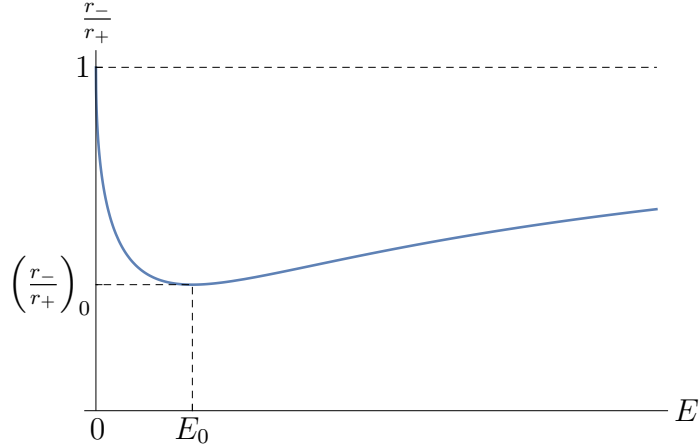

Following the discussion of section \ref{subsec:Boundary_Regions} the parameters $\omega$ and $\delta \varphi$ that specify the shape of the entangling curve at the catenoid limit acquire the values
\begin{align}
\omega_{\mathrm{catenoid}} &= 0 ,\\
\delta \varphi_{\mathrm{catenoid}} &= + \infty .
\end{align}
As all catenoids are characterized by the same degenerate $\omega$ and $\delta \varphi$, the corresponding entangling curve cannot be identified by these parameters for this class of surfaces. The catenoid minimal surface in global coordinates extends between angles
\begin{align}
\cot {\theta _ + } &= {\left( { - 1} \right)^n}\sinh \left( {2n{\mathop{\rm Re}\nolimits} {\delta _1}} \right) ,\\
\cot {\theta _ - } &= {\left( { - 1} \right)^n}\sinh \left( {2\left( {n + 1} \right){\mathop{\rm Re}\nolimits} {\delta _1}} \right) ,
\end{align}
which define two circles parallel to the equator that comprise the entangling curve. In \Poincare coordinates, the entangling curve comprises of two concentric circles with radii $r_+$ and $r_-$. In the following, the ratio ${{{r_ - }}}/{{{r_{_ + }}}} $ is used to characterize the form of the entangling curve in the case of catenoids. This ratio is given by
\begin{equation}
\frac{{{r_ - }}}{{{r_{_ + }}}} = {e^{2{\mathop{\rm Re}\nolimits} {\delta _1}}} 
\end{equation}
and it is plotted versus the integration constant $E$ in figure \ref{fig:catenoid_radii}.

The ratio ${{{r_ - }}}/{{{r_{_ + }}}} $ obeys $\mathop {\lim }\limits_{E \to 0} \frac{{{r_ - }}}{{{r_{_ + }}}} = \mathop {\lim }\limits_{E \to \infty } \frac{{{r_ - }}}{{{r_{_ + }}}} = 1$ and it acquires its minimum value
\begin{equation}
\left(\frac{r_-}{r_+}\right)_0 \simeq 0.367039
\end{equation}
at $E=E_0$. For ratios $\left(\frac{r_-}{r_+}\right) > \left(\frac{r_-}{r_+}\right)_0$ there are two catenoids anchored at the same entangling curve, while for $\left(\frac{r_-}{r_+}\right) < \left(\frac{r_-}{r_+}\right)_0$ there is none.

Unlike the general case, where the parameter $v$ has to take values in the whole real axis in order to span the minimal surface, in the catenoid limit the range of the coordinate $v$ becomes finite and specifically $v \in \left[0 , 2 \pi / \sqrt{3 e_2} \right)$. It is a direct consequence that the universal constant term in the area formula \eqref{eq:properties_area} becomes finite and specifically,
\begin{equation}
A_{\mathrm{catenoid}} = \Lambda L - 2\pi {\Lambda ^2}\sqrt {\frac{2}{E}} \left( {\frac{{E}}{3} {\omega _1} + 2\zeta \left( {{\omega _1}} \right)} \right) .
\end{equation}
In the case of catenoids it is convenient to define the quantity
\begin{equation}
a_0^{\rm{catenoid}} \left(E\right) := - 2\pi {\Lambda ^2}\sqrt {\frac{2}{E}} \left( {\frac{{E}}{3} {\omega _1\left(E\right)} + 2\zeta \left( {{\omega _1\left(E\right)}} \right)} \right) = 2\pi a_0 \left( E , 0 \right) ,
\end{equation}
which can be used to compare the area of catenoids corresponding the same entangling curve. The quantity $a_0^{\rm{catenoid}}$ has the same monotonicity properties as $a_0$.

\subsubsection{The Conical Limit}
\label{subsec:Conical}

The last boundary of the moduli space of the elliptic minimal surfaces is $\wp \left( a_1 \right) = e_2$ and $\wp \left( a_2 \right) = - 2 e_2$ for $E < 0$. Similarly to the catenoid limit, only one of the two \Lame eigenfunctions becomes real and periodic and the solution reduces to
\begin{equation}
Y = \frac{\Lambda }{{\sqrt { - 3{e_2}} }}\left( {\begin{array}{*{20}{c}}
{\sqrt {\wp \left( u \right) - {e_2}} \cosh \left( {\sqrt { - 3{e_2}} v} \right)}\\
{\sqrt {\wp \left( u \right) - {e_2}} \sinh \left( {\sqrt { - 3{e_2}} v} \right)}\\
{\sqrt {\wp \left( u \right) + 2{e_2}} \cos \left( {{\varphi _2}\left( u ; a_2 \right)} \right)}\\
{ - \sqrt {\wp \left( u \right) + 2{e_2}} \sin \left( {{\varphi _2}\left( u ; a_2 \right)} \right)}
\end{array}} \right) ,
\end{equation}
where $\wp \left( a_2 \right) = - 2 e_2$.

\begin{figure}[ht]
\centering
\begin{picture}(100,52)
\put(0,0){\includegraphics[width = 0.5\textwidth]{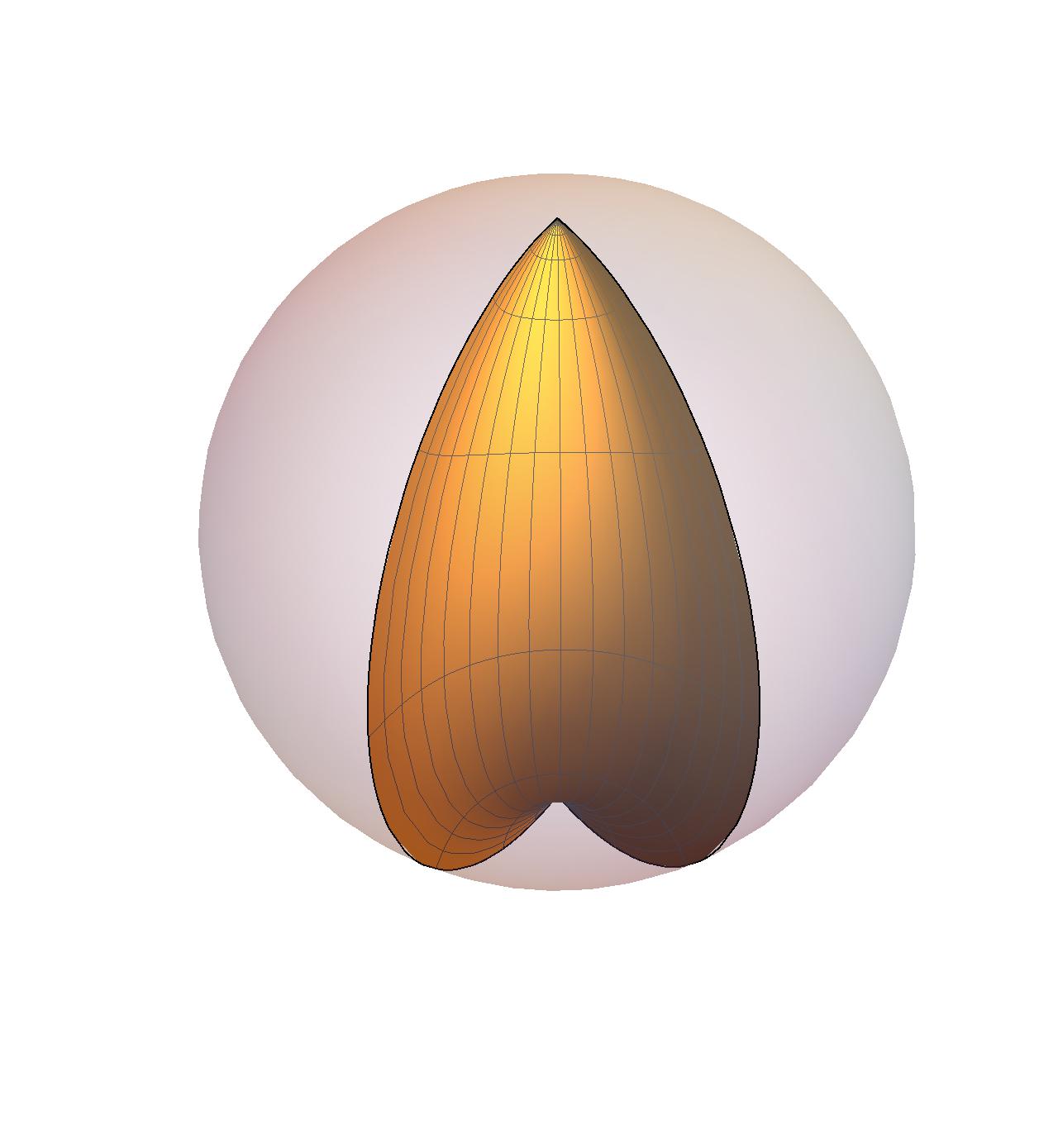}}
\put(50,10){\includegraphics[width = 0.5\textwidth]{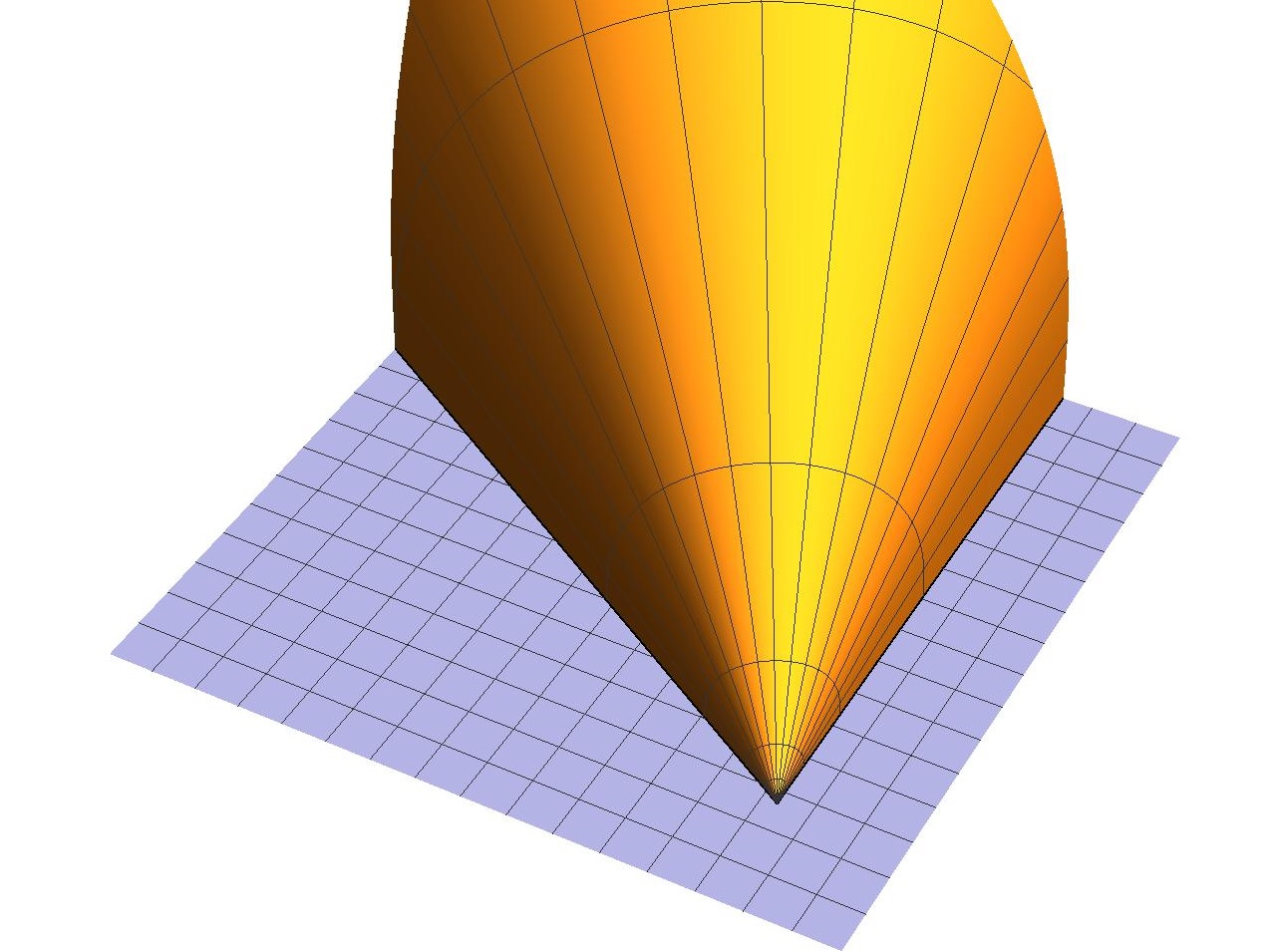}}
\end{picture}
\vspace{-45pt}
\caption{The conical minimal surface in global and \Poincare coordinates}
\label{fig:conical}
\end{figure}

Converting to global coordinates, we find
\begin{align}
r &= \frac{\Lambda }{{\sqrt { - 3{e_2}} }}\sqrt {\left( \wp \left( u \right) -e_2 \right){{\cosh }^2}\left( {\sqrt { - 3{e_2}v} } \right) +3{e_2}} ,\\
\theta  &= \tan^{-1} \left( \sqrt {\frac{{\wp \left( u \right) + 2{e_2}}}{{\wp \left( u \right) - {e_2}}}} \csch \left( {\sqrt { - 3{e_2}v} } \right) \right) ,\\
\varphi  &= - {\varphi _2}\left( {u;{a_2}} \right) ,
\end{align}
implying that the specific case of minimal surfaces can be written in the form
\begin{equation}
f\left( {r\sin \theta ,\varphi } \right) = 0 .
\end{equation}
In \Poincare coordinates we find
\begin{align}
z &= \Lambda \sqrt {\frac{{ - 3{e_2}}}{{\wp \left( u \right) - {e_2}}}} {e^{ - \sqrt { - 3{e_2}v} }} ,\\
r &= \Lambda \sqrt {\frac{{\wp \left( u \right) + 2{e_2}}}{{\wp \left( u \right) - {e_2}}}} {e^{ - \sqrt { - 3{e_2}v} }} ,\\
\varphi  &= {\varphi _2}\left( {u;{a_2}} \right) ,
\end{align}
implying that the minimal surface can be expressed in the form
\begin{equation}
f\left( {\frac{r}{z},\varphi } \right) = 0 .
\end{equation}
This expression describes a conical surface with the tip of the cone placed at the origin of the boundary plane. Figure \ref{fig:conical} depicts the conical minimal surfaces in global and \Poincare coordinates.

The parameters $\omega$ and $\delta \varphi$ specifying the entangling curve, at the limit of the conical minimal surfaces take the values
\begin{align}
\omega_{\mathrm{conical}} &= \infty , \\
\delta \varphi_{\mathrm{conical}} &= \pi - 2 {\mathop{\rm Im}\nolimits}  \delta_2 .
\end{align}

In the specific case of the conical minimal surfaces, the area formula \eqref{eq:properties_area} becomes problematic. The reason is the fact that the azimuthal angle $\varphi$ is a function solely of $u$. Thus, the substitution of $v$ with the azimuthal angle $\varphi$, performed to introduce an integration variable that is geometrically connected to the points of the entangling curve, unfortunately fails. In this case $v$ is related with the polar angle. An appropriate redefinition is $x = - 3 e_2 v$ and it yields
\begin{equation}
A_{\mathrm{conical}} = \Lambda L + \frac{2}{3} {\Lambda ^2} \left( { {\omega _1} + \frac{\zeta \left( {{\omega _1}} \right)}{e_2}} \right) \int_{ - \infty }^{ + \infty } {dx}. 
\end{equation} 
The universal term is diverging due to the non-smoothness of the entangling curve.

\section{Geometric Phase Transitions}
\label{sec:Phase_Transitions}
\subsection{Spiral Entangling Curves}
\label{subsec:PT_Helicoids}
In general the entangling curve for the elliptic minimal surfaces separates the boundary sphere to two regions. The only exception to this rule is the case of the catenoids, where the entangling curve is the union of two disjoint circles and consequently separates the boundary sphere to three regions. Thus, as long as we don't study catenoid elliptic minimal surfaces, there is no way to find two different minimal surfaces corresponding to the same entangling curve as a result of topological rearrangement of the matching of minimal surfaces and entangling curves.

However, one has to examine whether several of the elliptic minimal surfaces correspond to the same boundary curve. As shown in section \ref{subsec:Boundary_Regions}, the boundary curve is determined solely by the parameters $\omega$ and $\delta \varphi$. As the dependence of $\delta \varphi$ on the primary parameters $E$ and $\wp \left( a_1 \right)$ is quite complicated, the simpler way to determine whether there are minimal surfaces with the same entangling curve is plotting $\delta \varphi$ versus the energy constant along constant $\omega$ curves in the moduli space of solutions. Such constant $\omega$ curves have the form
\begin{equation}
\wp \left( {{a_1}} \right) = \frac{{{\omega ^2} + 2}}{{{\omega ^2} - 1}}\frac{E}{6} .
\end{equation}
The constant $\omega$ curves for $\omega < 1$ lie entirely in the $E>0$ region, whereas parameters $\omega > 1$ lie entirely in the $E<0$ region. The segment of each constant $\omega$ curve within the allowed region of parameters for elliptic minimal surface solutions has one endpoint being a helicoid with $E = E_h \left( \omega \right)$, where
\begin{equation}
E_h \left( \omega \right) = \frac{1}{\omega} - \omega ,
\end{equation}
whereas the other endpoint is always the $E = \wp \left( a_1 \right) = 0$ position.

\begin{figure}[ht]
\centering
\begin{picture}(100,96)
\put(0,44){\includegraphics[width = 0.5\textwidth]{helicoid_global.jpg}}
\put(50,44){\includegraphics[width = 0.5\textwidth]{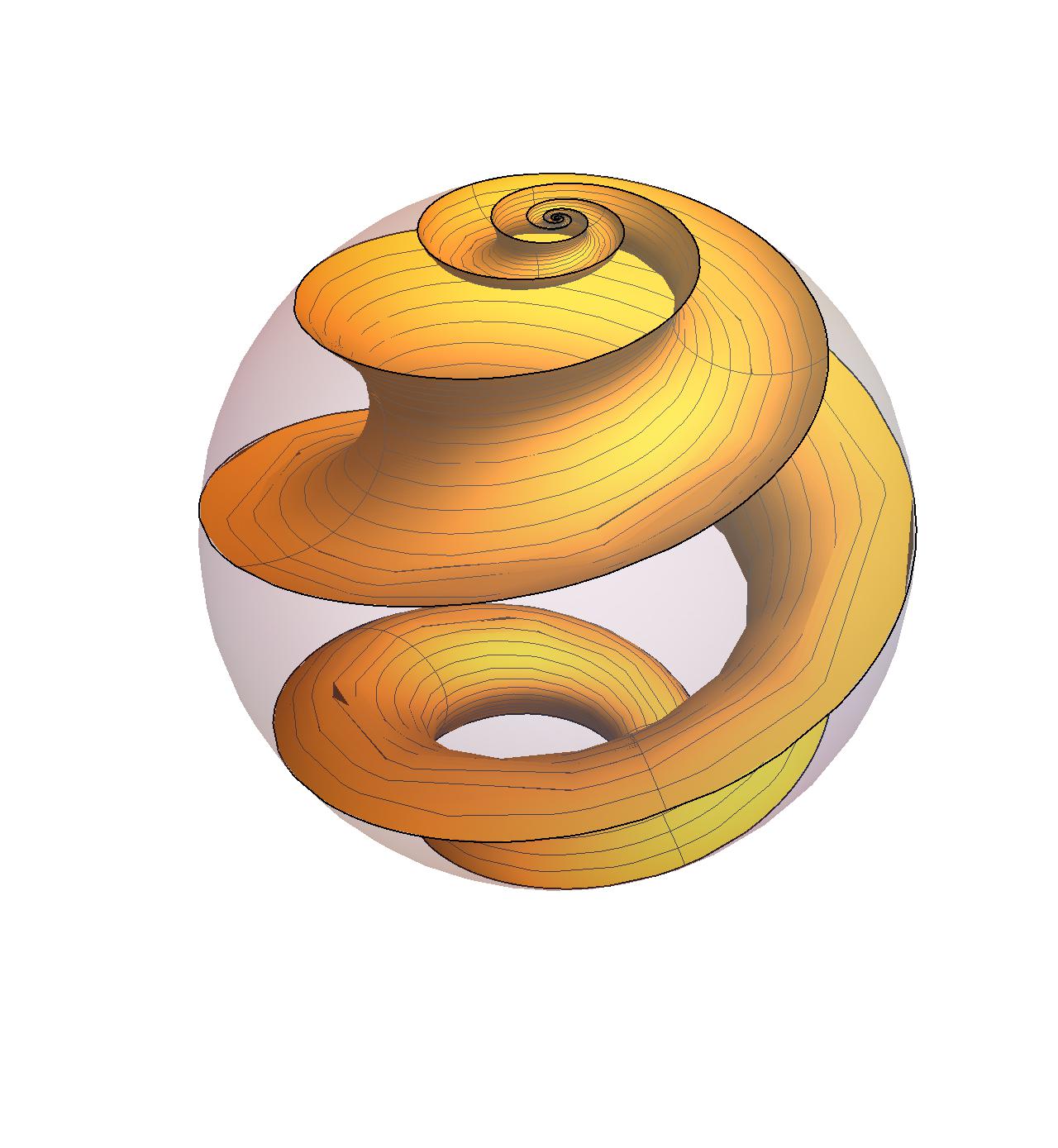}}
\put(0,10){\includegraphics[width = 0.5\textwidth]{helicoid_poincare.jpg}}
\put(50,10){\includegraphics[width = 0.5\textwidth]{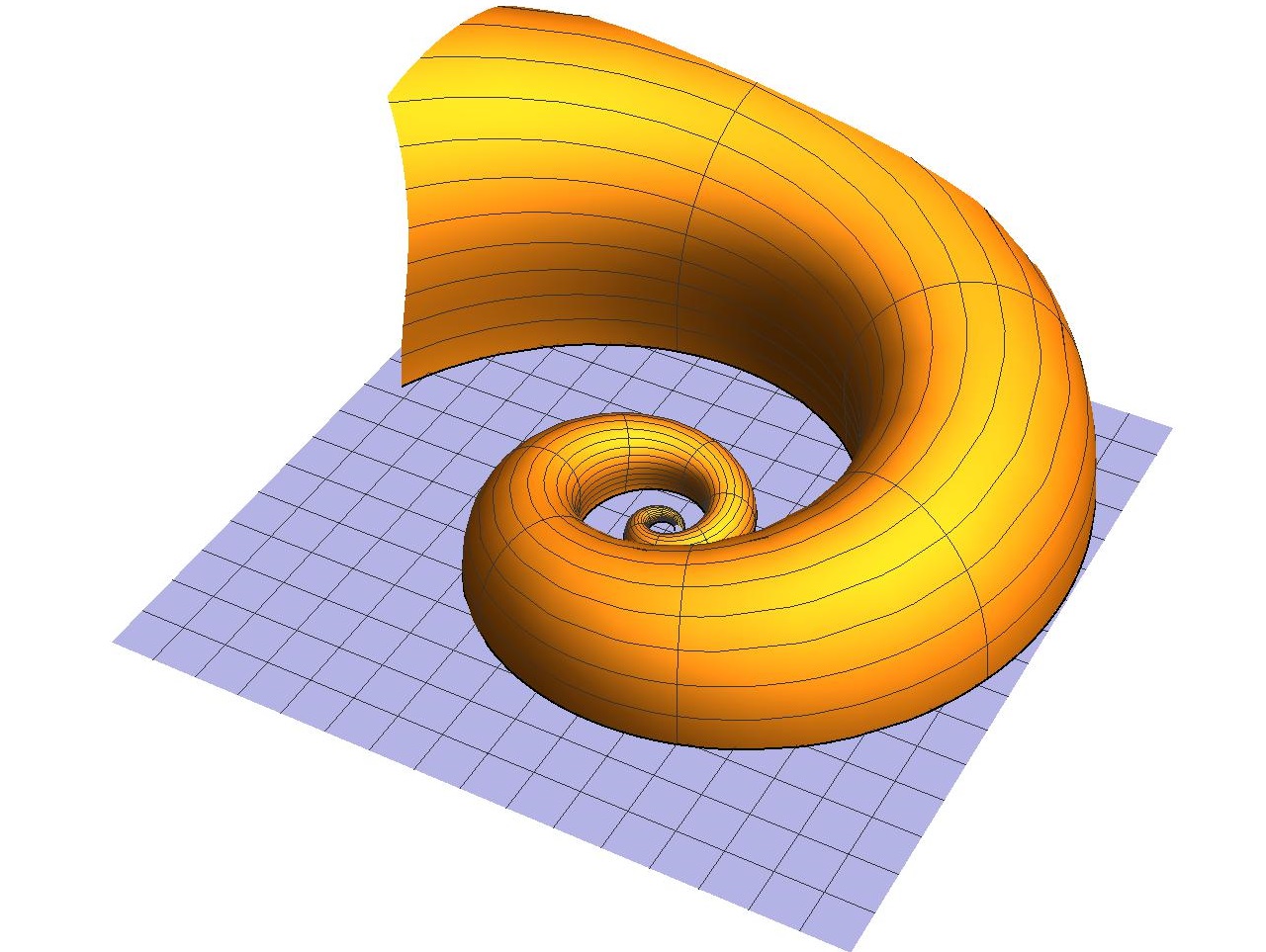}}
\end{picture}
\vspace{-45pt}
\caption{Two minimal surfaces corresponding to the same boundary curve defined by $\omega = 1 / 4$ and $\delta \varphi = \pi$}
\label{fig:helicoid_PT}
\end{figure}

It can be shown that $\delta \phi$ as function of $E$ and $\omega$ has the following properties,
\begin{equation}
\mathop {\lim }\limits_{E \to 0} \delta \varphi \left( {E,\omega } \right) = 0,\quad \mathop {\lim }\limits_{E \to {E_h}\left( \omega  \right)} \delta \varphi \left( {E,\omega } \right) = \pi 
\end{equation}
and
\begin{equation}
\begin{split}
\frac{\partial }{{\partial E}}\delta \varphi \left( {E,\omega } \right) &< 0, \quad \mathrm{for} \quad E < 0 ,\\
\frac{\partial }{{\partial E}}\delta \varphi \left( {E,\omega } \right) &> 0,\quad \mathrm{for} \quad 0 < E < {E_0} ,\\
\frac{\partial }{{\partial E}}\delta \varphi \left( {E,\omega } \right) &< 0,\quad \mathrm{for} \quad E > {E_0} .
\end{split}
\end{equation}
\begin{figure}[ht]
\centering
\begin{picture}(100,40)
\put(14.5,2.5){\includegraphics[width = 0.5\textwidth]{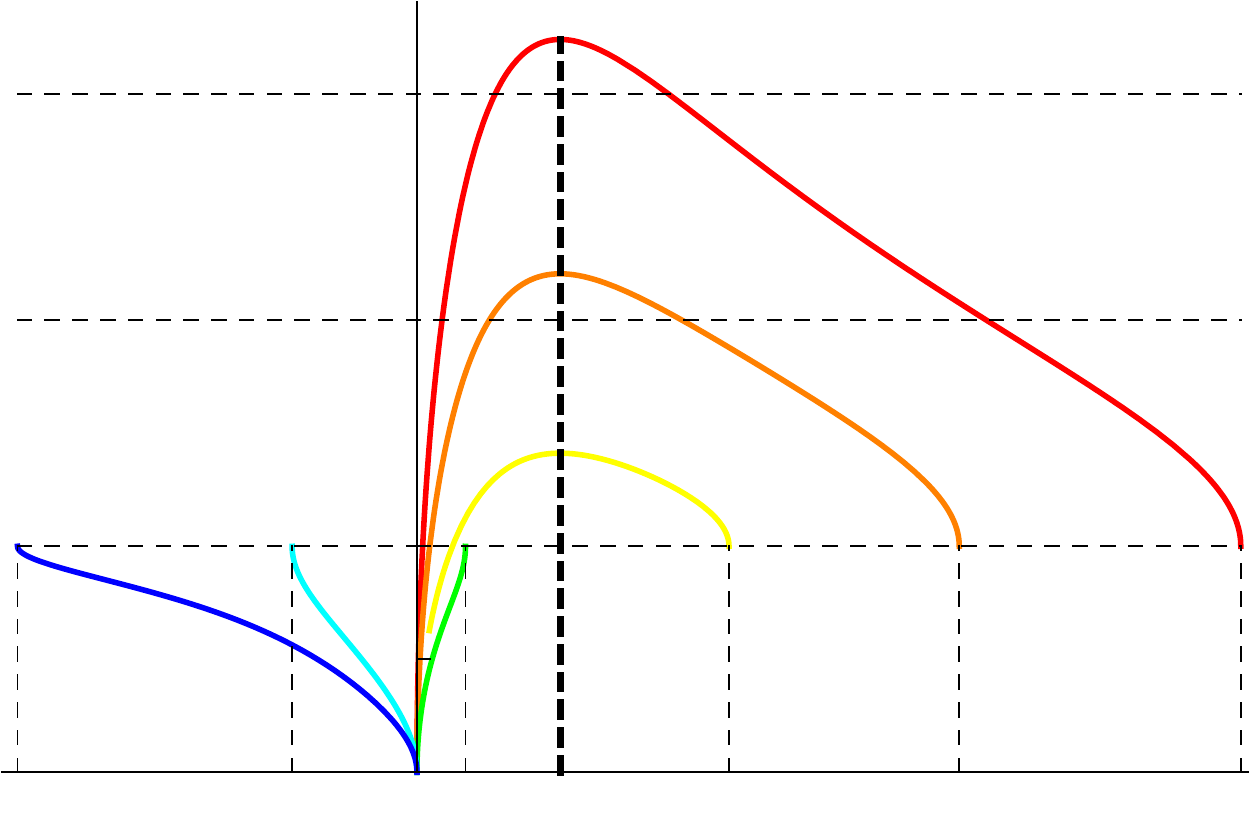}}
\put(75,4.75){\includegraphics[width = 0.08\textwidth]{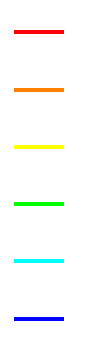}}
\put(81,31){$\omega = \omega_1 = 0.10$}
\put(81,26.25){$\omega = \omega_2 = 0.15$}
\put(81,21.5){$\omega = \omega_3 = 0.25$}
\put(81,16.75){$\omega = \omega_4 = 0.75$}
\put(81,12){$\omega = \omega_5 = 2$}
\put(81,7.25){$\omega = \omega_6 = 5$}
\put(11.5,16){\rotatebox{90}{$\delta \varphi$}}
\put(38,-2){$E$}
\put(11,1.5){$E_h(\omega_6)$}
\put(22,1.5){$E_h(\omega_5)$}
\put(40,1.5){$E_h(\omega_3)$}
\put(49,1.5){$E_h(\omega_2)$}
\put(60,1.5){$E_h(\omega_1)$}
\put(35.5,1.5){$E_0$}
\put(29,14.5){$\pi$}
\put(28,23.5){$2\pi$}
\put(28,32.5){$3\pi$}
\end{picture}
\vspace{-18pt}
\caption{The parameter space for elliptic minimal surface solutions}
\label{fig:parameters_boundary}
\end{figure}
The above properties are evident in figure \ref{fig:parameters_boundary}, which depicts the dependence of $\delta \varphi$ on $E$ for various values of $\omega$. We define the critical value $\omega_0$, for the $\omega$ parameter as
\begin{equation}
E_h \left( \omega_0 \right) = E_0 ,
\end{equation}
which implies that
\begin{equation}
\omega_0 \simeq 0.458787 .
\end{equation}
Notice that $\delta \varphi$ and $2\pi - \delta \varphi$ correspond to the same entangling curve. Following from the above properties of $\delta \varphi \left( {E,\omega } \right)$, we conclude that for a given entangling curve being characterized by $\omega_1$ and $\delta \varphi_1 \leq \pi$ there are the following possibilities for an elliptic minimal surface
\begin{enumerate}
\item When $\omega_1 > \omega_0 $, $\delta \varphi$ is a monotonous function of $E$ ranging in $\left[ 0 , \pi \right]$. Consequently, for every angle $\delta \varphi_1$ there is a unique $E$, and, thus, a unique minimal surface.
\item When $\omega_1 < \omega_0 $, $\delta \varphi$ is not one-to-one but it is an increasing function of $E$ for $E < E_0$ and a decreasing function of $E$ for $E > E_0$, and, thus, it acquires a maximum value equal to $\delta \varphi _{\max } = \delta \varphi \left( E_0 , \omega_1 \right)$. We may distinguish two cases:
\begin{enumerate}
\item $\pi < \delta \varphi_{\max} < 2 \pi$. In this case, if $\delta \varphi_1 > 2 \pi - \delta \varphi_{\max}$, there will be three distinct values of $E$ corresponding to an appropriate value of $\delta \varphi$ and consequently three distinct minimal surfaces. Let these values be $E_1$, $E_2$ and $E_3$, where $E_1 < E_2 < E_3$, then $\delta \varphi \left( E_1 , \omega_1 \right) = \delta \varphi_1$ and $\delta \varphi \left( E_{2,3} , \omega_1 \right) = 2 \pi - \delta \varphi_1$. The smaller the value of $E$, the more ``confined'' is the appearance of the surface as in the left part of figure \ref{fig:helicoid_PT}. There is only one exception to this rule when $\delta \varphi_1 = \pi$; in this case which there are exactly two minimal surfaces, one being a helicoid. On the other hand, if $\delta \varphi_1 < 2 \pi - \delta \varphi_{\max}$, there will be only one solution with $E < E_0$.
\item $\delta \varphi_{\max} > 2 \pi$. In this case, there are three distinct embedding minimal surfaces for all values of $\delta \varphi_1$ with the same properties as described in the case above. Depending on the value of $\delta \varphi_1$, there may exist more minimal surfaces, coming in pairs, with self-intersections, which correspond to $\delta \varphi = 2 \pi n \pm \delta \varphi_1 > 2 \pi$.
\end{enumerate}
\end{enumerate}

\begin{figure}[ht]
\centering
\begin{picture}(100,37)
\put(15,3){\includegraphics[width = 0.5\textwidth]{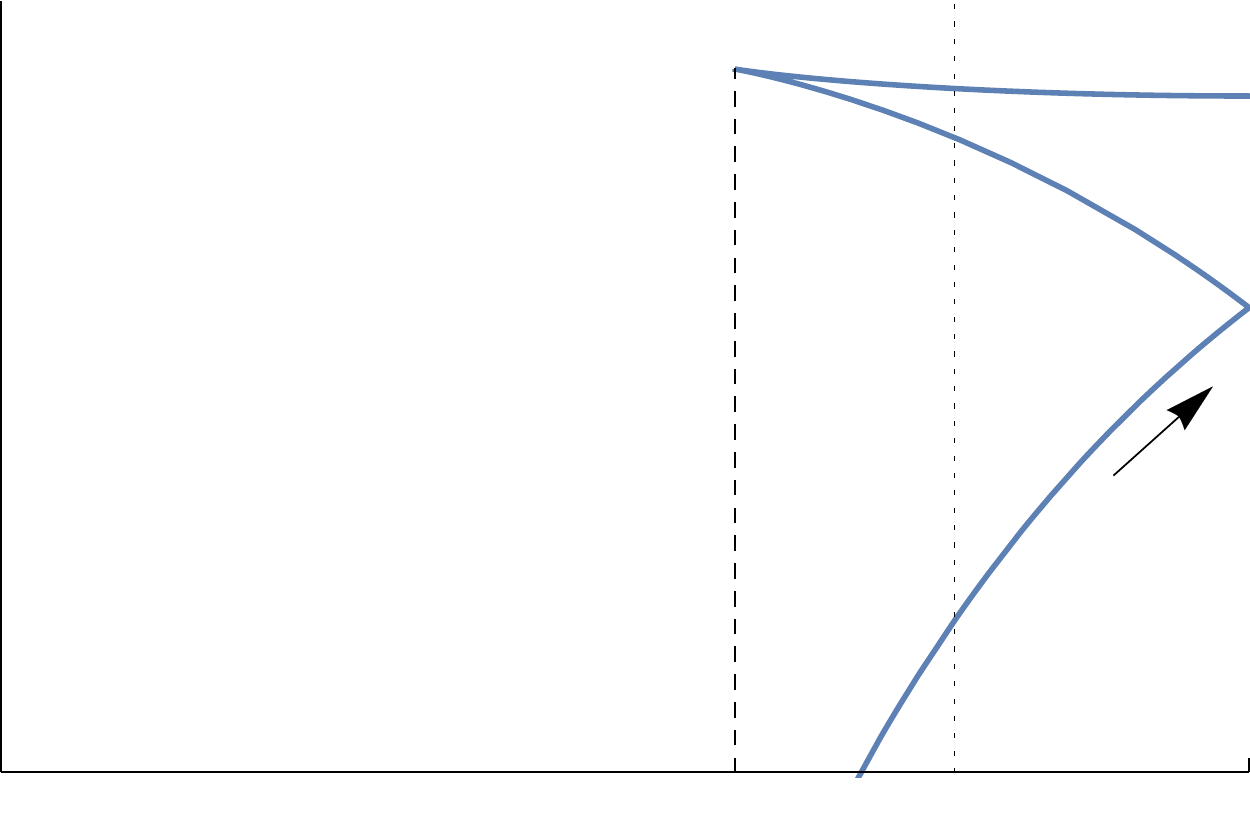}}
\put(66.75,4.25){$\min \left( \delta\varphi , 2 \pi - \delta\varphi \right)$}
\put(14.25,36.25){$a_0$}
\put(38,2){$2 \pi - \delta\varphi_{\max}$}
\put(64,2){$\pi$}
\put(63.5,18){$E$ increases}
\put(35,32){$E = E_0$}
\put(66,31){$E = E_h$}
\end{picture}
\vspace{-20pt}
\caption{The coefficient $a_0$ versus $\min \left( \delta\varphi , 2 \pi - \delta\varphi \right)$ for $\omega = 0.25$. For this value of $\omega$, $\pi < \delta\varphi_{\max} < 2\pi$. Points on a vertical line are surfaces with the same entangling curve.}
\label{fig:general_a0}
\end{figure}
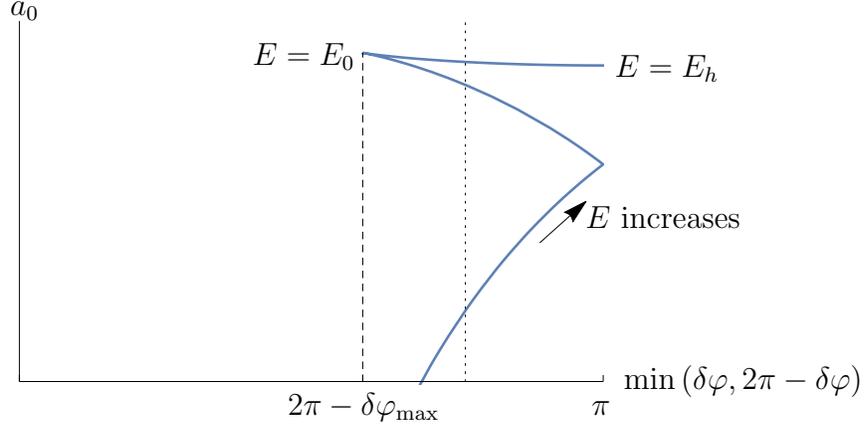

Summing up, if only embedding surfaces are considered, there exist at most three selections for a minimal surface with a given entangling curve. It is quite complicated to compare analytically the parameter $a_0$ for elliptic surfaces with the same entangling curve to find which one is the globally preferred. As shown in figure \ref{fig:general_a0}, it turns out that the globally preferred surface is always the one with the minimum value of $E$, which is the only one having $\delta \varphi < \pi$. 
As the entangling curve separates the boundary to two unequal regions, we could say in a humorous manner that the unstable minimal surfaces have lost their way and wrapped around the wrong region of the boundary.

A consequence of the above is the fact that helicoids with $\omega < \omega_0$ are globally unstable. This bound coincides with the bound for the parameter $\omega$ determined in \cite{Wang_helicoids} for the local stability of a helicoid. From a purely mathematical point of view, in this work we made progress determining $\omega_0$ analytically through equation \eqref{eq:elliptic_E0_analytical}. Moreover, we managed to find the stable minimal surface to which an unstable helicoid collapses.

As members of an associate family of minimal surfaces share the same local stability properties and since we showed that elliptic minimal surfaces with the same value for $E$ belong to such a family, the results of \cite{Wang_helicoids,Wang_catenoids} imply that all elliptic minimal surfaces with $E < E_0$ are locally stable, whereas those with $E > E_0$ are locally unstable. This implies that comparing the three elliptic minimal surfaces with the same entangling curve, the surface with $E=E_1$ is both locally and globally stable, the surface with $E=E_2$ is locally stable but globally unstable, while the surface with $E=E_3$ is both locally and globally unstable.

\subsection{Circular Entangling Curves}
\label{subsec:PT_Catenoids}
\begin{figure}[ht]
\centering
\begin{picture}(100,96)
\put(0,44){\includegraphics[width = 0.5\textwidth]{catenoid_global.jpg}}
\put(50,44){\includegraphics[width = 0.5\textwidth]{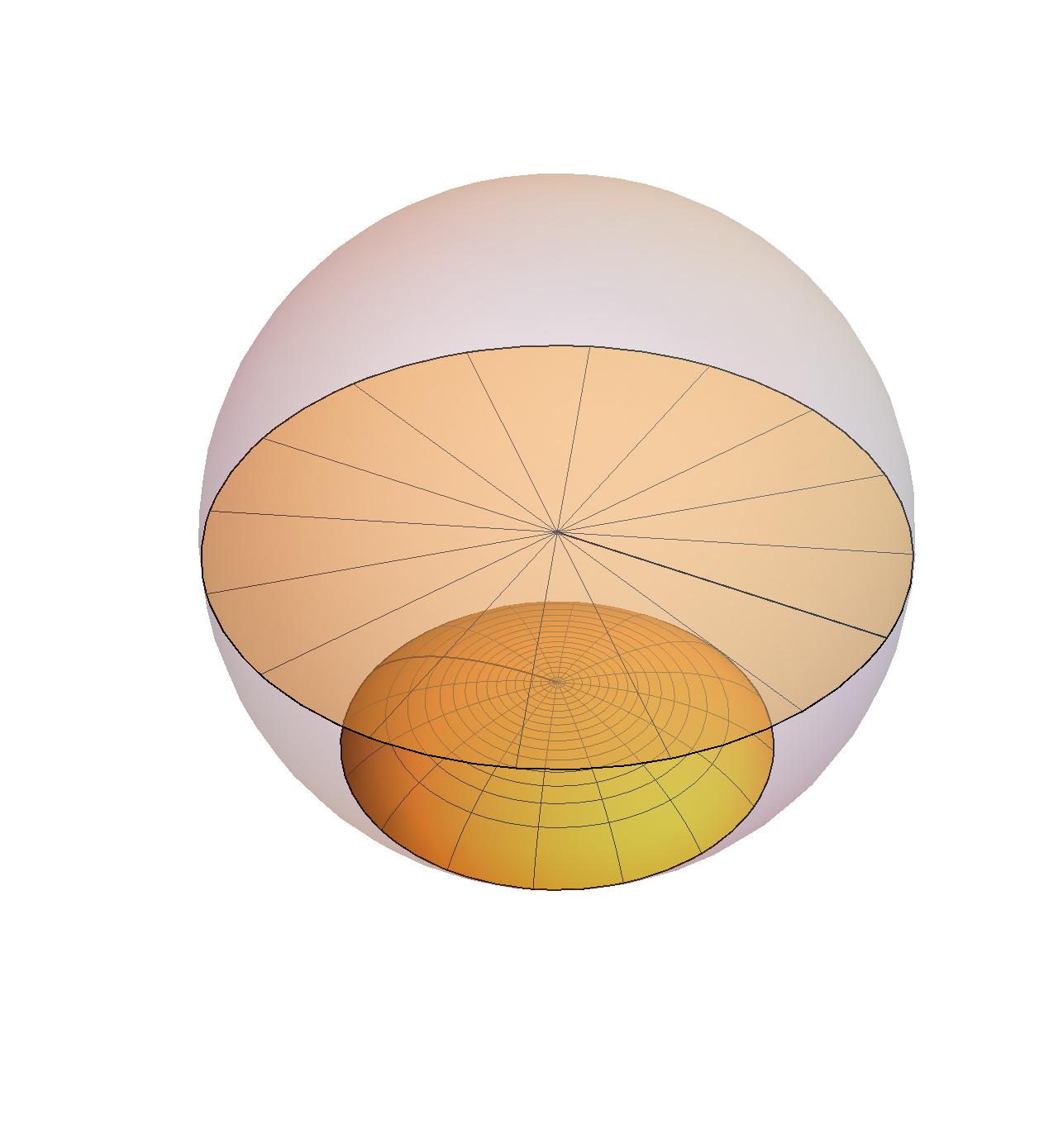}}
\put(0,10){\includegraphics[width = 0.5\textwidth]{catenoid_poincare.jpg}}
\put(50,10){\includegraphics[width = 0.5\textwidth]{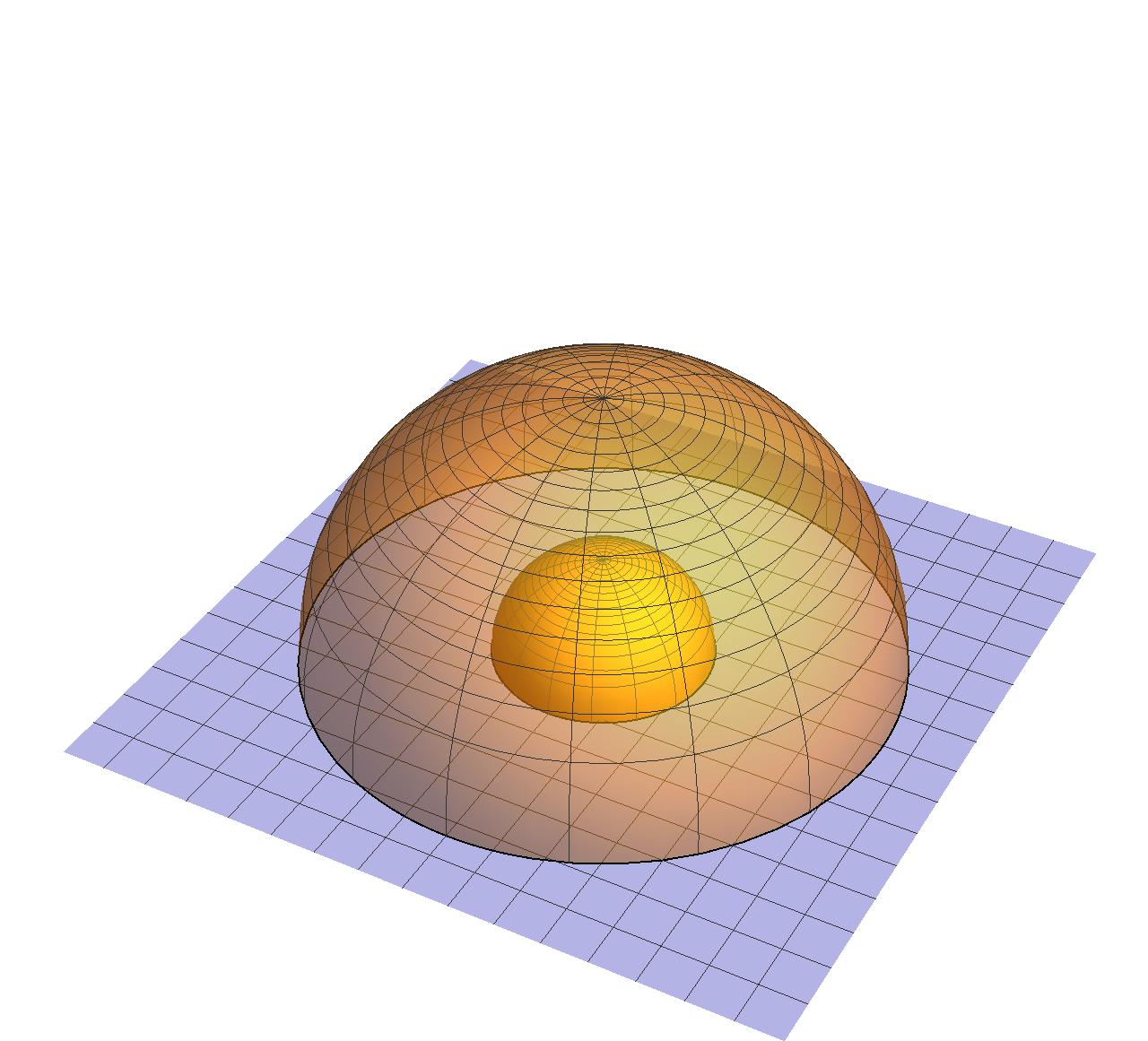}}
\end{picture}
\vspace{-45pt}
\caption{A catenoid and a Goldschmidt minimal surface corresponding to the same boundary conditions}
\label{fig:catenoid_PT}
\end{figure}

For catenoids, the boundary curve consists of two circles, which are parallel to the equator and separate the boundary sphere to three regions. As a result, there is the possibility of a geometric phase transition, between any of the two catenoid minimal surfaces corresponding to the same ratio of boundary circle radii and a Goldschmidt solution being the union of two disjoint surfaces each corresponding to a polar cap region, as those presented in \cite{Bakas:2015opa}. Figure \ref{fig:catenoid_PT} depicts a catenoid and a Goldschmidt minimal surface sharing the same boundary conditions.
Although the background geometry is different, 
the situation is similar to the usual problem of a soap bubble attached to two rings. Searching for a minimal surface in flat space that is anchored to two coaxial circles, there are three options. Two of those are portions of the catenoid and the third option is the Goldschmidt solution being the union of the two disks each being the minimal surface corresponding to a single circle boundary.

We remind that the area of the minimal surface corresponding to a polar cap region is $A = \Lambda L - 2\pi {\Lambda ^2}$ (see for example \cite{Bakas:2015opa}) and consequently, the area for the union of two such surfaces is given by
\begin{equation}
A = \Lambda L - 4\pi {\Lambda ^2} .
\end{equation}

\begin{figure}[ht]
\centering
\begin{picture}(100,37)
\put(25,3){\includegraphics[width = 0.5\textwidth]{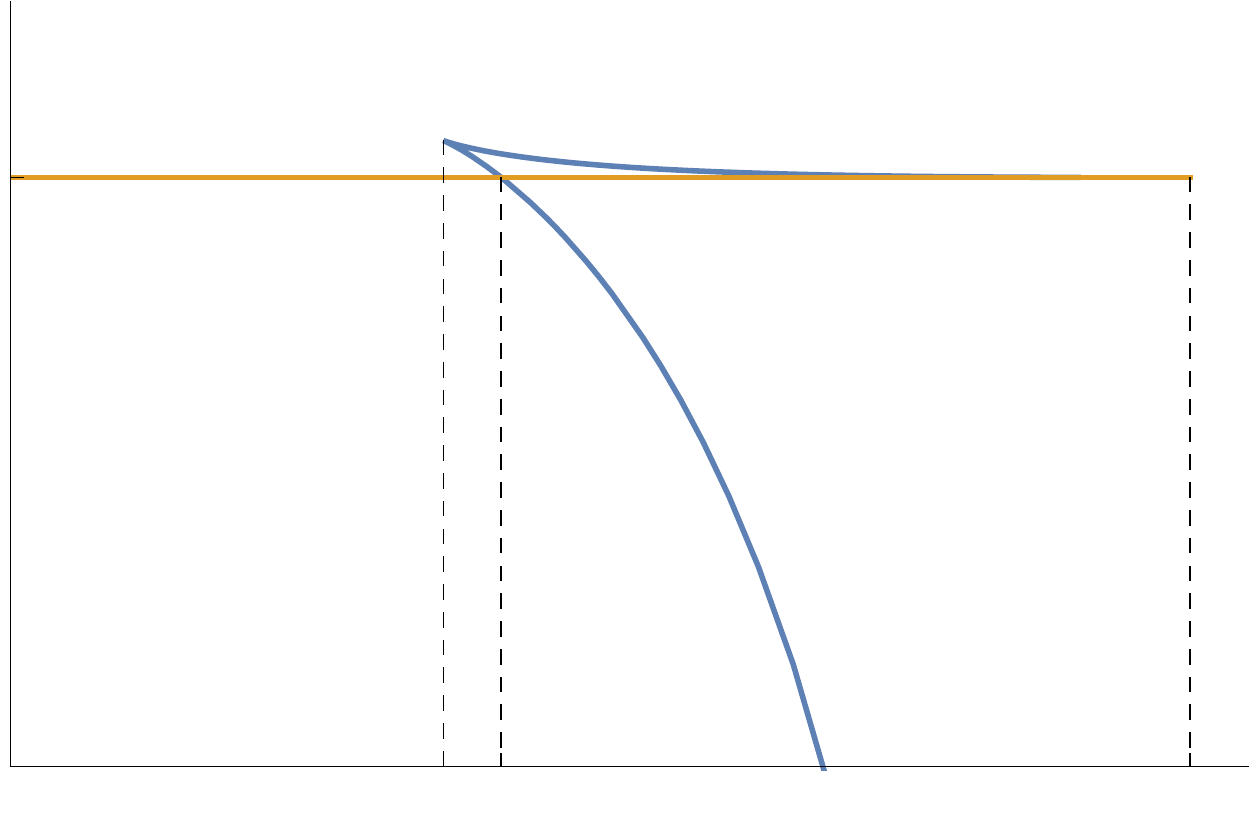}}
\put(75.75,4.25){$\frac{r_-}{r_+}$}
\put(24.25,36.25){$a_0$}
\put(19.5,27.75){$-4\pi$}
\put(36.25,0.5){$\left(\frac{r_-}{r_+}\right)_0$}
\put(44.25,0.5){$\left(\frac{r_-}{r_+}\right)_c$}
\put(72,1.5){1}
\put(54,18){$E < E_0$}
\put(50,30){$E > E_0$}
\end{picture}
\vspace{-20pt}
\caption{The coefficient $a_0^{\rm{catenoid}}$ as function of the ratio of the radii of the boundary circles. The singular point of the curve corresponds to $E = E_0$.}
\label{fig:catenoid_a0}
\end{figure}
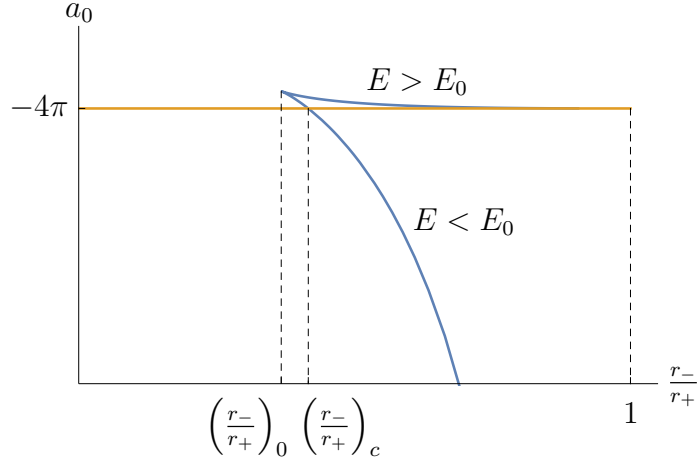

The catenoid is preferred to the two disjoint surfaces when $a_0^{\rm{catenoid}} < - 4 \pi$. This inequality holds when the integration constant $E$ is smaller than the critical value $E_c \simeq 0.760039$ satisfying
\begin{equation}
{\omega _1 \left( E_c \right)}\frac{E_c}{6} + \zeta \left( {{\omega _1 \left( E_c \right)}} \right) = \sqrt {\frac{E_c}{2}}  .
\end{equation}
Consequently, since $E_c < E_0$, when the ratio of the radii of the boundary circles is smaller than the critical value $\left(\frac{r_-}{r_+}\right)_c \simeq 0.416073$, the disjoint surfaces are the preferred solution, whereas, when the ratio of the radii is larger that this critical value, the catenoid corresponding to the smaller value of $E$ for the given ratio is preferred. The catenoid corresponding to the larger value of $E$ is never preferred in comparison to any of the other two options. Figure \ref{fig:catenoid_a0} depicts the dependence of the coefficient $a_0^{\rm{catenoid}}$ on the ratio $\left(\frac{r_-}{r_+}\right)$.

\section{Discussion}
\label{sec:Discussion}

We constructed a family of static minimal surfaces in AdS$_4$  starting from a specific class of solutions of the Pohlmeyer reduced system, namely the Euclidean cosh-Gordon equation. This specific class comprises of solutions depending on only one of the two isothermal coordinates parametrizing the minimal surface. For these solutions, the equations of motion for the embedding functions are reduced to four pairs of effective \Schrodinger problem with opposite eigenvalues, each pair consisting of a flat potential and an $n=1$ \Lame potential. An appropriate ansatz built on one eigenfunction corresponding to the finite band and another corresponding to the finite gap of the \Lame spectrum is shown to satisfy the geometric and Virasoro constraints of the problem, and, thus, provide a family of static minimal surfaces in AdS$_4$.

The family of elliptic minimal surfaces is a two-parameter family, having as special limits the helicoids, catenoids and conical minimal surfaces in H$^3$. This two-parameter family of solutions can be divided to one-parameter families each containing a single helicoid surface and either a catenoid or a conical surface and having the following properties: they are associate families of minimal surfaces and furthermore all their members correspond to a unique solution of the Pohlmeyer reduced system.

The general minimal surface corresponds to an entangling curve in the boundary being the union of two logarithmic spirals one being the rotation of the other by a given angle. It is shown that in general there may exist more than one elliptic minimal surfaces corresponding to the same boundary conditions, allowing geometric phase transitions between them. Conditions for the global and local stability of an elliptic minimal surface are derived. Interestingly, the relevant critical values of the surface parameters are connected to the energy that a point particle moving in one dimension under the influence of a hyperbolic sine potential must have so that its ``time of flight'' is maximum.

The constructed surfaces, being co-dimension two minimal surfaces in AdS$_4$, have particular interest in the framework of holographic duality, since their area is connected to the entanglement entropy in the boundary CFT through the Ryu-Takayanagi conjecture. Unlike the minimal surfaces typically used in the literature, namely those corresponding to a disk or an infinite strip region in the boundary, these surfaces are anchored to entangling curves characterized by non-trivial curvature. As such, they can provide a useful tool in the study of the relation between entanglement entropy and the geometric characteristics of the entangling curve \cite{Bueno:2015xda,Solodukhin:2008dh}. Furthermore, the geometric phase transitions discovered between different minimal surfaces can provide some light in the role of entanglement entropy as an order parameter in confinement/deconfinement phase transitions.

An important result in the program of holographic entanglement entropy is the equivalence of the first law of entanglement thermodynamics to Einstein equations at linear order \cite{Lashkari:2013koa,Faulkner:2013ica}. However, these results are based on calculations using ``semi-spherical'' minimal surfaces corresponding to spherical entangling curves, which are special in two ways: First, the entangling curve has constant curvature. Second, the minimal surface does not just have vanishing mean curvature, but both principal curvatures vanish; they are the analogue of a plane in hyperbolic space. Since the holographic calculation of the variations of entanglement entropy strongly depends on the geometric characteristics of the minimal surface, verification of this results making use of elliptic minimal surfaces will greatly support the idea of gravity being an emergent entropic force related to quantum entanglement statistics. Furthermore, such calculations are interesting in terms of the stress-energy/Cotton tensor duality appearing in AdS$_4$ metric perturbations and the appropriate prescription that has to be attached to Ryu-Tanayanagi conjecture, so that it is valid in the case of perturbations obeying non-Dirichlet boundary conditions \cite{Bakas:2015opa}. 

The presented techniques are generalizable to higher dimensions, where Pohlmeyer reduction results in multi-component integrable systems of the sinh-Gordon family. Unfortunately, such minimal surfaces will not be co-dimension two surfaces and consequently will not be interesting in the context of Ryu-Takayanagi conjecture, but only from a more mathematical point of view. On the contrary, generalizations of the constructed elliptic minimal surfaces in AdS$_4$ involving more general linear combinations of the $n=1$ \Lame eigenfunctions can lead to the construction of minimal surfaces with interesting geometric characteristics and potential applications in holographic entanglement entropy.

\acknowledgments
This work is dedicated to the memory of Prof. Ioannis Bakas, whose sudden loss filled with grief the scientific community. It is the outcome of uncountable hours of enjoyable discussions with him on the connection between geometry and physics and continues on the path of our previous joint work.
 
A preliminary account of the results was presented at the ``Workshop on Geometry and Physics'' held in Ringberg Castle, Germany, 20-25 November 2016. I thank the organizer Dieter L\"{u}st for his kind invitation and the participants for fruitful discussions. I would also like to thank M. Axenides and E. Floratos for useful discussions.

\appendix

\section{Useful Formulas for the Weierstrass Functions}
\label{sec:Weierstrass_functions}

The Weierstrass function $\wp$ is an elliptic (doubly periodic) function of one complex variable which satisfies the equation
\begin{equation}
{\left( {\frac{{d\wp }}{{dz}}} \right)^2} = 4{\wp ^3} - {g_2}\wp  - {g_3} .
\label{eq:Weierstrass_p_equation}
\end{equation}

The periods of $\wp$ are connected with the roots of the cubic polynomial
\begin{equation}
Q \left( y \right) = 4 y^3 - g_2 y -g_3.
\label{eq:elliptic_wp_cubic_polynomial}
\end{equation}
Let the three roots be $e_1$, $e_2$ and $e_3$. The absence of a quadratic term implies that the roots satisfy $e_1 + e_2 + e_3 = 0$. In the following we concentrate in the case all three roots are real. We order the roots so that $e_1 > e_2 > e_3$. Then, the fundamental periods of the function $\wp$ are a real one $2 \omega_1$ and an imaginary one $2 \omega_2$ which are related to the roots as follows,
\begin{equation}
{\omega _1} = \frac{{K\left( k \right)}}{{\sqrt {{e_1} - {e_3}} }},\quad {\omega _2} = \frac{{iK\left( {k'} \right)}}{{\sqrt {{e_1} - {e_3}} }},
\label{eq:elliptic_periods_D_pos}
\end{equation}
where $K\left( k \right)$ is the complete elliptic integral of the first kind and
\begin{equation}
k^2 = \frac{{{e_2} - {e_3}}}{{{e_1} - {e_3}}},\quad {k'}^2 = \frac{{{e_1} - {e_2}}}{{{e_1} - {e_3}}},\quad {k^2} + k{'^2} = 1 .
\label{eq:elliptic_moluli_D_pos}
\end{equation}


The Weierstrass function $\wp$ obeys the half-period relations,
\begin{equation}
\wp \left( {{\omega _1}} \right) = {e_1},\quad \wp \left( {{\omega _2}} \right) = {e_3},\quad \wp \left( {\omega _3} \right) = {e_2},
\label{eq:elliptic_half_period}
\end{equation}
where $\omega_3 := \omega_1 + \omega_2$.


As long as all three roots are real, equation \eqref{eq:Weierstrass_p_equation} has two real solutions in the real domain, 
\begin{align}
y_1 \left( x \right) &= \wp \left( x \right) ,\label{eq:elliptic_y_solution_unbound}\\
y_2 \left( x \right) &= \wp \left( x + \omega_2 \right) ,\label{eq:elliptic_y_solution_bound}
\end{align}
the first one being unbounded and ranging between $e_1$ and $+ \infty$ and the second one being bounded and ranging between $e_3$ and $e_2$. In the opposite case, there is only one unbounded real solution in the real domain.

The Weierstrass $\zeta$ function is a doubly quasi-periodic function, which is defined so that
\begin{equation}
\frac{{d\zeta }}{{dz}} = - \wp .
\label{eq:Weierstrass_zeta}
\end{equation}
Finally, the Weierstrass $\sigma$ function obeys the defining relation
\begin{equation}
\frac{1}{\sigma }\frac{{d\sigma }}{{dz}}= \zeta .
\label{eq:Weierstrass_sigma}
\end{equation}

The Weierstrass elliptic function $\wp$ is an even function of $z$, while Weierstrass functions $\zeta$ and $\sigma$ are odd functions of $z$,
\begin{align}
\wp \left( { - z} \right) &= \wp \left( z \right) ,\\
\zeta \left( { - z} \right) &=  - \zeta \left( z \right) ,\\
\sigma \left( { - z} \right) &=  - \sigma \left( z \right) .
\end{align}

As mentioned above, the functions $\zeta$ and $\sigma$ are not periodic. Under a shift of the complex variable $z$ in the lattice defined by the periods of $\wp$, they transform as
\begin{align}
\wp \left( {z + 2m{\omega _1} + 2n{\omega _2}} \right) &= \wp \left( z \right) ,\label{eq:Weierstrass_period_wp}\\
\zeta \left( {z + 2m{\omega _1} + 2n{\omega _2}} \right) &= \zeta \left( z \right) + 2m\zeta \left( {{\omega _1}} \right) + 2n\zeta \left( {{\omega _2}} \right) ,\label{eq:Weierstrass_period_zeta}\\
\sigma \left( {z + 2m{\omega _1} + 2n{\omega _2}} \right) &= {\left( { - 1} \right)^{m + n + mn}}{e^{\left( {2m\zeta \left( {{\omega _1}} \right) + 2n\zeta \left( {{\omega _2}} \right)} \right)\left( {z + m{\omega _1} + n{\omega _2}} \right)}} \sigma \left( {z} \right) .\label{eq:Weierstrass_period_sigma}
\end{align}
The quantities $\zeta \left( {{\omega _1}} \right)$ and $\zeta \left( {{\omega _2}} \right)$ obey the non-trivial relation
\begin{equation}
\omega _2 \zeta \left( {{\omega _1}} \right) - \omega _1 \zeta \left( {{\omega _2}} \right) = i\frac{\pi}{2}.
\label{eq:Weierstrass_zeta_special}
\end{equation}

The Weierstrass functions obey the homogeneity relations,
\begin{align}
\wp \left( {z;{g_2},{g_3}} \right) &= {\mu ^2}\wp \left( {\mu z;\frac{{{g_2}}}{{{\mu ^4}}},\frac{{{g_3}}}{{{\mu ^6}}}} \right) ,\label{eq:Weierstras_homogeneity_wp}\\
\zeta \left( {z;{g_2},{g_3}} \right) &= \mu \zeta \left( {\mu z;\frac{{{g_2}}}{{{\mu ^4}}},\frac{{{g_3}}}{{{\mu ^6}}}} \right) ,\\
\sigma \left( {z;{g_2},{g_3}} \right) &= \frac{1}{\mu }\sigma \left( {\mu z;\frac{{{g_2}}}{{{\mu ^4}}},\frac{{{g_3}}}{{{\mu ^6}}}} \right).
\end{align}
Choosing $\mu = i$, the homogeneity relations yield
\begin{align}
\wp \left( {z;{g_2},{g_3}} \right) &=  - \wp \left( {iz;{g_2}, - {g_3}} \right) ,\\
\zeta \left( {z;{g_2},{g_3}} \right) &= i\zeta \left( {iz;{g_2}, - {g_3}} \right) ,\\
\sigma \left( {z;{g_2},{g_3}} \right) &=  - i\sigma \left( {iz;{g_2}, - {g_3}} \right) ,
\end{align}
which imply that on the imaginary axis of the $z$ plane, $\wp$ is real, whereas $\zeta$ and $\sigma$ are imaginary.

As an elliptic function, $\wp$ possesses an addition formula. The functions $\zeta$ and $\sigma$ are not elliptic, however, they also possess similar properties,
\begin{align}
\wp \left( {z + w} \right) &=  - \wp \left( z \right) - \wp \left( w \right) + \frac{1}{4}{\left( {\frac{{\wp '\left( z \right) - \wp '\left( w \right)}}{{\wp \left( z \right) - \wp \left( w \right)}}} \right)^2} , \label{eq:Weierstrass_addition_wp}\\
\zeta \left( {z + w} \right) &= \zeta \left( z \right) + \zeta \left( w \right) + \frac{1}{2}\frac{{\wp '\left( z \right) - \wp '\left( w \right)}}{{\wp \left( z \right) - \wp \left( w \right)}} , \label{eq:Weierstrass_addition_zeta}\\
\wp \left( z \right) - \wp \left( w \right) &= - \frac{{\sigma \left( {z - w} \right)\sigma \left( {z + w} \right)}}{{{\sigma ^2}\left( z \right){\sigma ^2}\left( w \right)}}. \label{eq:Weierstrass_addition_sigma}
\end{align}
Applying the last formula in the special case $w$ coincides with any of the half-periods yields,
\begin{equation}
\wp \left( z \right) - {e_{1,3,2}} =  - \frac{{\sigma \left( {z + {\omega _{1,2,3} }} \right)\sigma \left( {z - {\omega _{1,2,3} }} \right)}}{{{\sigma ^2}\left( z \right){\sigma ^2}\left( {{\omega _{1,2,3} }} \right)}}.
\label{eq:Weierstrass_addition_sigma_special}
\end{equation}

Finally, the Weierstrass functions obey the following integral formula
\begin{equation}
\wp '\left( a \right)\int {\frac{{dz}}{{\wp \left( z \right) - \wp \left( a \right)}} = 2\zeta \left( a \right)z + \ln \frac{{\sigma \left( {z - a} \right)}}{{\sigma \left( {z + a} \right)}}} .
\label{eq:Weierstrass_integral}
\end{equation}

\end{document}